\documentclass[acmsmall, nonacm, screen]{acmart}
\setcopyright{none}
\usepackage[utf8]{inputenc}
\usepackage{amsfonts}
\usepackage{braket,microtype,amsmath,mathtools,graphicx,enumitem,booktabs,bm,xspace,float,mathdots,mleftright}
\usepackage{amsthm}
\usepackage{stmaryrd}
\usepackage{enumitem} %
\usepackage{textcomp}
\usepackage{xcolor}
\usepackage{multicol}
\usepackage{ifthen}
\usepackage{makecell}
\usepackage{wrapfig}
\usepackage{subcaption}
\hypersetup{hypertexnames=false}
\usepackage{tabularx}
\usepackage[T1]{fontenc}
\usepackage{textcomp}
\usepackage{colortbl}
\usepackage[english]{babel}
\usepackage{thm-restate}
\usepackage{quantikz}
\usepackage[breakable]{tcolorbox}
\usepackage{enumitem}
\usepackage{crossreftools}
\usepackage{semantic}
\usepackage{mathpartir}
\usepackage{dsfont}
\usepackage{array}
\usepackage{ragged2e}

\allowdisplaybreaks

\DeclareMathOperator{\tr}{tr}
\DeclareMathOperator{\arcosh}{arcosh}

\DeclarePairedDelimiter\abs{\lvert}{\rvert}
\DeclarePairedDelimiter\norm{\lVert}{\rVert}
\DeclarePairedDelimiter\floor{\lfloor}{\rfloor}
\DeclarePairedDelimiter\ceil{\lceil}{\rceil}
\DeclarePairedDelimiter\parens{\lparen}{\rparen}

\DeclarePairedDelimiter\bracks{\lbrack}{\rbrack}

\DeclarePairedDelimiter\Bracks{\lBrack}{\rBrack}
\DeclarePairedDelimiter\tuple{\langle}{\rangle}

\newcommand{\ketbra}[2]{\ket{#1} \!\! \bra{#2}}
\renewcommand{\proj}[1]{\ket{#1} \!\! \bra{#1}}
\newcommand{\ot}{\otimes}

\newcommand{\eps}{\varepsilon}

\newcommand{\cH}{\mathcal{H}}

\newcommand{\cE}{\mathcal{E}}
\newcommand{\cF}{\mathcal{F}}
\newcommand{\cP}{\mathcal{P}}
\newcommand{\cM}{\mathcal{M}}

\newcommand{\cI}{\mathcal{I}}
\newcommand{\cC}{\mathcal{C}}
\newcommand{\cS}{\mathcal{S}}

\newcommand{\cU}{\mathcal{U}}

\newcommand{\cL}{\mathcal{L}}
\newcommand{\cQ}{\mathcal{Q}}

\newcommand{\R}{\mathbb{R}}
\newcommand{\C}{\mathbb{C}}

\newcommand{\N}{\mathbb{N}}

\newcommand{\bigO}{\mathcal{O}}

\newcommand{\veca}{{\vec{a}}}
\newcommand{\vecq}{{\vec{q}}}
\newcommand{\vecr}{{\vec{r}}}

\newcommand{\vecv}{{\vec{v}}}
\newcommand{\vecx}{{\vec{x}}}
\newcommand{\vecy}{{\vec{y}}}
\newcommand{\vecz}{{\vec{z}}}

\newenvironment{proofcase}[1]{\medskip\emph{Case} #1:}{}

\usepackage{listings}
\lstdefinelanguage{cqpl}{
  classoffset=0,
  morekeywords={
    declare,ext,proc,uproc,locals,
    do,end,
    skip,call,if,then,repeat,for,in,with,
    call\_uproc\_and\_meas,meas,
    not,and,or
  },
  keywordstyle=\color{blue}\bfseries\ttfamily,
  classoffset=1,
  morekeywords={Bool,Fin},
  keywordstyle=\color{blue}\bfseries\ttfamily,
  classoffset=0,
  sensitive=true,
  morecomment=[l]{//},
  morecomment=[n]{/*}{*/},
  commentstyle=\color{gray}\ttfamily,
  morestring=[b]",
  columns=fullflexible,
  keepspaces=true,
  basicstyle=\footnotesize\ttfamily,
  numberstyle=\color{gray}\sffamily\tiny,
  mathescape=true,
  alsoletter={'},
}
\lstdefinestyle{haskellstyle}{
  language=Haskell,
  backgroundcolor=\color{black!0},
  commentstyle=\color{green!40!black},
  keywordstyle=\color{blue},
  stringstyle=\color{purple},
  basicstyle=\ttfamily\footnotesize,
  breakatwhitespace=false,
  breaklines=true,
  captionpos=b,
  keepspaces=true,
  showspaces=false,
  showstringspaces=false,
  showtabs=false,
  tabsize=2,
  mathescape=true,
}
\lstdefinestyle{nonum}{
    numbers=none,
}
\lstdefinestyle{tiny}{
    basicstyle=\tiny\ttfamily,
}

\usepackage[capitalize,nameinlink,noabbrev]{cleveref}
\numberwithin{equation}{section}

\newtheorem{theorem}{Theorem}
\newtheorem*{theorem*}{Theorem}

\newtheorem{problem}{Problem}[]
\newtheorem*{problem*}{Problem}
\crefname{problem}{Problem}{Problems}

\newtheorem*{example*}{Example}
\crefname{example}{Example}{Examples}

\newtheorem*{matrixsearchexample*}{Matrix Search Problem}
\crefname{matrixsearchexample}{Matrix Search Problem}{}
\newtheorem*{searchexample*}{Search Problem}
\crefname{searchexample}{Search Problem}{}
\newtheorem*{costtheorem*}{Cost Theorem (simplified)}
\crefname{costtheorem}{Cost Theorem (simplified)}{}

\makeatletter
\AddToHook{cmd/appendix/before}{%
\crefalias{section}{appendix}%
\crefalias{subsection}{appendix}%
\crefalias{subsubsection}{appendix}%
}
\makeatother

\newcommand{\Vars}{\textsf{Vars}}
\newcommand{\FreeVars}{\textsf{fv}}
\newcommand{\Vals}{\textsf{Vals}}

\newcommand{\aux}{\textsf{aux}}

\newcommand{\Distr}[1]{\textsf{Distr}(#1)}

\newcommand{\detstate}[1]{\ensuremath{\mathds{1}_{#1}}}
\newcommand{\DistrExp}[2]{\mathbb{E}_{#1}\bracks*{#2}}
\newcommand{\Prob}[2]{\text{Pr}_{#1}(#2)}
\newcommand{\TVDist}[2]{\textsf{TV}\parens*{#1, #2}} 
\newcommand{\probDistance}[2]{\Delta\parens*{#1, #2}}

\newcommand{\NotationBox}[2][]{%
\framebox{#2}
\hfill
\ifstrempty{#1}{}{\textit{(#1)}}
}
\newcommand{\NotationSize}{\footnotesize}

\newcommand{\OurFramework}{\texorpdfstring{\textsc{\textsf{Traq}}}{Traq}}

\newcommand{\ProtoLang}{\textsc{\textsf{Cpl}}}
\newcommand{\CQPL}{\textsc{\textsf{Qpl}}}

\newcommand{\UnitaryCompiler}{\ensuremath{\cU}}
\newcommand{\QuantumCompiler}{\ensuremath{\cQ}}

\newcommand{\CostCompiler}{\QuantumCompiler{}}
\newcommand{\UCostCompiler}{\UnitaryCompiler{}}

\newcommand{\CostName}{\textsc{\textsf{Havoc}}}
\newcommand{\ExpCostName}{\textsc{\textsf{ExpCost}}}
\newcommand{\UCostName}{\textsc{\textsf{Cost}}}

\newcommand{\CostMetricQ}{\ensuremath{\widehat{\ExpCostName}^{\QuantumCompiler{}}}}
\newcommand{\CostMetricHavoc}{\ensuremath{\widehat{\CostName{}}^{\QuantumCompiler{}}}}
\newcommand{\CostMetricU}{\ensuremath{\widehat{\textsc{\textsf{Cost}}}^{\UCostCompiler}}}

\newcommand{\CostExpr}{\cC}

\newcommand{\CQPLCost}{\ExpCostName}
\newcommand{\UQPLCost}{\UCostName}

\newcommand{\kwbluebf}[1]{\texttt{\textbf{\textcolor{blue}{#1}}}}

\newcommand{\kw}[1]{\kwbluebf{#1}} %
\newcommand{\PrimName}[1]{\kwbluebf{#1}} %
\newcommand{\kwthin}[1]{\kwbluebf{#1}} %
\newcommand{\kwbasic}[1]{\kwbluebf{#1}} %

\newcommand{\SetOfType}{\textsf{Type}}
\newcommand{\SetOfVals}{\textsf{Vals}}

\newcommand{\SetOfDistrExpr}{\textsf{DExpr}}
\newcommand{\SetOfExpr}{\textsf{Expr}}
\newcommand{\SetOfPartiallyAppliedFuns}{\textsf{PAExpr}}

\newcommand{\SetOfStmt}{\textsf{Stmt}}
\newcommand{\SetOfFunc}{\textsf{Funs}}

\newcommand{\SetOfUnitaryOps}{\textsf{UOps}}
\newcommand{\SetOfUnitaryCommands}{\textsf{UStmt}}
\newcommand{\SetOfProbCommands}{\textsf{PStmt}}
\newcommand{\SetOfProcedures}{\textsf{Proc}}

\newcommand{\BlankArg}{\_}
\newcommand{\is}{\leftarrow}
\newcommand{\rndarrow}{\ensuremath{<-\!\!\$}}

\newcommand{\Bool}{\kwthin{Bool}}
\newcommand{\Fin}[1]{\kwthin{Fin}\langle#1\rangle}
\newcommand{\BitVec}[1]{\kwthin{BoolVec}\langle#1\rangle}
\newcommand{\Arr}[2]{\kwthin{Vec}\langle#1, #2\rangle} %
\newcommand{\DistrType}[1]{#1} %

\newcommand{\BasicOp}[1]{\text{op}_{#1}}

\newcommand{\NotOp}{\kwbasic{not}}
\newcommand{\AndOp}{\texttt{\&\&}}
\newcommand{\OrOp}{\texttt{||}}

\newcommand{\XorOp}{\oplus}

\newcommand{\Bernoulli}{\kwbasic{bern}}
\newcommand{\Uniform}{\kwbasic{unif}}

\newcommand{\ArrIndex}[2]{#1.#2} %
\newcommand{\ArrUpdate}[3]{#1[#3 / #2]} %

\newcommand{\protosample}[2]{#1 \rndarrow~ #2} %
\newcommand{\protoif}[3]{\kwbasic{if}~ #1 ~\{ #2 \} ~\kwbasic{else}~ \{ #3 \}}

\newcommand{\protodef}[5]{\kwbasic{fn}~ #1 (#2) ~\kwbasic{do}~ #4 ; ~\kwbasic{return}~ #5 ~\kwbasic{end}}
\newcommand{\protoext}[1]{\kwbasic{ext fn}~ #1}

\newcommand{\Primitive}{\cP}
\newcommand{\GenericPrimitiveCall}{\cP_{\eps}[\vec{\lambda}]}
\newcommand{\primOneFun}[3]{#1} %

\newcommand{\wellTypedStmt}[3]{\vdash #3} %
\newcommand{\wellTypedExpr}[3]{\vdash #2 : #3} %
\newcommand{\wellTypedFun}[4]{\vdash #2 : #3 -> #4} %

\newcommand{\InterpCtx}{\hat{F}}
\newcommand{\Domain}[1]{\Bracks{#1}} %
\newcommand{\evalExprDet}[1]{\Bracks{#1}} %
\newcommand{\evalProb}[1]{\Bracks{#1}} %

\newcommand{\TickOf}[2][]{\ensuremath{\{ #2\ifstrempty{#1}{}{(#1)} \mapsto 1 \}}}
\newcommand{\QuantumTick}[1]{\ensuremath{\{ #1^U \mapsto 1 \}}} %
\newcommand{\ClassicalTick}[2][]{\ensuremath{\{ #2\ifstrempty{#1}{}{(#1)} \mapsto 1 \}}} %

\newcommand{\costu}[1]{\CostMetricU\bracks*{#1}} %
\newcommand{\costq}[2]{\CostMetricQ\bracks*{#1}\ifstrempty{#2}{}{(#2)}} %
\newcommand{\costqmax}[1]{\CostMetricHavoc\bracks*{#1}} %

\newcommand{\QueryCost}{\textsc{\textsf{Query}}}
\newcommand{\ExpQueryCost}{\textsc{\textsf{EQuery}}}
\newcommand{\SemParFuns}{\cS}

\newcommand{\UAlgPrimQueries}[3]{\QueryCost{}^{\cU}_{\cU#1_{#2}\ifstrempty{#3}{}{, #3}}}
\newcommand{\QAlgPrimQueriesU}[3]{\QueryCost{}^{\cU}_{\cQ#1_{#2}\ifstrempty{#3}{}{, #3}}}
\newcommand{\QAlgPrimQueriesQ}[3]{\QueryCost{}^{\cQ}_{\cQ#1_{#2}\ifstrempty{#3}{}{, #3}}}
\newcommand{\QAlgPrimQueriesExpU}[4]{\ExpQueryCost{}^{\cU}_{\cQ#1_{#2}\ifstrempty{#3}{}{, #3}}(\ifstrempty{#4}{}{#4})}
\newcommand{\QAlgPrimQueriesExpQ}[4]{\ExpQueryCost{}^{\cQ}_{\cQ#1_{#2}\ifstrempty{#3}{}{, #3}}(\ifstrempty{#4}{}{#4})}

\newcommand{\ErrProb}{\ensuremath{\widehat{\textsf{err}}^{\QuantumCompiler{}}}}
\newcommand{\ErrProbU}{\ensuremath{\widehat{\textsf{err}}^{\UnitaryCompiler{}}}}
\newcommand{\progerrprob}[1]{\ErrProb[#1]} %
\newcommand{\progerrprobU}[1]{\ErrProbU[#1]} %

\newcommand{\qpskip}{\kwbasic{skip}}

\newcommand{\SWAPGate}{\operatorname{SWAP}}
\newcommand{\COPY}{\operatorname{COPY}}

\newcommand{\CSWAPGate}{\operatorname{CSWAP}}
\newcommand{\PhaseOnZero}[2]{\ensuremath{\texttt{PhaseOnZero}}(#2)} %
\newcommand{\adjU}[1]{\texttt{Adj-}#1}
\newcommand{\ctrlU}[1]{\texttt{Ctrl-}#1}

\newcommand{\qpcallu}[2]{\kwbasic{call}~ #1(#2)}
\newcommand{\qpcalldagger}[2]{\kwbasic{call}^\dagger~ #1(#2)}
\newcommand{\qpunitary}[2]{#1 ~\texttt{*=}~ #2}

\newcommand{\CallUProcAndMeas}{\kwbasic{meas}} %
\newcommand{\qpcallandmeas}[2]{\CallUProcAndMeas{}~ #1(#2)}

\newcommand{\qpassign}[2]{#1 := #2}
\newcommand{\qprandom}[2]{#1 :={\!\!\$}~ #2}
\newcommand{\qpcall}[2]{\kwbasic{call}~ #1(#2)}

\newcommand{\qpif}[2]{\kwbasic{if}~ #1 ~ \{ ~ #2 ~ \}}
\newcommand{\qpifte}[3]{\kwbasic{if}~ #1 ~ \{ ~ #2 ~ \} ~\kwbasic{else}~ \{~ #3 ~\}}

\newcommand{\uqplprocdef}[3]{\kwbasic{uproc}~#1 (#2) \{ #3 \}}
\newcommand{\cqplprocdef}[4]{\kwbasic{proc}~#1 (#2) \{  #4 \}}
\newcommand{\extuproc}[1]{\kwbasic{ext uproc}~ #1}
\newcommand{\extcproc}[1]{\kwbasic{ext proc}~ #1}

\newcommand{\wellTypedQPL}[3]{\vdash #3} %

\newcommand{\UInterpCtx}{\hat{U}}
\newcommand{\CInterpCtx}{\hat{H}}

\newcommand{\evalQPL}[2][]{{\left[\mspace{-5mu}\left[\mspace{-5mu}\left[{#2}\right]\mspace{-4.5mu}\right]\mspace{-4.5mu}\right]^{#1}}} %

\newcommand{\evalCQPL}[1]{\evalQPL{#1}} %
\newcommand{\evalUQPL}[1]{\evalQPL[\mathsf{U}]{#1}} %
\newcommand{\evalUnitaryOp}[2]{\evalQPL[\textsf{U}]{#1}} %

\newcommand{\UnitaryEmbed}{\textsf{Utry}}
\newcommand{\unitaryEmbed}[1]{\UnitaryEmbed[#1]}

\newcommand{\quantChan}[1]{\cE_{#1}}

\newcommand{\unitaryDistance}[3]{\Delta^\textsf{U}_{#1}\parens*{#2,#3}}

\newcommand{\unitaryApproxEmbed}[5]{\unitaryDistance{#2;#3}{#1}{\unitaryEmbed{#5} \ot I_{#3}}
{\expandafter\ifx\expandafter\relax\detokenize{#4}\relax\else \le #4 \fi}%
}
\newcommand{\unitaryApproxEmbedStrong}[4]{\unitaryDistance{#2}{#1}{\unitaryEmbed{#4} \ot I_{#2}}
{\expandafter\ifx\expandafter\relax\detokenize{#3}\relax\else \le #3 \fi}%
}

\newcommand{\uqplcost}[1]{\UQPLCost{}[#1]} %

\newcommand{\cqplexpcost}[1]{\CQPLCost[#1]} %

\newcommand{\FreshVars}[1]{\textsf{fresh}(#1)}

\newcommand{\compileUQPL}[1]{\UnitaryCompiler[#1]} %

\newcommand{\compileCQPL}[1]{\QuantumCompiler[#1]} %

\newcommand{\UAlgPrim}[2][{\cP}]{\cU#1_{#2}}
\newcommand{\QAlgPrim}[2][{\cP}]{\cQ#1_{#2}}

\newcommand{\PrimAny}{\kw{any}}
\newcommand{\PrimAll}{\kw{all}}

\newcommand{\PrimCAny}{\ensuremath{\kw{any}_{\texttt{det}}}}
\newcommand{\PrimRAny}{\ensuremath{\kw{any}_{\texttt{rand}}}}

\newcommand{\PrimSearch}{\kw{search}}

\newcommand{\QSearch}{\textbf{QSearch}} %

\newcommand{\PrimSimon}{\kw{simon}}
\newcommand{\pcoll}{{p_\text{coll}}}

\newcommand{\SimonQueries}[3]{Q_\textbf{simon}(#1, #2, #3)} %

\newcommand{\PrimAmplify}{\kw{amplify}}
\newcommand{\pmin}{p_\text{min}}
\newcommand{\pgood}{p_\text{good}}

\newcommand{\protoamplify}[2][\PrimAmplify]{#1_{\pmin}[#2]} %

\newcommand{\ProtoLangSyntax}{
\[
  \begin{array}{rllrll}
    \SetOfType \ni \tau &::=&
    \Fin{N}
    ~|~ \Arr{n}{\tau}
    ~|~ \Bool
    ~|~ \BitVec{n}
    &
    \SetOfVals \ni v &::=&
    \N
    ~|~ \{0,1\}^{*}
    ~|~ [ v^{*} ]
    \\
    \SetOfDistrExpr{} \ni \mu & ::= &
    \Uniform{}_\tau
    ~|\; \Bernoulli{}[p]
    &
    \BasicOp{1} &::=&
      \NotOp{}
    \\
    \SetOfExpr \ni e &::=&
    x
    ~|\; v
    ~|~ \ArrIndex{x}{i}
    ~|~ \ArrUpdate{x}{i}{x'}
    ~|\; \BasicOp{n}(x_1, \ldots, x_n)
    & 
    \BasicOp{2} &::=&
      \texttt{=}
      ~|~ \texttt{<}
      ~|~ \texttt{+}
      ~|~ \texttt{*}
      ~|~ \AndOp{}
      ~|~ \OrOp{}
      ~|~ \XorOp{}
    \\
    \SetOfPartiallyAppliedFuns \ni \lambda &::=&
    \multicolumn{4}{l}{
    f(x_1, \ldots, x_n, \BlankArg{}^{*})
    }
    \\
    \textsf{Primitives} \ni \cP &::=&
    \multicolumn{4}{l}{
      \PrimAny ~|~ \PrimAll ~|~ \PrimSearch
      ~|~ \PrimAmplify_{\pmin}
      ~|~ \PrimSimon_{\pcoll}
    }
    \\
    \SetOfStmt \ni s &::=&
    \multicolumn{4}{l}{
    x <- e
    ~|~ \protosample{x}{\mu}
    ~|~ \protoif{b}{s_t}{s_f}
    ~|~ s_1 ; s_2
    ~|~ \vecy <- f(\vecx)
    ~|~ \vecy <- \cP_{\eps}[\vec{\lambda}]
    }
    \\
    \SetOfFunc{} \ni \cF &::=&
    \multicolumn{4}{l}{
    \protodef{f}{\vecx}{\vec{t}'}{s}{\vecy}
    ~|~ \protoext{f}
    }
\end{array} \]
}

\newcommand{\QPLFullSyntax}{
\[\def\arraystretch{1}
\begin{array}{rrll}
    &
    \SetOfUnitaryOps{} \ni U &::=&
        \texttt{X}
        \;|\; \texttt{Y}
        \;|\; \texttt{Z}
        \;|\; \texttt{H}
        \;|\; \texttt{CNOT}
        ~|~ \SWAPGate{}
        ~|~ \COPY{}
        ~|~ U_e
        ~|~ U_\mu
        ~|~ \PhaseOnZero{\tau}{\phi}
        \;|\; \adjU{U}
        \;|\; \ctrlU{U}
    \\
    &
    \SetOfUnitaryCommands{} \ni w &::=&
       \qpskip
       \;|\; \qpunitary{\vecq}{U}
       \;|\; w_1 ; w_2
       \;|\; \qpcallu{g}{\vecq}
       \;|\; \qpcalldagger{g}{\vecq}
    \\
    &
    \SetOfProbCommands{} \ni s &::=&
      \qpskip
      \;|\; \qpassign{x}{e}
      \;|\; \qprandom{x}{\mu}
      \;|\; s_1 ; s_2
      \;|\; \qpifte{b}{s_t}{s_f}
      \;|\; \qpcall{h}{\vecx}
      \;|\; \qpcallandmeas{g}{\vecx}
    \\
    &
    \SetOfProcedures{} &::=&
        \uqplprocdef{g}{\vecq}{w}
        ~|~ \cqplprocdef{h}{\vecx}{}{s}
        ~|~ \extuproc{g}
        ~|~ \extcproc{h}
\end{array} \]
}

\newcommand{\AndOr}{AND-OR}

\newcommand{\IsRowAllOnes}{\textsf{IsRowAllOnes}}
\newcommand{\HasAllOnesRow}{\textsf{HasAllOnesRow}}
\newcommand{\epstot}{\ensuremath{\eps_\text{total}}}

\newcommand{\traqimplementationfootnote}{Publicly available at https://github.com/qi-rub/traq.}

\newcommand{\auxUnitary}[1]{\aux_{#1}^U}

\newcommand{\FineExpCostName}{\textsc{\textsf{FExpCost}}}
\newcommand{\FineGrainedCostExpr}{\cC^{\textsf{Fine}}}
\newcommand{\cqplexpcostfine}[1]{\FineExpCostName[#1]} %
\newcommand{\SimplifyCost}{\textsf{Simpl}}

\title{\OurFramework{}: Estimating the Quantum Cost of Classical Programs}

\author{Anurudh Peduri}
\email{anurudh.peduri@rub.de}
\orcid{0000-0002-6523-7098}
\affiliation{%
  \institution{Ruhr~University~Bochum}
  \department{Chair for Quantum Information, Faculty of Computer Science}
  \city{Bochum}
  \country{Germany}
}

\author{Jam Kabeer Ali Khan}
\email{jamkhan@connect.hku.hk}
\orcid{0009-0007-8505-9667}
\affiliation{%
  \institution{MPI-SP}
  \city{Bochum}
  \country{Germany}
}

\author{Gilles Barthe}
\orcid{0000-0002-3853-1777}
\email{gilles.barthe@mpi-sp.org}
\affiliation{%
  \institution{MPI-SP}
  \city{Bochum}
  \country{Germany}
}
\affiliation{
    \institution{IMDEA Software Institute}
    \city{Madrid}
    \country{Spain}
}

\author{Michael Walter}
\orcid{0000-0002-3073-1408}
\email{michael.walter@lmu.de}
\affiliation{%
  \institution{Ludwig-Maximilians-Universit\"at M\"unchen}
  \department{Chair for Quantum Information Theory, Faculty of Physics and Faculty of Mathematics, Computer Science \& Statistics}
  \city{Munich}
  \country{Germany}
}
\affiliation{%
  \institution{Munich Center for Quantum Science and Technology (MCQST)}
  \city{Munich}
  \country{Germany}
}
\affiliation{%
  \institution{Ruhr~University~Bochum}
  \department{Chair for Quantum Information, Faculty of Computer Science}
  \city{Bochum}
  \country{Germany}
}
\affiliation{%
  \institution{University of Amsterdam}
  \department{Korteweg-de Vries Institute for Mathematics and QuSoft}
  \city{Amsterdam}
  \country{Netherlands}
}

\begin{document}
\begin{abstract}
Predicting practical speedups offered by future quantum computers has become a major focus of the quantum community.
Typically, such predictions involve numerical simulations supported by lengthy manual analyses and are carried out for one specific algorithm at a time.
In this work, we present \OurFramework{}, a principled approach towards estimating the quantum speedup of classical programs fully automatically.
It consists of a classical language that includes high-level primitives amenable to quantum speedups, a compilation to low-level quantum programs, and a source-level cost analysis with provable guarantees.
Our cost analysis upper bounds the complexity of the resulting quantum program and is sensitive to the input data of the program (in addition to providing worst-case costs).
\OurFramework{} is implemented as a Haskell package with an extensive evaluation.

\end{abstract}

\maketitle
\section{Introduction}
\label{sec:intro}
Quantum algorithms have the potential to offer significant speedups over their classical counterparts~\cite{ronnow2014defining}.
Examples of quantum speedups include Grover's unstructured search algorithm~\cite{grover1996,boyer98qsearch,Hoyer_2003}, quantum max/min-finding~\cite{durr99qmin,ahuja1999qmax}, quantum counting~\cite{Brassard1998counting}, Shor's factorization and discrete logarithm algorithms~\cite{shor1994,shor1997}, quantum algorithms for linear systems~\cite{hhl2009} and convex optimization problems~\cite{brandao2017quantum,van2017quantum}.
The Quantum Algorithm Zoo~\cite{QAzoo} provides a comprehensive collection of quantum algorithms, discussing their complexity and potential speedups.

A popular approach to leverage quantum speedups in existing classical programs is by replacing selected subroutines of the classical program with equivalent, more efficient, quantum subroutines.
This has been termed \emph{quantization} in some prior works~\cite{Szegedy2004walks,quantization2007} and we adopt this terminology in this work.
One can then estimate the benefits of the quantization by analyzing
the cost of the resulting quantum program. Quantization plays an
important role in cryptanalysis, where it is used to estimate the
security of cryptographic constructions against quantum
attacks~\cite{quantumpreimageattack2017,quantumsvp2025,quantumlatticesieves2020,quantumsecurityaes2019,David2024,cryptonestedsearch2024},
and in optimization, where it is used to estimate the
potential benefits of quantum computers for specific instances of
NP-hard optimization
problems~\cite{eshaghian2024hybridsatsolvers,Cade2023quantifyinggrover,wilkening2024quantumalgorithmsolving01,wilkening2025quantumsearchmethodquadratic,lefterovici2025beyondasymptoticqls,nannicini_fast_2022,qubra2023simplex}.
So far, quantization and the subsequent cost analysis have been done
manually, which is tedious and error-prone. Cost analysis is
particularly tedious for NP-hard optimization problems, as it is
more interesting to seek precise cost estimates for a specific input~\cite{buhrman2025formalframeworkquantumadvantage}; to
do so, one must track the values of intermediate computations.

\paragraph{This work}
We present \OurFramework{}, a principled framework
for automating the quantization of classical programs
and cost analysis of the resulting quantized programs in the query cost model, a model commonly used in
quantum computing and quantum cryptography to reason about a program
complexity. A main feature of \OurFramework{} is that cost analysis is
carried out at the source level before compilation to a quantum program; the reasons for this will be explained in \Cref{sec:motivating-example}.
We show that \OurFramework{} is sound,
i.e.\, the compiler produces approximately correct programs,
and the cost analysis returns an upper bound of the actual cost of the resulting quantum program.
Interestingly, the cost analysis relies on a separate, non-trivial error analysis.
Furthermore, we implemented \OurFramework{} as a Haskell package, and evaluated it on a representative set of case studies to assess its practicality.

\paragraph{Contributions}
$\OurFramework{}$ utilizes techniques from programming languages and quantum computing to enable an approach to estimating the quantum cost of classical programs, with the following key contributions:
\begin{itemize}
	\item Classical (probabilistic) source language~(\cref{sec:framework:protolang}) with \emph{high-level primitives}~(\cref{sec:setting:primitives}) that can be quantized via compiling~(\cref{sec:compiler}) to a quantum programming language.
	\item Source-level error analysis with formal guarantees on the correctness of our compiler~(\cref{sec:errors}).
	\item Source-level cost analysis to bound the input-sensitive expected quantum cost of compiled programs with formal guarantees~(\cref{sec:cost}).
	\item A Haskell package,\footnote{\traqimplementationfootnote{}} with a DSL to write source programs, with support for cost analyses and compilation, and an extensible library~(\cref{sec:Implementation}), and many primitives and case studies of programs~(\cref{sec:evaluation}).
\end{itemize}

\section{Preliminaries}
\label{sec:preliminaries}

This section provides the relevant background for describing probabilistic and quantum computation.
For brevity, we only describe concepts that are used in the main paper.
We give a more detailed exposition in \cref{app:background} in the supplementary material which also covers additional concepts used therein.

\subsection{States}
For programs, we denote the set of all variable names by $\Vars{}$, and the set of all possible values as $\Vals{}$.
The set of program states, which map variables to values, is denoted $\Sigma = \Vars{} \to \Vals{}$.
We assume for simplicity that the above sets are finite.

\subsection{Probabilistic Computing}
To a finite set $A$ (such as $\Sigma$), we associate a space of discrete probability distributions~$\Distr{A} \subset A \to [0, 1]$.
For a distribution $\mu \in \Distr{A}$, the probability of obtaining a value $a$ is denoted $\mu(a)$; any distribution satisfies $\sum_{a \in A} \mu(a) = 1$.
We equip distributions with a monadic structure,
with the \emph{delta distributions} $\detstate{a} \in \Distr{A}$ for $a \in A$ as the unit,
and the \emph{distribution expectation} $\DistrExp{\mu}{M}$ as the bind.
Given $\mu \in \Distr{A}$ and a \emph{probabilistic function} $M \colon A \to \Distr{B}$, the bind is defined as
\[
\DistrExp{\mu}{M} = \DistrExp{a \sim \mu}{M(a)} = \sum_{a \in A} \mu(a) M(a) \in \Distr{B}.
\]
The \emph{total variation distance} of two distributions $\mu, \mu' \in \Distr{A}$ is $\TVDist{\mu}{\mu'} = \tfrac12 \sum_a \abs{\mu(a) - \mu'(a)}$.
This induces a distance metric $\Delta$ on probabilistic functions $M, M' \colon A \to \Distr{B}$, defined as
\begin{equation}\label{eq:Delta}
	\probDistance{M}{M'} = \max_a \TVDist{M(a)}{M'(a)}.
\end{equation}

\subsection{Quantum Computing}\label{sub:quantum_computing}
We recall the basic formalism of quantum computing,
and refer to textbooks~\cite{nielsen2010quantum,YM08,wilde2013quantum} for a more detailed exposition.

To a finite set $A$, we associate a \emph{Hilbert space} $\cH_A$,
which is a finite-dimensional complex vector space with an inner product.
It has an orthonormal \emph{standard basis} (also called \emph{computational basis}) labelled by elements $a \in A$, denoted $\ket{a} \in \cH_A$.
Any vector $\ket\psi \in \cH_A$ can be written as a linear combination or \emph{superposition} $\ket\psi = \sum_a \psi_a \ket a$.
Given two spaces $\cH_A$ and $\cH_B$, the combined space is defined by the tensor product $\cH_A \ot \cH_B$, which can be identified with~$\cH_{A \times B}$.
We denote by $\cH = \cH_\Sigma$ the overall Hilbert space of all quantum variables; note that we can identify $\cH = \bigotimes_{v \in \Vars} \cH_\Vals$.
We denote identity operators by $I$, and the orthogonal projection onto the subspace~$\C \ket{a}$ by $\proj{a}$.
The notation $M^\dagger$ denotes the adjoint of a linear operator~$M$.

The state of all the quantum variables is described by a unit vector $\ket\psi \in \cH$.
This is often called a ``pure'' quantum state, which suffices for the discussion in the main part of this paper (in the supplementary material we use the more general notion of a ``mixed'' quantum state, or density matrix, which is necessary to describe the state of subsets of quantum variables).
There are two basic kinds of operations on quantum variables.
The first is to apply a unitary $U$, which is a linear operator satisfying $U^\dagger U = U U^\dagger = I$.
On applying $U$ on state $\ket\psi$, we obtain $U\ket\psi$.
The second is to measure in the standard (computational) basis.
When measuring a state $\ket\psi$, we obtain the output state $\ket{x}$ with probability $\abs{\braket{x|\psi}}^2$, along with the classical outcome~$x \in \Sigma$.
Both operations can also be applied to a subset of quantum variables in a natural way.

\subsection{Queries and Cost Model}
\label{sec:cost:model}

We use the \emph{query cost model}~\cite{buhrman2002querycomplexity} which provides a well-established proxy for time complexity and is agnostic of the details of the platform (hardware, gateset, etc.).
It is widely used in the context of bounding quantum speedups, from combinatorial optimization~\cite{Cade2023quantifyinggrover,cade22communitydetection} to quantum cryptanalysis of post-quantum cryptography~\cite{NISQStatus}.
Much of our design is general, but we leave it for future work to incorporate other, more detailed costs.

In the query model, program inputs are modeled by externally interpreted functions $f \colon X\to Y$ (for some finite sets $X, Y$) and one counts the number of queries to these functions.
We distinguish between two types of queries: classical and quantum.
\OurFramework{} counts the expected number of classical and quantum queries to each externally interpreted function for a given interpretation.

A \emph{classical query} is simply a call to the function~$f$.
This function could be implement in code or by loading data from a data structure or ROM/RAM.

A \emph{quantum query}, in contrast, is made by invoking a \emph{unitary}~$U_f$ (or its inverse~$U_f^\dagger$), with an action of the following form:
\begin{equation}\label{eq:U_f}
  U_f \ket{x} \ket{0} \ket{0}
  = \ket{x} \ket{f(x)} \ket{\psi_x}
\end{equation}
for every $x \in X$ where $\ket{x}$ is a standard basis quantum state,
and $\ket{\psi_x}$ are arbitrary quantum states on auxillary variables.
The power of quantum computation arises because $U_f$ can not only applied to basis vectors but also to superpositions: we have
\[ U_f \sum_x \alpha_x \ket{x}\ket0\ket0 = \sum_x \alpha_x \ket{x} \ket{f(x)} \ket{\psi_x}. \]
Quantum queries are typically realized succinctly by a quantum circuit (for example, the quantization of a classical circuit for~$f$), or by using a suitable data-structure like a QRAM/QROM~\cite{qram}.

The above notions are nicely compositional:
classical and quantum subroutines naturally give rise to functions and unitaries, respectively, that can be queried as above.

Some quantum algorithms require \emph{strong quantum queries} to a function~$f$.
By this we mean a unitary with the action
\begin{align}\label{eq:strong}
  \ket{x}\ket{y} \mapsto \ket{x}\ket{y \oplus f(x)}.
\end{align}
Compared to to~\eqref{eq:U_f} there are no auxiliary variables and the unitary is defined on all inputs (not just for $y=0$).
Strong quantum queries can be realized using one quantum query each to~$U_f$ and to~$U_f^\dagger$ (this is known as the \emph{compute-uncompute} pattern~\cite{aaronson2003uncompute}), hence they are already naturally incorporated in the model above.
One can similarly implement quantum phase queries.

These notions extend naturally to probabilistic functions $F \colon X \to \Distr{Y}$.
Classical queries simply call such a function to obtain a random result, while quantum queries are modelled by the following unitary (and its inverses):
\begin{equation}\label{eq:U_F}
  U_F \ket{x} \ket{0} \ket{0} \ket{0}
  = \sum_{y \in Y} \sqrt{F(x)(y)} \ket{x} \ket{y} \ket{y} \ket{\psi_{x, y}}.
\end{equation}
In quantum information language, $U_F$ is a unitary extension of the ``classical'' quantum channel corresponding to~$F$.
It has the following intuitive interpretation:
if we apply $U_F$ to an input quantum state and discard all but the second register, the result is a sample from the classical distribution~$F(x)$, for~$x$ obtained by measuring the input state.
Note that \eqref{eq:U_F} plainly generalizes \eqref{eq:U_f} from deterministic to probability functions (the second $\ket y$ can be removed by using a CNOT or absorbed into~$\ket{\psi_{x,y}}$).
As a special case, quantum sampling access to a probability distribution~$\mu \in \Distr Y$ is modeled by a unitary~that~acts~as
\begin{equation}\label{eq:U_mu}
  U_\mu \ket{0} \ket{0} \ket{0}
  = \sum_{y \in Y} \sqrt{\mu(y)} \ket{y} \ket{y} \ket{\psi_{x, y}},
\end{equation}
There is also a ``strong'' notion of a quantum query to a probabilistic function or probability distribution (which we will not need here).

\section{Overview of \OurFramework}
\label{sec:motivating-example}
\label{sec:overview}
This section gives a tour of the various aspects of \OurFramework{} using illustrative examples.
We provide a probabilistic source language, $\ProtoLang{}$: an imperative language with function calls.
This language provides \emph{high-level primitives} that serve as building blocks available to programmers for solving specific computational tasks.
We first discuss the compiler that quantizes programs, then the source-level cost analysis, followed by the formal guarantees.

\subsection{Quantization of Classical Programs}\label{sub:quantization_of_classical_programs}

We start with the following example to illustrate the quantization of classical programs:
\begin{problem}
  \label{ex:list-search}
  Given a bit string $L$ of size $N$, does it contain one or more~$1$s?
\end{problem}
\noindent We can model the input to this problem as a function
{\[ L \colon \Fin{N} \to \Bool \]}%
where the type $\Fin{N}$ has values $\{0 \ldots N - 1\}$, and $\Bool = \Fin{2}$.
Then we can solve this using a \PrimSearch{} primitive:
{\begin{equation}\label{program:list-search}\begin{aligned}
  &\kwbasic{fn}~ \textsf{ContainsOne}() -> \Bool ~\kwbasic{do}
\\ &\quad b, i <- \PrimSearch{}[L];
\\ &\quad \kwbasic{return}~ b
\\ &\kwbasic{end}
\end{aligned}\end{equation}}%
Here, the primitive \PrimSearch{} returns a $b \colon \Bool$ and $i \colon \Fin{N}$.
If there is a $1$ in the bit string $L$, then search returns $b = 1$ and some $i$ such that~$L(i) = 1$.
Otherwise, it returns $b = 0$ (and an arbitrary $i$).

\paragraph*{Quantization via Compilation.}
We quantize programs written in \ProtoLang{} using a quantum compiler $\QuantumCompiler{}[\cdot]$, which targets a quantum language \CQPL{}.
The primitives are implemented using quantum algorithms.
For example, the \PrimSearch{} primitive above is implemented using a version of \emph{Grover's search algorithm}~\cite{grover1996,boyer98qsearch}, which provides a quadratic speedup over classical search.
This allows programmers to benefit from quantum speedups by writing classical programs.

Most quantum algorithms for realizing primitives (e.g.\ Grover~\cite{grover1996,boyer98qsearch}, Simon~\cite{simon97}) are inherently \emph{probabilistic} in nature---they may not always return the same answer, and they can also fail with some probability.
In this work, we restrict ourselves to \emph{Monte Carlo} or bounded-runtime algorithms to implement each primitive, ensuring that all programs terminite in finite time, and hence have finite cost.
That is, the algorithm and the resulting quantized program always runs in some finite amount of time, but can produce incorrect results with some probability, which can be made as small as desired.
The compilation and therefore costs depend on the choice of maximum allowed error, or ``error budget'', for each primitive.
\OurFramework{} annotates each primitive with an $\eps \in (0, 1)$ denoting the maximum allowed error in its quantized implementation:
{\[ b, i <- \PrimSearch{}_\eps[L] \]}%
Then the compiler $\QuantumCompiler{}$ takes an $\eps$-annotated \ProtoLang{} program, and produces a \CQPL{} program.
It compiles the above \PrimSearch{} call to a quantum algorithm that \emph{approximately} implements the ideal functionality of search.
The error probabilities of such primitive calls will combine nontrivially to an error probability of the overall program.
\OurFramework{} computes this automatically and chooses the $\eps$-annotations of subroutine calls for a given overall error budget~$\epstot$.
We describe this in more detail at the end of \cref{sec:motivation:cost} and in \cref{sub:putting_it_all_together}.

\paragraph*{Quantum Queries and Composition.}
General quantum algorithms combine classical/probabilistic and quantum computation to realize the most efficient implementations.
E.g., best-in-class quantum search algorithms combine random sampling together with quantum iterations~\cite{Cade2023quantifyinggrover}.
The compiler~\QuantumCompiler{} utilizes this and will therefore in general produce programs that combine classical and quantum computation.
But as discussed in \cref{sec:cost:model}, quantum algorithms usually access their input by unitary quantum queries~\eqref{eq:U_f}.
To overcome this mismatch, we define a second compiler $\compileUQPL{\cdot}$ which compiles to a purely unitary program in our target quantum language.
This compiler allocates auxiliary quantum variables statically, as well as generates compute-uncompute patterns for primitives that require strong oracle queries.
We illustrate this with our second example:
\begin{problem}
  \label{ex:matrix-search-example}
  Given a $N \times M$ matrix $A$ of $0$s and $1$s, does it have a row containing all 1s?
\end{problem}

\noindent
This problem is inspired by the more general problem known as \AndOr{} trees which has received significant attention in the quantum computing literature~\cite{Hoyer_2003,ambainis2010andortrees,ambainis2006vts} as it reveals challenges in composing quantum subroutines.
Similarly as before, we model the input matrix as a function
{\[
  A \colon (\Fin{N}, \Fin{M}) \to \Bool,
\]}%
where the first argument is a row and the second the column index.
We can then solve the problem using a nested search algorithm:
{
\begin{equation}\label{program:matrix-search}
\begin{aligned}
&\kwbasic{fn} ~\IsRowAllOnes(i \colon \Fin{N}) -> \Bool ~\kwbasic{do}
\\ &\quad b <- \PrimAll{}[A(i, \BlankArg{})];
\\ &\quad \kwbasic{return}~ b
\\ &\kwbasic{end}
\\[.1cm] &\kwbasic{fn} ~\HasAllOnesRow() -> \Bool ~\kwbasic{do}
\\ &\quad b, i <- \PrimSearch{}[\IsRowAllOnes(\BlankArg{})];
\\ &\quad \kwbasic{return}~ b
\\ &\kwbasic{end}
\end{aligned}
\end{equation}
}%
The primitive \PrimAll{} is a variant of \PrimSearch{} that returns $1$ if, and only if, all entries of the input (here, the $i$-th row) are equal to one.
As explained above, \OurFramework{} annotates the above primitive calls as $\PrimAll{}_{\eps_1}$ and $\PrimSearch{}_{\eps_2}$ respectively.
Then it compiles the program as $\compileCQPL{\HasAllOnesRow}$.
Because the quantum compilation of $\PrimSearch{}_{\eps_2}$ requires quantum query access to its input, this compilation in turn invokes the \emph{unitary} compilation $\compileUQPL{\IsRowAllOnes}$.
In particular, the call to $\PrimAll{}_{\eps_1}$ needs to be compiled fully unitarily.
We employ a fully unitary version of Grover search due to \citet{zalka1999groverbasedquantumsearchoptimal}.
We see that each primitive requires two compilations: a general quantum and a purely unitary one.

\subsection{Cost Analysis}
\label{sec:motivation:cost}

\OurFramework{} estimates the cost of the quantized programs produced
by our compilation. Given a source program~$s$ and an initial state~$\sigma$,
it computes a bound on the expected cost of the quantized program~$\compileCQPL{s}$ of the form:
{
\begin{equation}\label{eq:overview:cost-bound}
 \cqplexpcost{\compileCQPL{s}}(\sigma) \le \costq{s}{\sigma} + \progerrprob{s} \cdot \costqmax{s}.
\end{equation}}%
Here, $\cqplexpcost{\cdot}$ on the left-hand side is the actual expected cost of a target $\CQPL$ program, while the right-hand side is our bound on it.
Intuitively, {$\CostMetricQ{}$} can be interpreted as a bound on the \emph{expected cost} when there are no errors in execution of the quantized program;
{$\ErrProb{}$} is an upper bound on the error probability of the quantized program,
and {$\CostMetricHavoc{}$} is the \emph{havoc cost} of the compiled program, which upper bounds the cost of executing the quantized program regardless of errors in execution (more on this terminology below).
In the presence of unitary sub-computations, the computation of these quantities will also involve a unitary cost bound {$\CostMetricU{}$}.
All the above quantities are source-level quantities (they are defined for a source program~$s$), and they can be computed using a source-level analysis (even though they refer to the compilation, they can be evaluated without prior compilation to $\CQPL{}$).
We explain each of them in more detail below.

\paragraph*{Unitary Cost Analysis.}
To upper bound the cost of the unitary compilation $\UnitaryCompiler{}[\cdot]$,
we define a cost bound {$\CostMetricU{}$}.
As we will see, unitary programs are quantum circuits without any control flow.
Therefore their cost is simply the sum of costs of each individual operation in the circuit.
This is defined inductively on the structure of the source program.
The unitary cost analysis is used both in the havoc and expected cost analyses.

\paragraph*{Havoc Cost Analysis.}
The havoc cost bound {$\CostMetricHavoc{}$} of a program is a coarse upper bound of program execution cost of the quantization $\compileCQPL{\cdot}$ that holds regardless of whether errors occur during execution.
This differs from (but upper bounds) the usual notion of worst-case cost, where the cost of a sequence is the sum of costs, assuming each step produces the right output for the next step.
Just like the unitary cost, the havoc cost admits a simple definition by induction on the structure of the source program.
For the bit-string search program~\eqref{program:list-search} for \cref{ex:list-search}, \OurFramework{} computes the following havoc cost bound:
{
\[
  \CostMetricHavoc{}[\textsf{ContainsOne}] =
  \CostMetricHavoc{}[b, i <- \PrimSearch{}_\eps[L]] = W(N, \eps) \cdot 2 \cdot \CostMetricU{}[L]
\]}%
where $W(N, \eps) = O(\sqrt{N} \log(1/\eps))$ is a fixed, known function bounding the havoc number of
queries to~$L$ made by the quantum compilation of~$\PrimSearch{}_\eps$ to find a solution with probability $1 - \eps$~\cite{grover1996}.
The factor $2$ is due to the compute-uncompute pattern required for implementing strong unitary access, and
{$\CostMetricU{}[L]$} represents the cost of a single unitary query to
the external function $L$.

Similarly, for the ($\eps$-annotated) matrix search program~\eqref{program:matrix-search} for \cref{ex:matrix-search-example}, \OurFramework{} computes the following havoc cost bound: {
\begin{align}
  \CostMetricHavoc{}[\HasAllOnesRow]
  &= W(N, \eps_2) \cdot 2 \cdot \CostMetricU{}[\IsRowAllOnes]
  \nonumber
  \\ &= W(N, \eps_2) \cdot 2 \cdot W_U(M, \eps_1) \cdot 2 \cdot \CostMetricU{}[A]
  \label{eq:matrix-example:havoc}
\end{align}}%
where $W_U(M, \eps)$ is a fixed, known function bounding the number of queries made by the unitary compilation of ${\PrimAll{}_{\eps_1}}$.

\paragraph*{Input-sensitive Cost Analysis.}
The average-case complexity of algorithms is often significantly better than their worst-case complexity.
For example, quantum search over a space of~$N$ elements with at least~$K$ solutions and failure probability~$\eps$ has an expected query complexity of $E(N, K, \eps) = O(\sqrt{N/K})$~\cite{boyer98qsearch} (if~$K > 0$), which improves over its worst case complexity of~$O(\sqrt{N} \log(1/\eps))$~\cite{grover1996}.
This is modeled in \OurFramework{} by the \emph{input-sensitive} cost bound $\CostMetricQ{}$, which bounds the \emph{expected cost} of the compiled program on some given input.
Here, input refers both to the initial state~$\sigma$ of the program, as well as to the interpretation of the externally-defined functions---such as the bit-string~$L$ for \cref{ex:list-search} and the matrix~$A$ for \cref{ex:matrix-search-example}---by which one models the program input in the query cost model (\cref{sec:cost:model}).
For instance, \OurFramework{} computes the following expected cost bound for the bit-string search program~\eqref{program:list-search} for \cref{ex:list-search}:
{\begin{equation}\label{eq:expected cost p1}
  \CostMetricQ{}[\textsf{ContainsOne}]() = \CostMetricQ{}[\PrimSearch{}_\eps[L]]() = E(N, K_L, \eps) \cdot 2 \cdot \CostMetricU{}[L]
\end{equation}}%
where $E(N, K, \eps) = O(\sqrt{N/K})$ is a fixed, known function bounding the expected number of quantum queries to the predicate made by the quantum search algorithm, when searching a space of $N$ elements, the predicate has~$K$ solutions, and the maximum allowed error probability is~$\eps$.
The analysis computes the parameter~$K_L$ in \cref{eq:expected cost p1} as the \emph{actual} number of $1$s in the interpretation of the input bit-string~$L$.

Similarly, \OurFramework{} computes the following bound for the nested matrix search program~\eqref{program:matrix-search} for \cref{ex:matrix-search-example}:
{
\begin{align}
  \CostMetricQ{}[\HasAllOnesRow]()
  &= E(N, K_A, \eps_2) \cdot 2 \cdot \CostMetricU{}[\IsRowAllOnes]
  \nonumber
  \\
  &= E(N, K_A, \eps_2) \cdot 2 \cdot W_U(M, \eps_1) \cdot 2 \cdot \CostMetricU{}[A]
  \label{eq:matrix-example:exp}
\end{align}}%
after annotating the primitives as $\PrimSearch{}_{\eps_2}$ and $\PrimAll_{\eps_1}$ as described before.
Here, the value $K_A$ is the \emph{actual} number of all-one rows, and is evaluated using the source semantics for a given interpretation of the input matrix~$A$:
{\[
  K_A = \abs{\{ i \in \{0 \ldots N - 1\} \mid \evalProb{\IsRowAllOnes}(i) = 1 \}}
\]}%

\paragraph*{Error Analysis.}
The final term to discuss in \cref{eq:overview:cost-bound} is $\ErrProb{}$.
As stated before, \OurFramework{} annotates the source program by assigning an $\eps$ to each primitive.
Then the term $\ErrProb{}$ upper-bounds the total probability of error of the quantized program.
It is an input-sensitive quantity that can be evaluated using the source-level semantics, similary to the expected cost bound.
Just like the havoc cost, it admits a simple definition by induction on the structure of the source program.
For the matrix search program~\eqref{program:matrix-search} for \cref{ex:matrix-search-example}, \OurFramework{} computes
\begin{equation}\label{eq:matrix-example:epstot}
  \ErrProb[\HasAllOnesRow] = \eps_2 + W(N, \eps_2) \sqrt{2 \eps_1}.
\end{equation}
We explain this in more detail later in \cref{sec:errors}.

\subsection{Putting it all together}\label{sub:putting_it_all_together}

Overall, for the matrix search program~\eqref{program:matrix-search}, given an interpretation of the $N \times M$ input matrix $A$,
\OurFramework{} combines \cref{eq:overview:cost-bound,eq:matrix-example:havoc,eq:matrix-example:epstot,eq:matrix-example:exp} to obtain the following upper bound on the expected number of quantum queries to~$A$:
{
\begin{equation}\label{eq:put together}
\begin{aligned}
  \parens[\Big]{E(N, K_A, \eps_2) + \epstot W(N, \eps_2)}
  \cdot 2 \cdot W_U(M, \eps_1) \cdot 2 \\
= O\parens[\Big]{\parens[\big]{\sqrt{N/{K_A}} + \epstot \sqrt{N} \log(1/\eps_2)} \sqrt{M} \log(1/{\eps_1})},
\end{aligned}
\end{equation}
}%
where we abbreviate $\epstot := \ErrProb[\HasAllOnesRow] = \eps_2 + W(N, \eps_2) \sqrt{2 \eps_1}$.
This bound is computed on the source level, symbolically in $\eps_1,\eps_2$, \emph{without} requiring prior compilation to~$\CQPL{}$.

When compiling, \OurFramework{} will pick concrete values for the annotations~$\eps_1,\eps_2$ for a given maximum overall error~$\epstot = \ErrProb$ set by the user, using a heuristic that splits the error budget equally among each step.
\OurFramework{} then compiles to \CQPL{} programs for the concrete values of each epsilon.
For illustration, for parameters $N=M=1000$ and given a maximum allowed error~$\epstot = 0.001$, \OurFramework{} picks $\eps_2 = \epstot/2$ and $\eps_1 = (\epstot/W(N, \eps_2))^2/8$ by splitting equally,
and then compiles the $\ProtoLang{}$ program~\eqref{program:matrix-search} to a \CQPL{} program of 1675 lines.
The time to compute the bound and compile the program is 3.2s.

While computing the quantum cost bound~\eqref{eq:put together} is more expensive than merely evaluating the source program on the given input, we note that our approach is more scalable than carrying out the quantum cost analysis directly on the compiled quantum programs.
Indeed, the latter would require costly classical simulations of quantum programs.
For instance, for the values considered above, we would need to simulate a quantum program of 900 qubits, which is beyond feasibility of general purpose methods.

\subsection{Formal Guarantees}\label{sub:formal_guarantees}

We now summarize the formal guarantees underpinning \OurFramework.
The first key property is that our source-level cost analysis provides a sound upper bound~\eqref{eq:overview:cost-bound} on the actual cost of the compiled program, as discussed above.
We state this as a theorem, see \cref{sec:cost} for more detail.

\begin{theorem*}
For every well-formed $\eps$-annotated \ProtoLang{} statement $s$, and well-formed input $\sigma$:
{
\[
 \cqplexpcost{\compileCQPL{s}}(\sigma) \le \costq{s}{\sigma} + \progerrprob{s} \cdot \costqmax{s}.
\]}
\end{theorem*}

\noindent
To prove the above theorem, we also need to reason about the correctness of the compilation.
Indeed, since our cost bound is defined on the source level, we need to ensure that the intermediate states of the quantized program are well-approximated by the source-level semantics.
More precisely, we bound the distance between the resulting probability distribution in terms of the error bound~$\ErrProb$:

\begin{theorem*}
For every well-formed $\eps$-annotated \ProtoLang{} statement $s$, {$\probDistance{\evalCQPL{\compileCQPL{s}}}{\evalProb{s}} \le \progerrprob{s}$}.
\end{theorem*}

\noindent Here, $\evalProb{\cdot}$ is the probabilistic semantics of the classical source language \ProtoLang{},
and $\evalCQPL{\cdot}$ is the probabilistic semantics of the target quantum language \CQPL{}.
This proof of this theorem naturally involves defining and proving the correctness of the unitary compilation, as we will see in \cref{sec:errors}.

\section{Languages and Compilation}
\label{sec:setting}
We now present \OurFramework{} in more detail.
In this section, we start by describing the notion of primitives (\cref{sec:setting:primitives}), followed by the source language \ProtoLang{}~(\cref{sec:framework:protolang}) and target language \CQPL{}~(\cref{sec:framework:targetlang}), and finally the quantizing compilation from \ProtoLang{} to \CQPL{}~(\cref{sec:compiler}).
This sets the stage for the cost analysis, which will be presented in \cref{sec:cost}.

\subsection{Primitives}
\label{sec:setting:primitives}
Our source language and compiler are built around the notion of \textit{primitives}: building blocks available to programmers for solving specific computational tasks.
Formally, primitives are second-order functions that accept (in general probabilistic) functions and return values (or tuples of values).
Primitives that are supported by \OurFramework{} include:
{
\[
\begin{array}{rcl}
  \PrimAny, \ \PrimAll &: & (\tau \rightarrow \Bool) \rightarrow \Bool
  \\
  \PrimSearch &: & (\tau\rightarrow \Bool) \rightarrow {(\Bool, \tau)}
  \\
  \PrimAmplify_{\pmin} &: & (() \to \DistrType{(\Bool, \tau)}) \rightarrow \DistrType{(\Bool, \tau)}
  \\
  \PrimSimon_{\pcoll} &:& (\BitVec{n} \rightarrow \BitVec{n}) \rightarrow \BitVec{n}
\end{array}
\]
}%
where $\tau, \tau'$ are arbitrary types,
$\Bool$ denotes the type of booleans,
$\BitVec{n}$ denotes bit strings of length~$n$,
and $()$ is the unit type, so $() \to \tau$ denotes a function with no inputs that outputs a value of type~$\tau$.

The first set of primitives are the \textit{search-like primitives} $\PrimAny, \PrimAll, \PrimSearch$,
which accept a deterministic boolean predicate $f:\tau \rightarrow \Bool$.
The primitive $\PrimAny$ returns $1$ iff there is at least one $x : \tau$ such that~$f(x) = 1$,
and similarly the primitive $\PrimAll$ returns $1$ iff for every $x : \tau$, $f(x) = 1$;
and otherwise they return $0$.
The primitive $\PrimSearch$ is similar to $\PrimAny$, but also returns a uniformly random $x : \tau$ satisfying $f(x) = 1$ (if one exists),
and otherwise a uniformly random $x$.

The primitive $\PrimAmplify_{\pmin}$ is used to increase or ``amplify'' the success probability of a probabilistic algorithm.
It accepts a probabilistic function $f$ that outputs a pair $\DistrType{(\Bool, \tau)}$, where the first element denotes success, and the second is an arbitrary value in $\tau$.
If $f$ succeeds with non-zero probability, it is assumed to succeed with probability at least $\pmin$.
In this case, $\PrimAmplify_{\pmin}[f]$ will succeed with probability~$1$, and output a value from the output distribution of~$f$ conditioned on success.
Otherwise, if $f$ succeeds almost never, $\PrimAmplify_{\pmin}[f]$ behaves the same as~$f$.
The precise semantics is given in \cref{app:prim-amplify}.
We note that amplify is a very general primitive that in particular can be used to implement the search-like primitives discussed above.

The primitive $\PrimSimon_\pcoll$ solves a period-finding problem with applications in cryptography~\cite{simon97,Kaplan2016}.
Similarly to Shor's period finding algorithm, it offers an \emph{exponential} quantum speedup.
The primitive accepts a deterministic function $f:\BitVec{n} \to \BitVec{n}$ which is assumed to satisfy two properties:
(1) it has a non-zero period $s \in \{0,1\}^n$ (i.e.\ $\forall x, f(x \oplus s)=f(x)$), and
(2) for every other $t \not\in \{0, s\}$, the fraction of $x$ such that we have a ``collision'' $f(x) = f(x \oplus t)$ is at most~$\pcoll$.
Then the primitive outputs the period $s$.
The precise semantics is given in \cref{app:prim-simon}.

\paragraph{Realization}
Each primitive has a precise classical source-level semantics that defines its ideal behaviour, described informally above and formally in \cref{app:prim-amplify,app:prim-simon,app:search-prims}, which is used in the cost analysis and formal guarantees.
As motivated in \cref{sub:quantization_of_classical_programs}, each primitive is realized through two quantum implementations: a general \emph{quantum} implementation that can arbitrarily combine classical and quantum computation and use control flow such as early-exits, and a \emph{unitary} implementation that is restricted to unitary quantum circuits.
These implementations are functionally equivalent to the reference semantics on all inputs that satisfy the promise of the primitive (up to a desired maximum error probability).
The design of \OurFramework{} is modular---new primitives can be added by providing a reference semantics and the two quantum implementations.

\subsection{Source Language \ProtoLang{}}
\label{sec:framework:protolang}

\begin{figure}
\small
\ProtoLangSyntax{}
\caption{Syntax of \ProtoLang{}. The types for variables are omitted for brevity.}
\label{def:proto:syntax}
\end{figure}

The source language \ProtoLang{} is a classical (probabilistic) language with access
to high-level primitives as described above.
The syntax of \ProtoLang{} is shown in \cref{def:proto:syntax}.
We model the source language based on our Haskell DSL, and therefore in functional style, accepting inputs and returning outputs.

\paragraph*{Types}
The type $\Fin{N}$ represents bounded integers in $[0, N-1]$,
and $\Arr{n}{\tau}$ represents sequences of length $n$ and element type $\tau$.
We use the shorthands $\Bool$ for $\Fin{2}$, and $\BitVec{n}$ for $\Arr{n}{\Bool}$.

\paragraph*{Expressions}
The set of deterministic expressions is denoted \SetOfExpr{},
which can be either a variable~$x$, a value~$v$, an array
access $\ArrIndex{x}{i}$, or an array update $\ArrUpdate{x}{i}{x'}$.
Array indices $i$ are compile-time constants, and must be within bounds.
The set of distribution expressions, denoted \SetOfDistrExpr{}, includes~$\Uniform{}_\tau$, the uniform distribution over some type~$\tau$, and~$\Bernoulli{}[p]$, the Bernoulli distribution with parameter~$p$, i.e.\, the distribution of a boolean random variable that equals $1$ with probability~$p$.
A partially-applied function expression ($\SetOfPartiallyAppliedFuns{}$) takes the form~$\lambda = f(x_1, \ldots, x_n, \BlankArg{}^{*})$, where~$f$ is a function with at least $n$ arguments, where the first $n$ arguments are fixed to variables~$x_1, \ldots, x_n$.

\paragraph*{Statements}
The set of statements is denoted \SetOfStmt{}.
Statements can either be an assignment of a deterministic expression,
sampling from a distribution,
a conditional branch,
a sequence of two statements,
a function call,
or a primitive call.
We use vector notation $\vecx$, $\vec y$, $\vec\lambda$ as shorthand notation for tuples of some suitable length.
A primitive call $\vecy <- \cP_\eps[\vec\lambda]$ evaluates the primitive $\cP$ on a sequence of partially-applied functions $\vec{\lambda}$, and stores the results in $\vecy$.
As discussed in \cref{sec:motivating-example}, each such call is additionally annotated by \OurFramework{} with an $\eps \in (0, 1)$ denoting the maximum allowed failure probability that will be chosen during compilation.

\paragraph*{Programs}
Programs consist of a list of named functions $\cF$, each of which is either a definition and an external function.
A function definition~$\kwbasic{fn}$ consists of a function identifier $f$, parameter names~$\vecx$, a body statement $s$, and return variable names $\vec{y}$.
An external declaration~$\kwbasic{ext fn}$ solely consists of a function identifier.
The types for variables are omitted for brevity; we assume there is a global typing context $\Gamma$ that maps variables to their types.
We use $\Phi$ to denote the \emph{function context} for the program, which maps function names $f$ to their source code.

\paragraph*{Semantics}
$\ProtoLang{}$ programs are said to be \textit{well-formed} under a set of constraints.
First, programs must be \textit{safe} (e.g., no out-of-bounds array accesses).
Second, they must be \textit{well-typed} under typing rules that we present in \Cref{app:protolang:typing}.
Third, all recursion must be statically terminating, which can be enforced with a well-founded order on each function, requiring functions to call only smaller functions, or themselves, on structurally smaller arguments.
Finally, each primitive use in a \textit{well-formed} program must respect the primitive's promise as introduced earlier---for example, functions passed as arguments to $\PrimSearch$ should be deterministic, functions passed to the $\PrimSimon$ primitive should be nearly two-to-one, and so forth.
In this work we assume that all programs are well-formed.
We leave automatic well-formedness checking for future work.
Under these constraints, programs admit a standard set-theoretic denotational semantics.
In particular, the semantics of any \ProtoLang{} statement~$s \in \SetOfStmt$ is given by a probabilistic function
\begin{align}\label{eq:protolang:semantics}
  \evalProb{s} : \Sigma \to \Distr{\Sigma}
\end{align}
from deterministic program states to distributions over program states.
We define this in detail in \cref{app:protolang:semantics}.

\subsection{Target Language \CQPL{}}
\label{sec:framework:targetlang}

Our target language \CQPL{} is a quantum programming language based on a language introduced by \citet{selinger_qpl}.
In contrast to \ProtoLang{}, its semantics is imperative and arguments are passed by reference.
The syntax of \CQPL{} is shown in \Cref{def:cqpl:syntax-main}.
Its statements and procedures are split into two fragments:
classical and unitary.
The unitary fragment operates on quantum variables, using quantum operators, while the classical fragment operates on classical variables.

\paragraph{Unitary Operators}
The set of unitary operators is denoted by \SetOfUnitaryOps.
The language supports a set of widely used one- and two-qubit quantum gates, including the Pauli gates (X, Y, Z), the Hadamard gate (H), and the controlled-NOT gate (CNOT).
The $\COPY$ unitary acts on two tuples of the same type, and copies the first into the second (in the standard basis) if the latter is zero-initialized: $\COPY{}\ket{\sigma}\ket{0} = \ket{\sigma}\ket{\sigma}$.
The $\SWAPGate$ unitary swaps the two tuples of variables.
The unitary~$U_e$ evaluates a deterministic expression~$e$ into a zero-initialized output variable: $U_e \ket{\sigma}\ket0 = \ket\sigma \ket{\Bracks{e}(\sigma)}$,
where~$\Bracks{e}$ denotes the semantics of $e$.
Similarly, the unitary $U_\mu$ maps the zero state to a superposition
with amplitudes the square roots of the probabilities of $\mu$:
$U_\mu\ket{0, 0} \triangleq \sum_a \sqrt{\mu(a)} \ket{a, a}$.
This is the quantum analogue of a random sample, and indeed we obtain a sample from the distribution~$\mu$ by discarding the second variable (which has the same effect as measuring the first variable).
See also the discussion in \cref{sec:cost:model}.
The unitary $\PhaseOnZero{\tau}{\phi}$ for an angle $\phi \in [0, 2\pi]$ applies a phase $e^{i\phi}$ on $\ket0$, while leaving all other basis states unchanged.
For any unitary~$U$, the operators $\adjU{U}$ and $\ctrlU{U}$ correspond to $U^\dagger$ and $\proj0 \ot I + \proj1 \ot U$ respectively.

\paragraph{Unitary Statements}
The set of unitary statements is denoted \SetOfUnitaryCommands{}.
It includes applying a unitary~$U \in \SetOfUnitaryOps$ on quantum variables~$\vecq$, a sequence of two unitary statements, or calling a unitary procedure~$g$ ($\kwbasic{call}$) or its
inverse ($\kwbasic{call}^\dagger$) on some quantum variables $\vecq$.

\paragraph{Classical Statements}
The set of classical (probabilistic) statements is denoted \SetOfProbCommands{}.
This includes assigning an expression to a variable, sampling from a probability distribution, a sequence of two classical statements, a conditional branch, a call to a classical procedure~$h$ (\kwbasic{call}).
Finally, the  \CallUProcAndMeas{} statement enables the crucial interaction between the probabilistic and the unitary fragment:
it is a probabilistic statement that accepts some classical variables~$\vecx$, invokes a unitary procedure~$g$ with quantum variables initialized in the standard basis state $\ket{\vecx}$, measures the quantum variables after the execution of~$g$, and saves the measurement outcome back in the variable~$\vecx$.

\paragraph{Procedures}
Similar to statements, we have two types of procedures, unitary and probabilistic ones,
and each can either be defined or external.
Unlike for $\ProtoLang$ functions, $\CQPL$ procedures take their arguments by reference and have no return value.
A unitary procedure definition \kwbasic{uproc} consists of a procedure name~$g$, quantum variable parameter names~$\vecq$, and a body unitary statement~$w$.
A probabilistic procedure definition \kwbasic{proc} consists of a procedure name $h$, classical variable parameter arguments~$\vecx$, and a body probabilistic statement~$s$.
External unitary and probabilistic procedures (\kwbasic{ext proc}, \kwbasic{ext uproc}) only consist of a procedure name.

\paragraph{Programs}
A \CQPL{} program is a list of procedures.
We denote the \emph{procedure context} of a program by~$\Pi$, which is a mapping from procedure names to their source code.

{\crefname{figure}{Fig.}{Figs.}\begin{figure}[t]
\small{}
\QPLFullSyntax{}
\caption{Syntax of \CQPL{}. The type and (classical) expression syntax are common with \ProtoLang{} and omitted, see \cref{def:proto:syntax}.}
\label{def:cqpl:syntax-main}
\end{figure}}

\paragraph*{Semantics}
$\CQPL{}$ programs are said to be \textit{well-formed} under a set of constraints.
First, programs must be \textit{safe} (e.g., no out-of-bounds array accesses).
Second, they must be \textit{well-typed} under typing rules that we present in \Cref{app:cqpl:typing}.
Third, all recursion must be statically terminating, which can be enforced with a well-founded order on each function, requiring functions to call only smaller functions, or themselves, on structurally smaller arguments.
Under these constraints, programs admit a denotational semantics.
In particular, we associate to any probabilistic statement~$s\in\SetOfProbCommands$ and any unitary statement~$w\in\SetOfUnitaryCommands$ the following denotational semantics:
\begin{align}\label{eq:cqpl:semantics}
  \evalCQPL{s}, \evalCQPL{w} : \Sigma \to \Distr{\Sigma},
  \quad
  \evalUQPL{w} \in \cL(\cH)
\end{align}
For a probabilistic statement~$s$, $\evalCQPL{s}$ is obtained in a straightforward way by treating each statement as a probabilistic function.
For unitary statements~$w$, we have \emph{two} natural semantics:
a unitary semantics~$\evalUQPL{w}$, which is given by a unitary operator on the Hilbert space~$\mathcal H$ and used crucially in our analysis, and a probabilistic semantics~$\evalCQPL{w}$, which has the following natural definition: given~$\sigma \in \Sigma$, initialize the quantum variables corresponding to~$\Vars{}$ in the standard basis state~$\ket\sigma$ and all auxiliary quantum variables in the~$\ket0$ state, then apply the unitary semantics of~$w$ to this quantum state, and finally measure the quantum variables corresponding to~$\Vars{}$ to obtain a distribution over~$\Sigma$.
The precise definition of the semantics of \CQPL{} is given in \cref{app:cqpl:semantics}.

\paragraph*{Cost and Cost Expressions}
\CQPL{} programs also admit a standard \emph{cost semantics}, which models the actual query cost of \CQPL{} programs when run in a given initial state.
Formally, to any probabilistic statement $s \in \SetOfProbCommands{}$ and any unitary statements~$w \in \SetOfUnitaryCommands{}$ we associate
\begin{align}\label{eq:cqpl:costs}
  \cqplexpcost{s} : \Sigma \to \CostExpr{}, \quad \uqplcost{w} : \CostExpr{},
\end{align}
where $\CostExpr{}$ denotes the set of cost expressions.
A \emph{cost expression} is a mapping from declared procedure identifiers to numbers of calls, reflecting the query cost model introduced in \Cref{sec:cost:model}.
For a probabilistic \CQPL{} statement~$s$, the expected cost~$\cqplexpcost{s}$ maps an input~$\sigma$ to the \emph{expected cost} of running the probabilistic statement~$s$ on~$\sigma$.
In contrast, any unitary \CQPL{} statement~$w$ corresponds to a quantum circuit without any control flow (a~quantum straight-line program), and hence its cost is independent of the input; thus $\uqplcost{w}$ is simply a cost expression.
The full formal details are given in \cref{cqpl:cost}.

\subsection{Compiler}
\label{sec:compiler}
To quantize \ProtoLang{} programs, we define a quantum compiler \QuantumCompiler{} that maps \ProtoLang{} functions and statements to \CQPL procedures and statements, respectively.
Primitives are compiled into quantum algorithms, which obtain their advantage by querying their input in superposition, using unitary quantum queries of the form of \cref{eq:U_f}.
To support such unitary quantum queries to subroutines, as explained in the overview in \cref{sec:overview}, we also require another compiler \UCostCompiler{}, which maps \ProtoLang{} statements and functions to the unitary fragment of \CQPL{}, respectively, handling allocation of auxiliary quantum variables as well as the \emph{compute-uncompute} pattern~\cite{watrous2008quantumcomputationalcomplexity} whenever necessary.
We first discuss the unitary compiler~$\UCostCompiler$, as it contains the quantum core of \CQPL{}, and then present the general quantum compiler~$\QuantumCompiler$, which calls out to the former when compiling primitive invocations (see below and \cref{fig:compiler:quantum}).

\begin{figure}[t]
\NotationSize{}
\NotationBox{\UnitaryCompiler}
\begin{align}
  \compileUQPL{\protodef{f}{\veca}{}{s}{\vecr}}
  &\;=\; \uqplprocdef{f^U}{\veca, \vecr, \vecz}{
    \qpunitary{\veca, \veca'}{\COPY{}};~
    \compileUQPL{s}
  }
  \nonumber\\
  \compileUQPL{\protoext{f}}
  &\;=\; \extuproc{f^U}
  \nonumber\\
  \compileUQPL{x <- e}
  &\;=\; \qpunitary{\FreeVars{}(e), x'}{U_e};~ \qpunitary{x, x'}{\SWAPGate};
  \nonumber\\
  \compileUQPL{\protosample{x}{\mu}}
  &\;=\;
    \qpunitary{x', x''}{U_{\mu}};~
    \qpunitary{x, x'}{\SWAPGate};~
  \nonumber\\
  \compileUQPL{s_1; s_2}
  &\;=\; \compileUQPL{s_1}; \compileUQPL{s_2}
  \nonumber\\
  \compileUQPL{\protoif{b}{s_t}{s_f}}
  &\;=\; 
  \begin{aligned}
    &
    \qpunitary{\vecy_t, \vecy_t'}{\COPY{}};~
    \compileUQPL{s_t};~
    \qpunitary{\vecy_t, \vecy_t'}{\SWAPGate{}};~
    \\&
    \qpunitary{\vecy_f, \vecy_f'}{\COPY{}};~
    \compileUQPL{s_f};~
    \qpunitary{\vecy_f, \vecy_f'}{\SWAPGate{}};~
    \\&
    \qpunitary{b, \vecy_t, \vecy'_t}{\CSWAPGate};~
    \\&
    \qpunitary{b}{X};~
    \qpunitary{b, \vecy_f, \vecy'_f}{\CSWAPGate};~
    \qpunitary{b}{X};
  \end{aligned}
  \nonumber\\
  \compileUQPL{\vecy <- f(\vecx)}
  &\;=\;
    \qpcallu{f^U}{\vecx, \vecy', \vecz'};~ %
    \qpunitary{\vecy, \vecy'}{\SWAPGate}
  \nonumber\\
  \compileUQPL{\vecy <- \GenericPrimitiveCall{}}
  &\;=\;
    \qpcall{\UAlgPrim{\eps}[f^U_1, \ldots]}{\vecx^{(1)}, \ldots, \vecy', \vecz'};~ %
    \qpunitary{\vecy, \vecy'}{\SWAPGate{}}
  \label{eq:compile:prim:unitary}
\end{align}
\caption{Unitary compiler $\UnitaryCompiler$.
The symbols $\vecz$, $\veca'$, $x'$, $x''$, etc.\ are defined in the text.
In the compilation of primitive invocations in the last line, we abbreviate the partially-applied function arguments by $\lambda_i = f_i(\vecx^{(i)}, \BlankArg{}^{*})$.}
\label{fig:compiler:unitary}
\end{figure}

\paragraph*{Unitary Compilation}
The unitary compiler $\UnitaryCompiler{}$ is defined inductively in \cref{fig:compiler:unitary}.
In the following, $\FreshVars{x}$ denotes a fresh set of auxiliary quantum variables with the same types as~$x$.
Function definitions~$f$ are compiled to unitary procedures~$f^U$, which take as arguments quantum variables corresponding to the function's parameters and return values (recall that arguments in $\CQPL$ are passed by reference), as well as auxiliary quantum variables $\vecz$ used by the body as will be explained momentarily.
We denote by $\auxUnitary{f} \coloneqq \vecz$ the set of auxiliary quantum variables used by the compilation of a function~$f$.
The procedure first copies the parameters in the standard basis to a fresh set of auxiliary variables $\veca' = \FreshVars{\veca}$.
External functions are compiled to external unitary procedures.

We now describe the unitary compilation of statements (see \cref{fig:example:unitary:compilation} for an illustrative example).
For compiling a basic expression $x <- e$,
we first apply the unitary~$U_e$ to the free variables~$\FreeVars(e)$ of the expression and to a fresh auxiliary variable $x' = \FreshVars{x}$, which will be zero-initialized, and then swap the result into $x$.
Note that this is well-defined even when $\FreeVars(e)$ contains $x$.
For a sampling statement $\protosample{x}{\mu}$,
we similarly apply the unitary~$U_\mu$ that prepares a superposition corresponding to~$\mu$ on $x' = \FreshVars{x}$, $x'' = \FreshVars{x}$ (which are zero-initialized), and swap $x'$ into $x$.
The variable $x''$ ensures that a probabilistic computation is kept classical:
the quantum variable~$x$ is entangled with the auxiliary variable~$x''$;
as a consequence, the reduced state of $x$ is no longer in a superposition but follows the classical distribution~$\mu$.
This is the standard way of realizing classical distributions in unitary quantum circuits, without requiring any measurement.
A sequence is compiled to a sequence of the individual compilations.
Conditional statements are compiled using if-conversion: we run both statements, and select the right result based on the control bit.
Here $\vecy_t, \vecy_f$ denotes the set of variables that are modified in the branches $s_t, s_f$ respectively,
and $\vecy'_t = \FreshVars{\vecy_t}, \vecy'_f = \FreshVars{\vecy_f}$ are zero-initialized auxiliary variables.
For each branch, we first copy the variables about to be modified, run the compilation of the branch, and then swap back the old values.
At the end, we conditionally swap back the computed results, controlled on $b$.
For function calls $\vecy <- f(\vecx)$, the compilation is similar to basic expressions:
we call the compiled procedure $f^U$ on the input, fresh outputs $\vecy' = \FreshVars{\vecy}$, and fresh auxiliary variables $\vecz' = \FreshVars{\auxUnitary{f}}$,
and then swap out the output variables $\vecy$ for $\vecy'$.

\begin{figure}[t]\NotationSize{}
\begin{subfigure}[c]{0.3\textwidth}
\begin{align*}
     & \kwbasic{fn}~ f(b, x, y) ~\kwbasic{do}~
  \\ &\quad \kwbasic{if}~ b ~\{
  r <- x + y;
  \}
  \\ &\quad \kwbasic{else}~ \{
  r <- x - y;
  \} 
  \\ &\quad \kwbasic{return}~ r
  \\ & \kwbasic{end}
\end{align*}
\end{subfigure}
\begin{subfigure}[c]{0.25\textwidth}
\centering
\[ \xrightarrow{\hspace{0.1\textwidth} \UnitaryCompiler{} \hspace{0.1\textwidth}} \]
\end{subfigure}
\begin{subfigure}[c]{0.3\textwidth}
\begin{align*}
  & \kwbasic{uproc}~ f^U(b, x, y, r, z, r_t, z') ~ \{
  \\&\quad \qpunitary{r, z}{\SWAPGate{}};~ \qpunitary{x, y, r}{U_{+}};
  \\&\quad \qpunitary{r, r_t}{\SWAPGate{}};~
  \\&\quad \qpunitary{r, z'}{\SWAPGate{}};~ \qpunitary{x, y, r}{U_{-}};
  \\&\quad \qpunitary{b, r, r_t}{\CSWAPGate{}};
  \\& \}
\end{align*}
\end{subfigure}
\caption{Illustration of a \ProtoLang{} function~$f$ and its unitary \CQPL{} compilation~$f^U$.
In the unitary compilation, both branches of the conditional statement are run on fresh auxiliary quantum variables.
Before the final step, the variable $r$ has the result of the \kwbasic{else} branch, and the variable $r_t$ has the result of the \kwbasic{then} branch.
The final line swaps $r$ with $r_t$, controlled on $b$, to store the correct result in $r$.}
\label{fig:example:unitary:compilation}
\end{figure}

Each $\eps$-annotated primitive $\cP_\eps$ compiles to a \CQPL{} unitary procedure $\UAlgPrim{\eps}$ that implements the unitary quantum algorithm underlying the primitive and can make calls to the unitary compilations~$f_i^U$ of its function arguments~$f_i$.
The compilation is shown in \cref{eq:compile:prim:unitary}.
The partially-applied arguments $\vecx^{(i)}$ are passed to $\UAlgPrim{\eps}$, along with the return variables~$\vecy$, the auxiliary variables used by the~$f^U_i$, and any additional auxiliary variables used by the algorithm (denoted $\auxUnitary{\cP}$).
Similar to expressions and function calls, we swap out the output variables $\vecy$ with fresh auxiliary variables $\vecy' = \FreshVars{\vecy}$, to ensure each computation is run on zero-initialized output variables.

\paragraph*{Quantum Compilation}
The quantum compiler~\QuantumCompiler{} is defined inductively in \cref{fig:compiler:quantum}.
Function definitions are compiled to procedure definitions, which take the function's parameters and return variables as arguments (recall that arguments in $\CQPL$ are passed by reference). %
External functions are compiled to external procedures.
Assignments and sampling statements compile directly to the corresponding statements in the probabilistic fragment.
A sequence of statements compiles to a sequence of individual compilations.
Similarly, conditionals compile to conditionals in the target language, by compiling the statements in each branch.
A call to a $\ProtoLang{}$ function~$f$ compiles to a call to the \CQPL{} procedure $f$. %

\begin{figure}[t]
\NotationSize{}
\NotationBox{\QuantumCompiler}
\begin{align}
  \compileCQPL{\protodef{f}{\veca}{}{s}{\vecr}}
  &\;=\; \cqplprocdef{f}{\veca, \vecr}{}{\compileCQPL{s}}
  \nonumber\\
  \compileCQPL{\protoext{f}}
  &\;=\; \extcproc{f}
  \nonumber\\
  \compileCQPL{x <- e}
  &\;=\; \qpassign{x}{e}
  \nonumber\\
  \compileCQPL{\protosample{x}{\mu}}
  &\;=\; \qprandom{x}{\mu}
  \nonumber\\
  \compileCQPL{s_1;s_2}
  &\;=\;
  \compileCQPL{s_1};
  \compileCQPL{s_2}
  \nonumber\\
  \compileCQPL{\protoif{b}{s_t}{s_f}}
  &\;=\;
  \kwbasic{if}~ b ~\{~ \compileCQPL{s_t} ~\}
  ~\kwbasic{else} ~\{~ \compileCQPL{s_f} ~\}
  \nonumber\\
  \compileCQPL{\vecy <- f(\vecx)}
  &\;=\; \qpcall{f}{\vecx, \vecy}
  \nonumber\\
  \compileCQPL{\vecy <- \GenericPrimitiveCall{}}
  &\;=\; \qpcall{\QAlgPrim{\eps}[f_1, f^U_1, \ldots]}{\vecx^{(1)}, \ldots, \vecy}
  \label{eq:compile:prim:quantum}
\end{align}
\caption{Quantum compiler $\QuantumCompiler$. In the compilation of primitive invocations in the last line, we abbreviate the partially-applied function arguments by $\lambda_i = f_i(\vecx^{(i)}, \BlankArg{}^{*})$, as in \cref{fig:compiler:unitary}.
The quantum procedure $\QAlgPrim{\eps}$ implementing the primitive call gets access to the quantum compilation~$f_i$ \emph{as well as to the unitary compilation}~$f^U_i$ of the function arguments, and it can make calls to the unitary fragment of the language as desired.
}
\label{fig:compiler:quantum}
\end{figure}

Similar to the unitary case, each $\eps$-annotated primitive $\cP_\eps$ compiles to a \CQPL{} procedure $\QAlgPrim{\eps}$ that implements the quantum algorithm underlying the primitive.
It may call out to the unitary fragment of the language as desired and can therefore make calls to both the quantum and unitary compilations of its function arguments.
The compilation of a primitive invocation is shown in \cref{eq:compile:prim:quantum}, where $f_i$ and $f^U_i$ are the quantum and unitary compilation of the function arguments~$f_i$, respectively.
The partially-applied arguments $\vecx^{(i)}$ are passed to $\QAlgPrim{\eps}$, along with the return variables~$\vecy$.

\section{Source-Level Error Analysis}
\label{sec:errors}
Having presented the compilation from $\ProtoLang{}$ to $\CQPL{}$ programs and the semantics of \ProtoLang{} and \CQPL{} programs, we now discuss the correctness of this compilation.
As explained, most quantum algorithms that offer speedups are Monte Carlo algorithms whose output may be incorrect with some probability.
In particular, this is true for the quantum algorithms used to implement primitives.

As explained in the compiler, each primitive is annotated with a parameter~$\eps$ that specifies the maximum allowed error of its compiled implementation, as compared to the ideal semantics.
The errors of individual calls add up in nontrivial ways, resulting in some overall error in the whole program.
We capture this overall error using a source-level error analysis.
More specifically, for both the unitary and the general quantum compiler, we define an error bound that can be computed from the source program and prove the following key property: the compilation preserves the semantics of source programs, for any input, up to total variation distance error that is upper-bounded by the source-level error bound.
This can be stated succinctly using the distance metric~$\Delta$ defined in \cref{eq:Delta}; see \cref{thm:uqpl:compile:semantics:simpler,thm:cqpl:compile:semantics}.

\subsection{Probabilistic versus Unitary Semantics}
In \cref{sec:framework:targetlang} we associated with any unitary \CQPL{} statement~$w$ two natural semantics: a probabilistic semantics~$\evalCQPL{w}$ and a unitary semantics~$\evalUQPL{w}$.
The former captures the behavior when the statement is run on classical inputs and the output is measured, which allows us to compare naturally with the semantics of the probabilistic source language~$\ProtoLang$.
However, only the latter captures the general behavior of the statement when run on arbitrary superpositions, such as in the context of quantum queries.

To analyze the error incurred by primitive implementations that makes calls to unitary compilations of subroutines, we therefore need relate these semantics and their associated error measures ($\Delta$ for the probabilistic semantics, the operator norm for the unitary semantics).
We prove a key robustness theorem which summarizes this analysis in a modular way.
Before stating the result, recall that quantum queries to probabilistic functions are modeled by unitaries as in \cref{eq:U_F}.
We say that a unitary~$U = U_{XE \to YE'}$ is a \emph{unitary extension} of a probabilistic function $f \colon X \to \Distr{Y}$ if it acts as in \cref{eq:U_F}.
If $U$ is a unitary extension of a probabilistic function~$f'$ such that $\Delta(f,f') \leq \eps$, we say that it is an \emph{$\eps$-close unitary extension of~$f$}.
Then our result states that if a unitary quantum algorithm implements a certain classical functionality~$P[f]$ up to some small error when making calls to any \emph{ideal} unitary extension of some~$f$, it still does so when instead given an imperfect unitary implementation of~$f$ (provided the quantum algorithm does not make too many calls to the subroutine).
Formally:

\begin{restatable}[Robustness of unitary quantum algorithms]{theorem}{ThmApproxPrimUnitaryImpl}
\label{thm:quantum-approx-subroutine-subst}
Consider a unitary quantum algorithm $W[U, U^\dagger]$ that makes~$L$ calls to a unitary~$U$ and its inverse~$U^\dagger$ such that, whenever~$U$ is a unitary extension of a probabilistic function $X \to \Distr{Y}$, $W[U, U^\dagger]$ is a unitary extension of a probabilistic function $Z\to\Distr{R}$.
Suppose that~$f,P[f]$ are two probabilistic functions such that, for every unitary extension~$U$ of~$f$, $W[U, U^\dagger]$ is an $\eps$-close unitary extension of~$P[f]$.
Then, for any $\tilde\eps$-close unitary extension~$\tilde U$ of~$f$, $W[\tilde U, {\tilde U}^\dagger]$ is still an~$(\eps+ L\sqrt{2\tilde\eps})$-close unitary extension of~$P[f]$.
\end{restatable}

\noindent
Crucially, even though the quantum algorithm has access to quantum queries (unitary implementations) of the subroutine, the error bound is stated purely in terms of the distance measure~$\Delta$ of the corresponding probabilistic function.
We prove this using quantum information techniques and in particular the continuity theorem of \citet{kretschmann2007cont}, which states that if two quantum channels have a trace-norm distance at most~$\eps$, then they have isometric extensions which are at most~$\sqrt{2\eps}$ far in operator-norm distance.
The proof of \cref{thm:quantum-approx-subroutine-subst} is given in \cref{proof:ThmApproxPrimUnitaryImpl}.

\subsection{Error Analysis and Correctness of Unitary Compilation}

The unitary compilation $\UnitaryCompiler{}$ produces a unitary quantum program that approximately implements the source program.
This unitary program may use some auxiliary workspace variables, and it is essential for the compiler to handle these auxiliary variables correctly~\cite{unqomp2021}.

\begin{figure}[t]
\NotationSize{}
\NotationBox{\ErrProbU}
\begin{align}
  \progerrprobU{\protodef{f}{\veca}{}{s}{\vecr}}
  &= \progerrprobU{s}
  \nonumber\\
  \progerrprobU{\protoext{f}}
  &= 0
  \nonumber\\
  \progerrprobU{x <- e}
  &= \progerrprobU{\protosample{x}{\mu}}
  = 0
  \nonumber\\
  \progerrprobU{s_1; s_2}
  &= \progerrprobU{s_1} + \progerrprobU{s_2}
  \nonumber\\
  \progerrprobU{\protoif{b}{s_t}{s_f}}
  &= \progerrprobU{s_t} + \progerrprobU{s_f}
  \nonumber\\
  \progerrprobU{\vecy <- f(\vecx)}
  &= \progerrprobU{\Phi[f]}
  \nonumber\\
  \progerrprobU{\vecy <- \GenericPrimitiveCall{}}
  &= \eps
  + \sum_{i=1}^k \UAlgPrimQueries{\cP}{\eps}{i} \cdot \sqrt{2 \cdot \progerrprobU{\Phi[f_i]}}
  \label{eq:error-analysis:unitary:prim}
\end{align}
\caption{
The unitary error bound \ErrProbU{} bounds the error of the unitary compilation of \ProtoLang{} programs (\cref{thm:uqpl:compile:semantics:simpler}).
In the last line, we abbreviate the partially-applied function arguments by $\lambda_i = f_i(...)$;
$\UAlgPrimQueries{\cP}{\eps}{i}$ is a bound on the number of calls made by the unitary quantum algorithm~$\UAlgPrim{\eps}$ to the unitary compilation~$f_i^U$ of the $i$-th subroutine, and
$\Phi$ denotes the function context, which maps any function to its source code.}
\label{fig:error-analysis:unitary}
\end{figure}

The unitary error bound is defined by induction on the structure of the source program (\cref{fig:error-analysis:unitary}).
Function definitions are assigned the error bound of their body, while external functions are assumed to have zero error (only for notational simplicity).
Elementary \ProtoLang{} statements can be compiled perfectly to unitary \CQPL{} statements and hence incur zero error.
For composite statements, we bound the error by the sum of the errors of the parts; this is the case even for conditional statements, since in the unitary compilation both branches can be taken in superposition.
For primitive calls, there are two sources of error: the failure probability~$\eps$ of the unitary algorithm~$\UAlgPrim[{\cP}]{\eps}$ implementing the primitive~$\cP$, and the error due to the compilation of each function argument~$f_i$.
The latter depends on the error between a single call to the unitary compilation~$f_i^U$ and an ideal quantum query to~$f_i$, which can be bounded inductively in terms of the unitary error bound $\progerrprobU{\Phi[f_i]}$, and the number of calls made by~$\UAlgPrim[{\cP}]{\eps}$ to~$f_i^U$, denoted $\UAlgPrimQueries{\cP}{\eps}{i}$.
If we combine the error bounds for these two sources of errors using \cref{thm:quantum-approx-subroutine-subst} we obtain \cref{eq:error-analysis:unitary:prim}.
The following theorem confirms the soundness of this analysis (see \cref{eq:protolang:semantics,eq:cqpl:semantics} for the definition of the semantics):

\begin{restatable}[Unitary Error Analysis]{theorem}{ThmUnitaryCompilePreservesSemanticsSimpler}
\label{thm:uqpl:compile:semantics:simpler}
For every well-formed \ProtoLang{} statement $s$, the distance between its probabilistic semantics and the probabilistic semantics of its unitary compilation is bounded by the unitary error bound:
$\probDistance{\evalCQPL{\compileUQPL{s}}}{\evalProb{s}} \le \progerrprobU{s}$.
\end{restatable}
\begin{proof}[Proof Sketch]
We prove this by induction on $s$.
We use the triangle inequality for the operator norm to bound the error of a sequence of statements.
For primitives calls, we use \cref{thm:quantum-approx-subroutine-subst} described above, and the triangle inequality to replace one function argument at a time.
\end{proof}
\noindent The full proof is provided in \cref{proof:ThmUnitaryCompilePreservesSemantics}.

\subsection{Error Analysis and Correctness of Quantum Compilation}
Similarly as in the unitary case, the quantum error bound is defined inductively on the structure of the source program, see \cref{fig:error-analysis:quantum}.
There are two key differences.
First, for conditional \ProtoLang{} statements, we can bound the error by the maximum (rather than the sum) of the errors of the branches, since it is compiled by the quantum compiler to a conditional statement in the probabilistic fragment of \CQPL{}.
Second, the quantum algorithm~$\QAlgPrim[{\cP}]{\eps}$ implementing a primitive~$\cP$ can call not only the unitary compilation, but also the quantum compilation of its function arguments~$f_i$.
The latter can be incorporated in a straightforward way, see \cref{eq:error-analysis:quantum:prim}.
We can then prove the following theorem, which was already announced in \cref{sub:formal_guarantees}.

\begin{restatable}[Quantum Error Analysis]{theorem}{ThmQuantumCompilePreservesSemantics}
\label{thm:cqpl:compile:semantics}
For every well-formed \ProtoLang{} statement $s$, the distance between its probabilistic semantics and the probabilistic semantics of its quantum compilation is bounded by the quantum error bound:
$\probDistance{\evalCQPL{\compileCQPL{s}}}{\evalProb{s}} \le \progerrprob{s}$.
\end{restatable}
\begin{proof}[Proof Sketch]
We prove this by induction on $s$.
For a sequence of statements, we use the union bound to bound the overall error as the sum of individual errors.
Primitive calls are implemented by probabilistic~\CQPL{} procedures which can in turn call out to unitary~\CQPL{} procedures; we use a union bound in the probabilistic fragment and \cref{thm:quantum-approx-subroutine-subst} in the unitary fragment, similarly as above.
\end{proof}

\noindent
The full proof is provided in \cref{proof:ThmQuantumCompilePreservesSemantics}.

\begin{figure}[t]
\NotationSize{}
\NotationBox{\ErrProb}
\begin{align}
  \progerrprob{\protodef{f}{\veca}{}{s}{\vecr}}
  &= \progerrprob{s}
  \nonumber\\
  \progerrprob{\protoext{f}}
  &= 0
  \nonumber\\
  \progerrprob{x <- e}
  &= \progerrprob{\protosample{x}{\mu}}
  = 0
  \nonumber\\
  \progerrprob{s_1; s_2}
  &= \progerrprob{s_1} + \progerrprob{s_2}
  \nonumber\\
  \progerrprob{\protoif{b}{s_t}{s_f}}
  &= \max(\progerrprob{s_t}, \progerrprob{s_f})
  \nonumber\\
  \progerrprob{\vecy <- f(\vecx)}
  &= \progerrprob{\Phi[f]}
  \nonumber\\
  \progerrprob{\vecy <- \GenericPrimitiveCall{}}
  &= \eps
  + \sum_{i=1}^k \QAlgPrimQueriesU{\cP}{\eps}{i} \cdot \sqrt{2 \cdot \progerrprobU{\Phi[f_i]}}
  + \sum_{i=1}^k \QAlgPrimQueriesQ{\cP}{\eps}{i} \cdot \progerrprob{\Phi[f_i]}
  \label{eq:error-analysis:quantum:prim}
\end{align}
\caption{The quantum error bound \ErrProb{} bounds the error of the quantum compilation of \ProtoLang{} programs (\cref{thm:cqpl:compile:semantics}).
As before, we abbreviate the partially-applied function arguments by $\lambda_i = f_i(...)$;
$\QAlgPrimQueriesU{\cP}{\eps}{i}$ and $\QAlgPrimQueriesQ{\cP}{\eps}{i}$ are bounds on the number of calls made by the quantum algorithm~$\QAlgPrim{\eps}$ to the unitary and quantum compilations of the $i$-th subroutine, respectively, and $\Phi$ denotes the function context.}
\label{fig:error-analysis:quantum}
\end{figure}

\section{Source-level Cost Analysis}
\label{sec:cost}
We now present our source-level cost analysis.
It takes as input a \ProtoLang{} program and bounds the cost of the program's \CQPL{} compilation, as produced by the compilers~\CostCompiler{} and~\UCostCompiler{}, defined in \cref{sec:compiler}.
The cost analysis is comprised of three quantities:
\begin{itemize}
\item The \emph{unitary} cost bound~\CostMetricU{}, which relates to the unitary compilation.
\item The \emph{expected} cost bound~\CostMetricQ{}, which can be intuitively interpreted as an upper bound on the expected cost of the quantum compilation \emph{assuming no errors} in execution. This will in general depend on the program's input.
\item The \emph{havoc} cost bound~\CostMetricHavoc{}, which upper-bounds the worst-case cost of the quantum compilation \emph{even in the presence of errors}.
\end{itemize}
The actual cost of the unitary compilation is directly upper-bounded by the unitary cost bound (\cref{thm:uqpl:compile:cost}).
To obtain an upper bound on actual expected cost of the quantum compilation, we must combine the expected and havoc cost bounds with the source-level error analysis of \cref{sec:errors} (\cref{thm:cqpl:compile:cost}).
We emphasize that all three quantities can be computed on the level of the source program, without requiring any costly simulation of quantum circuits.

\subsection{Cost Expressions}\label{sub:cost_expressions}
Recall from \cref{sec:framework:targetlang} that the cost of a \CQPL{} program is naturally captured by a \emph{cost expression}, defined as a mapping from external procedure identifiers to the number of calls made.
When a \ProtoLang{} program is compiled to \CQPL{}, each source-level external function $f$ is compiled to a classical procedure declaration~$f$ and a unitary procedure declaration~$f^U$.
These correspond to classical and quantum queries to~$f$, respectively, reflecting the cost model introduced in \Cref{sec:cost:model}.
For example, the cost expression~$\{f^U \mapsto 3, g \mapsto 5\}$ corresponds to three unitary quantum queries to~$f$ and five classical queries to~$g$.
Cost expressions have a natural monoidal structure by point-wise addition and they can be compared point-wise.

Our cost analysis computes cost bounds that are similarly given by cost expressions.
We use the notation $\QuantumTick{f}$ for the cost expression corresponding to one unitary query to $f$; it is the singleton mapping that sends~$f^U$ to one and all other procedure identifiers to zero.
Analogously, we write $\ClassicalTick{f}$ for the singleton mapping corresponding to one classical query to~$f$.

\subsection{Unitary Cost Analysis}
\label{sec:cost:unitary}

\begin{figure}[t]
\NotationSize{}
\NotationBox{$\CostMetricU$}
\begin{align}
  \costu{\protodef{f}{\veca}{}{s'}{\vecr}}
  &= \costu{s'}
\nonumber \\
  \costu{\protoext{f}} &= \QuantumTick{f}
\nonumber \\
  \costu{x <- e}
  &= \costu{\protosample{x}{\mu}} = 0
\nonumber \\
  \costu{s_1; s_2}
  &= \costu{s_1} + \costu{s_2}
\nonumber \\
  \costu{\protoif{b}{s_t}{s_f}}
  &= \costu{s_t} + \costu{s_f}
\nonumber \\
  \costu{\vecy \is f(\vecx)}
  &= \costu{\Phi[f]}
\nonumber \\
  \costu{\vecy <- \GenericPrimitiveCall}
  &= \sum_{i = 1}^k \UAlgPrimQueries{\cP}{\eps}{i} \cdot \costu{\Phi[f_i]}
  \label{eq:cost-analysis:unitary:prim}
\end{align}

\caption{The unitary cost bound $\CostMetricU$ bounds the cost of the unitary compilation of \ProtoLang{} programs (\cref{thm:uqpl:compile:cost}).
In the last line, we abbreviate the partially-applied function arguments by $\lambda_i = f_i(...)$;
$\UAlgPrimQueries{\cP}{\eps}{i}$ is a bound on the number of calls made by the unitary quantum algorithm~$\UAlgPrim{\eps}$ to the unitary compilation~$f_i^U$ of the $i$-th subroutine, and
$\Phi$ denotes the function context, which maps any function to its source code.}
\label{fig:cost-metric-unitary}
\label{def:unitary-cost}
\end{figure}

The unitary cost bound is defined by induction on the structure of the source program, see \cref{fig:cost-metric-unitary}.
For convenience, function definitions are assigned the cost of their body.
External functions are counted as one quantum query, as discussed in the cost model~(\cref{sec:cost:model}).
Elementary \ProtoLang{} statements are compiled to elementary unitary \CQPL{} statements and hence have zero cost.
For composite statements, we bound the cost by the sum of the cost of the parts; this is the case even for conditional statements, since in the unitary compilation both branches can be taken in superposition.
For a primitive $\cP_\eps$, its unitary algorithm $\UAlgPrim{\eps}$ invokes each function argument $f_i$.
The total cost is given in \cref{eq:cost-analysis:unitary:prim}, where $\UAlgPrimQueries{\cP}{\eps}{i}$ denotes an upper bound on the number of calls this algorithm makes to the unitary compilation of $f_i$.
The latter is given by a formula for each primitive, which also accounts for uncomputation, i.e.\ if a primitive requires \emph{strong unitary access} to its function $f$, making $Q$ queries to it, then the query expression accounts for this as $2Q$ unitary queries to $f$.
The following theorem confirms the soundness of this cost analysis:
\begin{restatable}[Unitary Cost Analysis]{theorem}{ThmUnitaryCompilePreservesCost}
\label{thm:uqpl:compile:cost}
For every well-formed \ProtoLang{} statement $s$,
the cost of its unitary compilation is upper-bounded by the unitary cost bound:
$\uqplcost{\compileUQPL{s}} \le \costu{s}$.
\end{restatable}

\noindent We prove this by structural induction on the source program $s$.

\subsection{Input-sensitive Quantum Cost Analysis}

The input-sensitive quantum cost analysis comprises two cost bounds: $\CostMetricQ{}$ and $\CostMetricHavoc{}$, which were briefly motivated in \cref{sec:motivation:cost}.

\paragraph{Havoc Cost Analysis.}
Similarly as in the unitary case, the quantum havoc cost bound $\CostMetricHavoc{}$ is defined inductively on the structure of the source program, see \cref{fig:cost-metric-havoc}.
There are two key differences: conditionals are bounded by the maximum of the branches, and primitives may call both the unitary and quantum compilation of its function arguments $f_i$.

\begin{figure}[t]
\NotationSize{}
\NotationBox{$\CostMetricHavoc{}$}
\begin{align}
  \costqmax{\protodef{f}{\veca}{}{s'}{\vecr}}
  &= \costqmax{s'}
\nonumber \\
  \costqmax{\protoext{f}}
  &= \ClassicalTick{f}
\nonumber \\
  \costqmax{x <- e}
  &= \costqmax{\protosample{x}{\mu}}
  = 0
\nonumber \\
  \costqmax{\vecy \is f(\vecx)} &= \costqmax{\Phi[f]}
\nonumber \\
  \costqmax{s_1; s_2}
  &= \costqmax{s_1} + \costqmax{s_2}
\nonumber \\
  \costqmax{\protoif{b}{s_t}{s_f}}
  &= \max(\costqmax{s_t}, \costqmax{s_f})
\nonumber \\
  \costqmax{\vecy <- \GenericPrimitiveCall}
  &= \sum_{i = 1}^k \parens*{
    \QAlgPrimQueriesU{\cP}{\eps}{i} \cdot \costu{\Phi[f_i]}
    +
    \QAlgPrimQueriesQ{\cP}{\eps}{i} \cdot \costqmax{\Phi[f_i]}
  }
  \label{eq:cost-analysis:havoc:prim}
\end{align}
\caption{The havoc cost bound $\CostMetricHavoc{}$ bounds the havoc cost of the quantum compilation of \ProtoLang{} programs (used in \cref{thm:cqpl:compile:cost}).
In the last line, we abbreviate the partially-applied function arguments by $\lambda_i = f_i(...)$;
$\QAlgPrimQueriesU{\cP}{\eps}{i}$ and $\QAlgPrimQueriesQ{\cP}{\eps}{i}$ are bounds on the number of calls made by the quantum algorithm~$\QAlgPrim{\eps}$ to the unitary compilation~$f_i^U$ and quantum compilation $f_i$ of the $i$-th subroutine, respectively, and
$\Phi$ denotes the function context, which maps any function to its source code.}
\label{fig:cost-metric-havoc}
\end{figure}

\paragraph{Expected Cost Analysis.}
Th expected cost bound $\CostMetricQ$ provides a more informative estimate of the quantum cost because it depends on the input as well as the semantics of \ProtoLang{}.
It is defined inductively on the structure of the program in \cref{fig:cost-metric-quantum}, but additionally takes the state of the program as an input.
For primitives, we use fine-grained expected query cost formulas to each function argument as shown in \cref{eq:cost-analysis:expcost:prim}.
In particular, these must be upper-bounded by the worst-case havoc formulas, i.e.\ for every $f_i$:
\[
\sum_{\vecv \in \Vals} \QAlgPrimQueriesExpQ{\cP}{\eps}{i}{\SemParFuns{}, \vecv} \le \QAlgPrimQueriesQ{\cP}{\eps}{i}
\quad\text{and}\quad
\QAlgPrimQueriesExpU{\cP}{\eps}{i}{\SemParFuns{}} \le \QAlgPrimQueriesU{\cP}{\eps}{i}.
\]

\begin{figure}[t]
\NotationSize{}

\NotationBox{$\costq{\cF}{} : \Vals \to \CostExpr{}$}
\begin{align*}
  \costq{\protodef{f}{\veca}{}{s'}{\vecr}}{\vecv}
  &= \costq{s'}{\{ \veca : \vecv \}}
\\
  \costq{\protoext{f}}{\vecv} &= \ClassicalTick[{\vecv}]{f}
\end{align*}

\NotationBox{$\costq{s}{} : \Sigma \to \CostExpr{}$}
\begin{align}
  \costq{x <- e}{\sigma}
  &= \costq{\protosample{x}{\mu}}{\sigma}
  = 0
\nonumber \\ \costq{s_1; s_2}{\sigma}
  &= \costq{s_1}{\sigma} +
  \DistrExp{\sigma' \sim \evalProb{s_1}(\sigma)}{
    \costq{s_2}{\sigma'}
  }
\nonumber \\ \costq{\protoif{b}{s_t}{s_f}}{\sigma}
  &= \begin{cases}
    \costq{s_t}{\sigma} & ~\text{if}~ \sigma(b) = 1
    \\
    \costq{s_f}{\sigma} & ~\text{if}~ \sigma(b) = 0
  \end{cases}
\nonumber \\
  \costq{\vecy <- f(\vecx)}{\sigma)}
  &= \costq{\Phi[f]}{\sigma(\vecx)}
\nonumber \\
  \costq{\vecy <- \GenericPrimitiveCall}{\sigma}
  &= ~~
  \sum_{i = 1}^k ~
  \parens*{
  \begin{aligned}
  &\QAlgPrimQueriesExpU{\cP}{\eps}{i}{\SemParFuns} \cdot \costu{\Phi[f_i]}
  \\
  &+ \sum_{\vecv \in \Vals}
    \QAlgPrimQueriesExpQ{\cP}{\eps}{i}{\SemParFuns{}, \vecv} \cdot \costq{\Phi[f_i]}{\sigma(\vecx^{(i)}), \vecv}
  \end{aligned}
  }
  \label{eq:cost-analysis:expcost:prim}
\end{align}
\caption{The input-sensitive expected cost bound $\CostMetricQ$.
In the last line, we abbreviate the partially-applied function arguments by $\lambda_i = f_i(...)$;
the formula $\QAlgPrimQueriesExpU{\cP}{\eps}{i}{\SemParFuns}$ bounds on the expected number of calls made by the quantum algorithm~$\QAlgPrim{\eps}$ to the unitary compilation~$f_i^U$ of the $i$-th subroutine, and $\QAlgPrimQueriesExpQ{\cP}{\eps}{i}{\SemParFuns{}, \vecv}$ bounds the expected number of calls to the quantum compilation~$f_i$ with input $\vecv$, for $\SemParFuns{}$ the tuple of semantics of each partially applied function, i.e., $\SemParFuns{}_i(\vecv') = \evalProb{\Phi[f_i]}(\sigma(\vecx^{(i)}), \vecv')$;
and $\Phi$ denotes the function context, which maps any function to its source code.}
\label{fig:cost-metric-quantum}
\label{def:quantum-cost}
\end{figure}

We can then prove the following theorem, which was already announced in \cref{sub:formal_guarantees}:
\begin{restatable}[Input-sensitive Quantum Cost Analysis]{theorem}{ThmQuantumCompilePreservesCost}
\label{thm:cqpl:compile:cost}
For every well-formed \ProtoLang{} statement $s$, and well-formed input $\sigma$:
the cost of its quantum compilation is upper-bounded by the following cost bound:
\[
 \cqplexpcost{\compileCQPL{s}}(\sigma) \le \costq{s}{\sigma} + \progerrprob{s} \cdot \costqmax{s}.
\]
\end{restatable}
\begin{proof}[Proof Sketch]
We prove this by induction on $s$.
We use the \cref{thm:cqpl:compile:semantics} to bound the deviation of semantics of the source program and the target program, and in the branches where it deviates, we bound the cost by the havoc cost (i.e.\ worst case).
For primitives, we use the bounds on the expected number of calls to each $f_i$ to compute the total expected cost.
\end{proof}

The full proof is provided in \cref{proof:ThmQuantumCompilePreservesCost}.

\section{Implementation}
\label{sec:Implementation}

We implemented $\OurFramework$ as a Haskell package\footnote{\traqimplementationfootnote{}} with $\approx$ 12k LOC,
which includes the source-language $\ProtoLang{}$ as a DSL with static typing,
a compilation to the target-language $\CQPL{}$,
and cost analysis of source programs, which evaluates the source programs using the probabilistic semantics.
A more detailed exposition is given in \cref{app:Implementation} in the supplementary material.

\paragraph{Programs.}
Source programs are represented using a polymorphic AST~\cite{treesthatgrow2017}, parametrized by the set of primitives $\cP$:
e.g.\ \lstinline[style=haskellstyle]!data Expr $\cP$ = ... | PrimCall $\cP$!
and \lstinline[style=haskellstyle]!data Program $\cP$!.
This polymorphism also supports annotation: a program of the type \lstinline[style=haskellstyle]!Program (Ann Double $\cP$)! has primitives annotated with a failure probability.

\paragraph{Error Budgets.}
\OurFramework{} automates the task of computing individual error budgets for each primitive call, given a total error budget.
It first annotates the program with symbolic errors using symbols $\eps_i$ for each primitive.
Then it uses the havoc cost analysis to obtain an expression for the total error, and picks the individual error budgets to satisfy the chosen error budget.
Once these have been chosen, \OurFramework{} can carry out the expected-cost analysis, as well as compile to \CQPL{}.

The implementation currently uses a basic heuristic to split the error budget equally to each primitive call and its arguments.
We leave it to future work to incorporate more sophisticated strategies; for example, one might use nonlinear optimization in the spirit of \cite{Meuli_2020} to choose individual error budgets that minimize the overall cost bound, subject to the desired total error budget.

\paragraph{Cost Analysis.}
\OurFramework{} implements the cost analyses described in \cref{sec:cost}.
In fact, it provides a generic \texttt{CostModel} typeclass, and instantiates it for the query cost model discussed throughout.
We leave it to future work to instantiate it for more fine-grained cost models, such as circuit~size~or~depth.
For the cost bounds $\CostMetricU{}$ and $\CostMetricHavoc{}$, it takes an $\eps$-annotated program and a choice of cost model, and outputs a bound in that model.
For the input-sensitive cost bound $\CostMetricQ{}$, it takes an $\eps$-annotated program along with inputs --- the tuple of inputs to the main function, as well as a map from external procedure names to Haskell functions --- and similarly outputs a bound.

\paragraph{Compilation.}
In \cref{sec:compiler}, we discussed the requirements for the compiler to correctly compile primitive calls.
In particular, the unitary compiler in the tool ensures that each function argument~$f^U_i$ is called correctly by fixing the partially-applied arguments to~$\vecx^{(i)}$.
It also appropriately passes the auxiliary variables for each $f^U_i$ through.
For the primitives whose realization requires \emph{strong unitary access} to its function argument, as in \cref{eq:strong}, the implementation compiles its function arguments using the compute-uncompute pattern (\cref{sec:cost:model}).

\paragraph{Extensibility.}
\OurFramework{} supports adding new primitives with ease.
We enable this by implementing a growing polymorphic AST inspired by prior Haskell work on extensible ASTs~\cite{datatypesalacarte2008,treesthatgrow2017}.
We provide \textit{typeclasses} to allow adding functionality to each new primitive. That is, to add a new primitive, the user needs to implement only the relevant typeclasses, rather than changing the core framework.
For example, the class \texttt{ExpCost} provides the function $\CostMetricQ$:
\begin{lstlisting}[style=haskellstyle, numbers=none]
    class ExpCost $\cP$ where expCost :: $\cP$ -> $\Sigma$ -> Cost
\end{lstlisting}
Then one can add instances for primitives annotated with a failure probability:
\begin{lstlisting}[style=haskellstyle, numbers=none]
    instance (Eval $\cP$) => ExpCost (AnnFail $\cP$) where expCost (AnnFail $\eps$ p) $\sigma$ = ...
\end{lstlisting}

\section{Evaluation}
\label{sec:evaluation}

We evaluate \OurFramework{} from key perspectives.
In \cref{sub:eval prims}, we discuss the \emph{scope} of quantum algorithms that can be incorporated into the framework, indicate how new ones can be added, and survey the ones that already have been incorporated.
In \cref{sub:eval case studies} we consider \emph{case studies} to evaluate how \OurFramework{} can succinctly express %
a variety of programs from the literature that make use of these quantum algorithms.
Finally, in \cref{sub:eval performance} we comment on the \emph{scalability} of \OurFramework{} as compared to the baseline performance of classical simulation of quantum programs.

\subsection{Primitives in \OurFramework{}}\label{sub:eval prims}
Primitives in \OurFramework{} are modeled as higher-order functions.
Accordingly, any quantum algorithm that can be used to implement functionality with classical input and output can be incorporated.
This fits a broad range of quantum algorithmic subroutines (including ones that at first glance appear genuinely quantum, such as amplitude amplification) and end-to-end quantum programs.
We demonstrate this by incorporating a representative range of quantum algorithms into \OurFramework{}.

Adding a primitive to \OurFramework{} involves three aspects: a classical semantics that describes its ideal functionality, concrete bounds on the number of calls to each of its function arguments (for both classical and quantum queries), and the concrete quantum and unitary algorithms that realize the primitive for any desired error bound.
We import and adapt results from the quantum algorithms literature that analyze the complexity and correctness of quantum algorithms for each primitive.
\Cref{tab:primitives:full-list} lists the \emph{primitives} implemented in \OurFramework{}, and their simplified cost expressions.
We briefly explain each of these primitives, along with the algorithms that realize them.
More detailed analyses and concrete cost bounds for these primitives are provided in the supplementary material.

\begin{table}[t]
\centering
\renewcommand{\arraystretch}{0.9}
\small
\begin{tabular}{lcc}
  \textbf{Primitive} $(\cP[\lambda])$ &
  $\QAlgPrimQueriesExpU{\cP}{\eps}{}{\evalProb{\lambda}}$ &
  $\UAlgPrimQueries{\cP}{\eps}{}$ \\
  \toprule

  $\PrimAny{} / \PrimAll{} / \PrimSearch{}$ &
  \makecell[l]{
    $\bigO(\sqrt{N/K})$~\cite{boyer98qsearch,Cade2023quantifyinggrover}
    \\ \tiny \quad where $K = \abs*{\{ v \mid \evalProb{\lambda}(v) = \detstate{1} \}}$
  } &
  $\bigO(\sqrt{N} \log(1/\eps))$~\cite{zalka1999groverbasedquantumsearchoptimal,Cade2023quantifyinggrover} \\
  \midrule

  $\primOneFun{\PrimAmplify_{\pmin}}{f}{\vecx}$ &
  \makecell[l]{
    $2 \cdot \bigO(\frac1{\sqrt{\pgood}})$~\cite{boyer98qsearch,Cade2023quantifyinggrover}
    \\ \tiny \quad where $\pgood = \Pr_{\evalProb{\lambda}()}(1)$
  } &
  $\bigO(\frac1{\sqrt{\pmin}} \ln(2/\eps))$~\cite{yoder14FPAA} \\
  \midrule

  $\primOneFun{\PrimSimon_{\pcoll}}{f}{\vecx}$ &
  \multicolumn{2}{c}{$\bigO(n + \log(1/\eps))$~\cite{Kaplan2016}}
  \\
  \midrule

  \PrimName{min}/\PrimName{max}/\PrimName{argmin}/\PrimName{argmax} &
  \multicolumn{2}{c}{$\bigO(\sqrt{N}\log(1/\eps))$~\cite{durr99qmin,Cade2023quantifyinggrover}}\\
  \bottomrule
\end{tabular}
\vspace{-2mm}
\caption{\small
Query bounds for the primitives implemented in \OurFramework{}.
The functionality of the primitives is discussed in \cref{sub:eval prims}; each accepts a single (partially applied) function argument.
For simplicity, we show a big-O expression cost expressions here (for small $\eps$), but our tool implements concrete non-asympotic costs bounds.
}
\vspace{3mm}
\label{tab:primitives:full-list}
\end{table}

\subsubsection{Search}
Our first class of primitives are search-like primitives: \PrimAny{}, \PrimAll{}, and \PrimSearch{}, all of which accept a single deterministic boolean function~$f$ called the \emph{predicate}.
Then \PrimAny{} checks whether any element in its domain satisfies~$f$, \PrimAll{} checks whether all such elements satisfy $f$, and \PrimSearch{} returns a uniformly random element satisfying~$f$ (if such an element exists).

\paragraph*{Algorithms and Cost}
Quantum algorithms for search go back to \citet{grover1996}.
The variants by \citet{boyer98qsearch} and \citet{zalka1999groverbasedquantumsearchoptimal} are particularly suitable; a non-asymptotic analysis has been given by \citet{Cade2023quantifyinggrover}.
Instead of implementing these directly, we find it convenient to desugar the search-like primitives to the more general \PrimAmplify{} primitive, which we discuss next.
The resulting cost expressions are shown in \cref{tab:primitives:full-list}, where $N$ is the size of the search space and $K$ the number of solutions (i.e., elements satisfying the predicate).
See \Cref{app:search-prims} for detailed analysis and proofs.

\subsubsection{Amplification}
We provide a primitive \PrimAmplify{} to arbitrarily increase or ``amplify'' the success probability of a subroutine.
Classically this can be achieved by \emph{rejection sampling}; the \emph{quantum amplitude amplification} technique provides a quadratic speedup.
Formally, \PrimAmplify{} accepts a probabilistic function~$f$, called the \emph{sampler}~$f$, which returns random pairs, consisting of a sample~$x$ from some domain and a boolean~$b$ indicating whether the sample is \emph{good}.
Then, \PrimAmplify{} outputs a good sample if one exists, and otherwise it outputs $b = 0$.

\paragraph{Constraint}
The primitive call requires a bound~$\pmin$ on the probability~$\pgood$ of a good sample being drawn from the sampler.
Formally, we assume that either $\pgood = 0$ or $\pgood \ge \pmin$.
That is, if the sampler ever returns a good sample, it must do so with at least probability $\pmin$.

\paragraph{Quantum Algorithms and Cost}
The quantum algorithm used is a modified version of the quantum search by \citet{boyer98qsearch}, where we replace the uniformly random initial state with the one produced by a unitary extension of the sampler~$f$.
We adapt the non-asymptotic analysis of \cite{Cade2023quantifyinggrover} appropriately.
The unitary algorithm uses the \emph{fixed-point amplitude amplification} algorithm and its analysis by \citet{yoder14FPAA}.
The (simplified) cost expressions are shown in \cref{tab:primitives:full-list};
see \Cref{app:prim-amplify} for detailed analysis and proofs.

\subsubsection{Simon}
The original Simon's problem~\cite{simon97} is the following period-finding problem:
Given a two-to-one function $f:\{0,1\}^n \to \{0,1\}^n$ with non-zero \emph{period} $s\in \{0,1\}^n$  (i.e.\ $\forall x, f(x \oplus s)=f(x)$), find $s$.
The primitive \PrimSimon{} accepts~$f$ as a predicate and outputs the period.
This basic problem underlies quantum attacks on certain symmetric cryptosystems (as we will see in the case studies).

\paragraph{Constraint.}
The function $f$ must be deterministic, have a non-zero period~$s$, and be at least approximately two-to-one in the following sense:
(2) for every $t \not\in \{0, s\}$, there can be at most a $\pcoll$-fraction of $x \in \{0,1\}^n$ such that $f(x) = f(x \oplus t)$, where $\pcoll$ is specified in the primitive call.

\paragraph{Algorithms and Cost.}
The quantum algorithm uses~$O(n)$ rounds, each of which consist of running a unitary quantum circuit that makes one superposition query to~$f$ and measuring $n$ qubits to obtain a vector orthogonal to~$s$ with good probability.
The algorithm then outputs a vector orthogonal to all of them, which will be $s$ with high probability.
The precise algorithm, query complexity, and failure probability bound is proven by \citet[Theorem~1]{Kaplan2016}.
The unitary algorithm proceeds similarly, implementing the measurements and classical postprocessing in terms of a unitary extension.
The (simplified) cost expressions are shown in \cref{tab:primitives:full-list}; see \cref{app:prim-simon} for detailed analysis and proofs.

\subsubsection{Min/Max-finding}
$\OurFramework{}$ supports primitives for optimization: $\PrimName{max}$, $\PrimName{argmax}$, $\PrimName{min}$, and $\PrimName{argmin}$.
Each of these primitives accepts a deterministic function~$f$.
The primitives $\PrimName{max}$ and $\PrimName{min}$ return the maximum or minimum value of $f$, respectively, whereas $\PrimName{argmax}$ and $\PrimName{argmin}$ return an input~$x \in \tau_1$ that maximizes or minimizes~$f$, respectively.

\paragraph{Algorithms and Cost.}
Since the domain is unstructured, this can be tackled by iterated quantum search with increasing thresholds.
We use the quantum algorithm \textbf{QMax} and the corresponding cost analysis of \citet[Corollary 1]{Cade2023quantifyinggrover}.
The (simplified) cost expressions are shown in \cref{tab:primitives:full-list}.

\subsection{Case studies in \OurFramework{}}\label{sub:eval case studies}

Now we demonstrate that \OurFramework{} is able to express various quantized programs from the literature, to enable automated cost analysis.
\Cref{tab:case-studies} shows a list of case-studies implemented in \OurFramework{}, with the corresponding lines of code (LoC) of the \OurFramework{} implementation and the primitives used.

\begin{table}[t]
  \centering
  \small
  \begin{tabular}{llrl}
    \setlength{\tabcolsep}{4pt}
    \textbf{Program} &
    \textbf{Domain} &
    \textbf{LoCs} &
    \textbf{Primitives} \\
    \midrule

    \textsc{Triangle Finding}~\cite{Buhrman_2005}  &
    \textsc{Search} &
    43  &
    \PrimSearch{} \\

    \textsc{Farthest Points}~\cite{ClusteringAlgorithms} &
    \textsc{Search} &
    39 &
    $\PrimName{argmax}$ \\

    \textsc{Matrix Search}~\cite{Hoyer_2003,ambainis2010andortrees}  &
    \textsc{Search} &
    18 &
    \PrimSearch{}, \PrimAll{} \\

    \textsc{Depth-3 nand}~\cite{Hoyer_2003,ambainis2010andortrees}  &
    \textsc{Search} &
    28  &
    \PrimAll{} \\

    \hline

    \textsc{Max-K-Sat}~\cite{Cade2023quantifyinggrover} &
    \textsc{Optimization} &
    33  &
    $\PrimName{argmax}$/\PrimSearch{} \\

    \textsc{0/1 Knapsack}~\cite{wilkening2024quantumalgorithmsolving01} &
    \textsc{Optimization} &
    98  &
    \PrimAmplify{} \\

    \hline

    \textsc{3-Round Feistel Attack}~\cite{Kaplan2016}  &
    \textsc{Cryptanalysis} &
    64  &
    \PrimSimon{} \\

    \textsc{Even-Mansour Attack}~\cite{Kaplan2016} &
    \textsc{Cryptanalysis} &
    25  &
    \PrimSimon{} \\

    \bottomrule
  \end{tabular}
  \vspace{-2mm}
  \caption{\small Overview of case studies discussed in \cref{sub:eval case studies}.}
  \label{tab:case-studies}
\end{table}

\subsubsection{Search}
The first set of case studies uses search-like algorithms to find a solution over a larger space.
We first describe two algorithms that use a single primitive call, and then describe a nested algorithm.

\paragraph{Triangle Finding. } Given an undirected graph $G=(V,E)$, we want to find distinct vertices $(x,y,z)$, such that $(x, y)$, $(y, z)$, and $(x, z)$ are edges in the graph. This problem is shown to have a quantum speedup by \citet{Buhrman_2005}. This speedup uses quantum search, first to find edge $(a, b)$, and then again to find a vertex $c$, such that $(a, b, c)$ forms a triangle.
In $\OurFramework{}$, this is achieved by using the $\PrimSearch$ primitive.

\paragraph{Farthest Points. } The problem is to find a pair of points $(p_1,p_2)$ with the largest distance in a set of points $D$;
One approach to solve this problem is to implement a subroutine \textbf{quant\_find\_max}~\cite[Algorithm 1]{ClusteringAlgorithms}, by repeatedly using the $\PrimSearch$ primitive with increasing thresholds in each iteration, until the maximum is reached.
We instead implement the algorithm in $\OurFramework{}$ directly using the $\PrimName{argmax}$ primitive, which has concrete bounds.

\paragraph{NAND Trees.}
We study a class of Boolean formula evaluation problems called \AndOr{} trees or NAND trees: evaluate a Boolean formula tree where each node is a NAND gate, and the leaves are Boolean variables.
Seminal results by \citet{Hoyer_2003} and \citet{ambainis2004quantum} give a quadratic speedup using nested quantum search.
The matrix search problem in \cref{ex:matrix-search-example} is a special case of a depth-2 tree.
We implemented a depth-$k$ NAND tree for constant $k$ in \OurFramework{},
using \PrimAll{} to implement an AND gate, and negating the output to obtain NAND.

\subsubsection{Optimization}
We also implemented two examples of local optimization algorithms for NP-hard problems described in the literature~\cite{Cade2023quantifyinggrover,wilkening2024quantumalgorithmsolving01}.

\paragraph{Max-k-SAT}
Max-$k$-SAT is an optimization variant of the satisfiability problem, where given a $k$-SAT instance with $m$ clauses, we must find an assignment that satisfies the \emph{maximum} number of these clauses.
This problem is NP-Hard for~$k\geq2$.
We consider the more general \emph{weighted} version where each clause has some positive weight, and we want to maximize the sum of weights of satisfied clauses.
\Citet{Cade2023quantifyinggrover} describes a local \emph{hill-climbing} algorithm for this: start with a random assignment, and repeatedly search over the $d$-neighbourhood of the current assignment for an improvement (i.e.\ satisfying more clauses).
The \emph{$d$-neighbourhood} is the set of all assignments that differ from the current one in at most $d$ variables.
We implemented the two variants of the above algorithm described in \cite{Cade2023quantifyinggrover} (for $d = 1$ and a fixed number of iterations):
(1) \emph{simple} - pick any better neighbour using \PrimSearch{}, and
(2) \emph{steep} - pick the best neighbour using \PrimName{argmax}.

\paragraph{0/1 Knapsack.}
In the 0/1 knapsack problem, we are given $n$ items with weights $w_i$ and values $v_i$, and we need to pick a subset of total weight at most a capacity $c$, maximizing the total value.
This problem is also NP-hard.
\Citet{wilkening2024quantumalgorithmsolving01} describe a local quantum search algorithm that uses \PrimAmplify{} in each iteration.
They describe a biased sampler that they call \emph{quantum tree generator}, which flips whether each item is picked or not with some fixed probability, and only picks it if it fits in the capacity.
This sampler therefore outputs a new subset of items, and a flag is the new value is larger; quantum amplitude amplification then yields a good assignment with high probability.
\Citet{wilkening2024quantumalgorithmsolving01} also dequantize their quantum algorithm to arrive at a simple \emph{classical tree generator} algorithm for 0/1 knapsack.
Here we find that, intuitively, one can also go the other way around:
we implement the classical tree generator in \OurFramework{} and obtain its quantization fully automatically from our compiler.

\subsubsection{Cryptanalysis}
Another practical use of quantum subroutines to obtain speedups is in cryptanalysis~\cite{Kaplan2016,GroverMeetsSimon,cryptonestedsearch2024}.
We consider attacks on two common cryptographic schemes from the literature.
The \emph{three-round Feistel scheme} is a secure pseudo-random permutation~\cite{luby1988feistel}.
The \emph{Even-Mansour construction} builds a block cipher from a public permutation~\cite{jofc-1997-14081}.
We implement the quantum attacks first described by \citet{kuwakado2010feistel,kuwakado2012evanmansour}, and adapted to the approximate promise setting by \citet{Kaplan2016}.
Both these attacks construct an almost periodic function from the respective scheme, and use the \PrimSimon{} primitive to compute the period and extract the secret.

\subsection{Comparison of \OurFramework{} with classical simulation}\label{sub:eval performance}
\begin{table}[t]
\centering
\small
\begin{tabular}{lll}
\midrule
Program & Max. Input Size & Qubits \\
\midrule
Matrix Search & $1600 \times 1600$ & 1308 \\
Depth-3 NAND & $120 \times 120 \times 120$ & 2368 \\
Max-3-SAT Hill-climbing & $n = 70$ variables & 179 \\
Triangle Finding & $n = 80$ vertices & 1103 \\
\end{tabular}
\caption{Maximum problem size such that \OurFramework{} can compute input-sensitive costs within a time cutoff of~$10$s, as well the number of qubits used by the compiled program.
We note that simulating quantum circuits with more than $100$ qubits is typically classically intractable.
}
\label{tab:scalability:qubit-count}
\end{table}

Our implementation uses source-program evaluation to perform input-sensitive cost analysis.
To demonstrate its scalability for large input sizes, we run our tool on the case study programs above.
We run \OurFramework{} on each program with random inputs of increasing sizes, till a time cutoff of 10s.
The experiments were run on a Linux laptop with an Intel Core i7-1270P (12 cores, 16 threads) and 16 GB RAM.
\Cref{tab:scalability:qubit-count} describes the results, which show that our input-sensitive cost analysis scales well with problem sizes, and can estimate costs in regimes where the classical simulation of the compiled quantum programs is fully intractable.

\section{Related Work}
\label{sec:related_work}
This section discusses related work on quantum languages and cost analysis, as well as techniques from classical analysis that inspired our work.

\paragraph{Quantum Programming Languages}
There exist numerous quantum programming languages at higher level of abstraction beyond quantum circuits~\cite{qml2005,quipper2013,Steiger2018projectq,qsharp2018,scaffcc,PolytimeQuantumPL,chardonnet2025hyrql,tower2022}.
Among those relevant to our work are Silq~\cite{silq2022} which offers a strong type system with safe uncomputation, and
Qunity~\cite{Voichick_2023,Mints_2025} which unifies both classical and quantum semantics, and allows nested quantum subroutines.
In contrast, \OurFramework{} gives a \emph{classical} (probabilistic) language which we quantize automatically through primitives,
so the programmer does not need to deal with quantum variables, uncomputation, etc.

\paragraph*{Quantum Compilation and Optimization}
We use classical-quantum target $\CQPL{}$, inspired by \citet{selinger_qpl}, as a compilation target for \ProtoLang{}.
Our focus was to provide a cost analysis with provable guarantees,
but in the future it could be of interest to target more expressive languages and produce concrete programs, including implementing circuits for data loading.
A few useful target languages are Qunity~\cite{Voichick_2023,Mints_2025}, OpenQASM~\cite{openqasm3}, QIRO~\cite{qiro2022}, and QSSA~\cite{qssa2022},
which can enable running compiled programs hardware.
Another useful direction is to connect the compiled programs to quantum circuit optimizers~\cite{SynthCircOptimizers,VOQC,Sivarajah_2020_tket},
and to tools that support better strategies for uncomputation~\cite{qurts2025,unqomp2021,uncomp2024} to use fewer qubits and gates.

\paragraph{Quantum Cost Analysis: theory}
There has been considerable work on quantum cost analysis of various algorithms.
Expected quantum costs have been studied for specific algorithms: various quantum search implementations~\cite{Cade2023quantifyinggrover}, with applications to hill-climbing~\cite{Cade2023quantifyinggrover} and community detection~\cite{cade22communitydetection};
SAT~\cite{eshaghian2024hybridsatsolvers,brehm2024assessingfaulttolerantquantumadvantage,Brehm2023walks},
knapsack~\cite{wilkening2024quantumalgorithmsolving01,wilkening2025quantumsearchmethodquadratic},
as well as linear systems~\cite{lefterovici2025beyondasymptoticqls}
and the simplex algorithm for linear programming~\cite{nannicini_fast_2022,qubra2023simplex},
identifying subroutines such as search, max-finding, and linear systems.
Another common application is the cost analysis of nested quantum search for
cryptanalysis~\cite{quantumlatticesieves2020,quantumpreimageattack2017,quantumsvp2025,quantumsecurityaes2019,David2024}.
There is also work in cryptanalysis to design attacks by combining period finding and quantum search~\cite{GroverMeetsSimon}.
All these prior analyses are manual and application-specific, which \OurFramework{} makes a first attempt towards automating.

More recent related work presents an algorithmic framework for the cost analysis of generic nested search algorithms~\cite{cryptonestedsearch2024}, subsuming several of the above works.
This uses an algorithmic technique called variable-time search~\cite{ambainis2023vts,ambainis2006vts}, which produces more efficient quantum programs.
There is also work exploring divide-and-conquer frameworks in the quantum setting~\cite{childs2022divconq}, providing an analogue to the classical paradigm.
It would be interesting to integrate these results in future versions of \OurFramework{}.

\paragraph{Quantum Cost Analysis: practice}
Frameworks such as Cirq~\cite{cirq}, Qiskit~\cite{Qiskit}, Qualtran~\cite{qualtran2024}, Quipper~\cite{quipper2013} enable resource estimation of large quantum circuits.
\cite{yuan2024TComplexity} gives a cost analysis and optimizing compiler for T-complexity of unitary programs,
and QuRA~\cite{Colledan2025QuRA} is a type system for Quipper to automatically estimate gate and qubit costs.
These tools estimate the worst-case costs, whereas \OurFramework{} can estimate input-sensitive costs, and in the presence of errors.
The Scaffold compiler~\cite{scaffcc} instruments classical-quantum programs, but input-sensitive costs require classical simulation of quantum programs (or quantum execution, which is presently infeasible already for moderately small quantum programs).
In contrast, \OurFramework{} supports source-level analysis, which avoids compilation and costly simulation of quantum programs.

In addition, Hoare-style weakest-precondition logic has been used to reason about expected runtimes of quantum programs~\cite{Liu2022qweakest,Avanzini2022QExpTrans}, inspired by similar approaches in the probabilistic setting~\cite{kaminski2018wp_ert,BatzKKMV23}.
probabilistic setting.
There is also work on quantum abstract interpretation~\cite{QuantumAbstractInterpretation,Jorrand_Perdrix_2009},
and logic for reasoning about approximately correct quantum circuits~\cite{yu2025saqrqclogicscalableapproximate,10.1145/3290344}.
These techniques could be useful to estimate input-sensitive parameters required by \OurFramework{} in the cost analysis.

\paragraph*{Cost-aware compilation}
Several works study formally the interactions between compilation and
cost.  The Cerco project~\cite{AmadioABBCGMMMPPRCST13} uses
source-level analysis to compute space and time bounds for generated
assembly, with guarantees proven w.r.t. machine-code cost
model. Similarly, \cite{Carbonneaux0RS14} prove stack-space bounds
for CompCert \cite{compcert} generated machine code using a
quantitative Hoare logic and a certified transformer that turns
source-level bounds into valid machine-code bounds. Furthermore,
The Jasmin compiler was instrumented with leakage
transformers to infer idealized cost bounds of compiled programs from
source-level analysis~\cite{BartheGLP21}.
Our work is inspired by prior work on accuracy-aware compilers~\cite{Misailovic2022}.

\section{Conclusion and Outlook}
\label{sec:conclusion_and_outlook}

We presented \OurFramework{}, a principled approach to analyze the
input-dependent expected costs of quantized classical programs. Our framework
provides \ProtoLang{} language with high-level primitives amenable to quantum
speedups, a compilation to classical-quantum programs, and a corresponding
source-level cost analysis, which upper-bounds the expected
cost of the compiled programs with provable guarantees.

There are many interesting directions for future work, and we discuss a few of them below.
First, it would be useful to implement a program logic to verify and guarantee that the promise of each primitive call is satisfied by its function arguments.
Such a verification could be either automated or delegated to an interactive theorem prover.
Second is computing concrete bounds such as gate complexity with formal guarantees, to give a more realistic understanding of the quantum advantage.
Third, in the paper we describe a simpler compiler with the focus on cost analysis, but it would be useful to improve the compilation by
utilizing quantum circuit optimizations~\cite{SynthCircOptimizers, VOQC, Sivarajah_2020_tket, Quartz, Pointing_2024,CompilingWithoutHelperQubits,uncomp2024}
to produce more efficient quantum programs.
We can also explore more efficient algorithms to realize the primitives, such as variable-time quantum search~\cite{ambainis2006vts,ambainis2023vts,cryptonestedsearch2024} which yields optimal expected quantum complexity for the search problem.
Lastly, we could explore computing expected cost without running the classical programs, e.g.\ by using heuristics to estimate input-dependent parameters, or using a program logic to derive such parameters.
When using heuristics, we must also adapt our formal guarantees to account for the approximation errors in the heuristics.

\begin{acks}
We thank Stacey Jeffery, Ugo Dal Lago, Ina Schaefer, and Jordi Weggemans for interesting related discussions.

MW and AP acknowledge the \grantsponsor{BMBF}{German Federal Ministry of Education and Research}{https://www.bmbf.de/} (QuBRA, \grantnum{BMBF}{13N16135}) and the \grantsponsor{BMFTR}{German Federal Ministry of Research, Technology and Space}{https://www.bmbf.de/} (QuSol, \grantnum{BMFTR}{13N17173}).
MW also acknowledges support by the \grantsponsor{ERC}{European Research Council}{https://erc.europa.eu/} through an ERC Starting Grant (SYMOPTIC, \grantnum{ERC}{101040907}) and the \grantsponsor{DFG}{German Research Foundation}{https://www.dfg.de/} under Germany's Excellence Strategy - \grantnum{DFG}{EXC 2092 CASA - 390781972}.
\end{acks}

\bibliographystyle{ACM-Reference-Format}
\bibliography{references}

\clearpage
\appendix

\section{Additional Background}
\label{app:background}
This section provides a more detailed background to concepts of probabilistic and quantum computing that are used in the formal theory and proofs of our framework.
We also prove \cref{thm:quantum-approx-subroutine-subst}, which is a key quantum information result used to prove the soundness of error analysis.

\subsection{Probabilistic Computing}
Recall that to a finite set $A$ (such as $\Sigma$), we associate a space of discrete probability distributions~$\Distr{A} \subset A \to [0, 1]$.
For a distribution $\mu \in \Distr{A}$, the probability of obtaining a value $a$ is denoted $\mu(a)$; any distribution satisfies $\sum_{a \in A} \mu(a) = 1$.
We denote by $\detstate{a} \in \Distr{A}$ the \emph{delta distribution} for $a \in A$.
Given a distribution $\mu \in \Distr{A}$ and a \emph{probabilistic function} $M \colon A \to \Distr{B}$, we define the \emph{distribution expectation} $\DistrExp{\mu}{M} \in \Distr{B}$ as
\[
  \DistrExp{\mu}{M} = \DistrExp{a \sim \mu}{M(a)} = \sum_{a \in A} \mu(a) M(a).
\]
Given two probabilistic functions $F \colon X \to \Distr{Y}$ and $G \colon Y \to \Distr{Z}$, their \emph{composition} $G \circ F \colon X \to \Distr{Z}$ is then defined as
\[ (G \circ F)(x) = \DistrExp{F(x)}{G} = \DistrExp{y \sim F(x)}{G(y)}. \]
The \emph{total variation distance} of two distributions $\mu, \mu' \in \Distr{A}$ is $\TVDist{\mu}{\mu'} = \tfrac12 \sum_a \abs{\mu(a) - \mu'(a)}$.
This is a metric, hence satisfies the triangle inequality, and it is also contractive: for any probabilistic function~$F \colon A \to \Distr{B}$, we have $\TVDist{\DistrExp{\mu}{F}}{\DistrExp{\mu'}{F}} \leq \TVDist{\mu}{\mu'}$.
The total variation distance induces a distance~$\Delta$ on probabilistic functions $F, F' \colon A \to \Distr{B}$, defined as
\[
  \probDistance{F}{F'}
\coloneqq \max_{x \in A} \TVDist{F(x)}{F'(x)}
= \max_{\mu \in \Distr{A}} \TVDist{\DistrExp{\mu}{F}}{\DistrExp{\mu}{F'}}
\]
The second formula follows by convexity.
The above is also a metric, hence it satisfies the triangle inequality.
It is also compatible with composition in the following sense:

\begin{lemma}\label{lem:sequence-approx-prob}
For any two probabilistic functions $F, F' : X \to \Distr{Y}$ and $G, G' : Y \to \Distr{Z}$
\[
	\probDistance{G \circ F}{G' \circ F'} \le \probDistance{G}{G'} + \probDistance{F}{F'}.
\]
\end{lemma}
\begin{proof}
We have
\begin{align*}
  \probDistance{G \circ F}{G' \circ F'}
\leq \probDistance{G \circ F}{G \circ F'} + \probDistance{G \circ F'}{G' \circ F'}
\leq \probDistance{F}{F'} + \probDistance{G}{G'},
\end{align*}
where we first used the triangle inequality and then the fact that the total variation distance is contractive under probabilistic functions, along with the second formula for~$\Delta$ given above.
\end{proof}

We show a result that compares the expectation of a positive random variable for nearby distributions.

\begin{lemma}
\label{lem:prob:deviation-of-expectation}
Let $\mu, \mu' \in \Distr{\Sigma}$ be two probability distributions, and let $T \colon \Sigma \to \R_{\geq0}$ be a positive random variable.
Then,
\[
  \DistrExp{\mu'}{T} \le \DistrExp{\mu}{T} + \TVDist{\mu}{\mu'} \cdot \max_{\sigma \in \Sigma} T(\sigma).
\]
\end{lemma}
\begin{proof}
Define the subset $\Sigma^{+} = \{ \sigma \in \Sigma \mid \mu'(\sigma) \ge \mu(\sigma) \}$, and let $\Sigma^{-} = \Sigma \setminus \Sigma^{+}$ be its complement.
Let~$w^{+} = \sum_{\sigma \in \Sigma^{+}} (\mu'(\sigma) - \mu(\sigma))$,
and~$w^{-} = \sum_{\sigma \in \Sigma^{-}} (\mu'(\sigma) - \mu(\sigma))$.
Then $w^{+} + w^{-} = 1 - 1 = 0$,
and by definition $w^{+} - w^{-} = 2 \TVDist{\mu}{\mu'}$.
Together, we see that $w^{+} = \TVDist{\mu}{\mu'}$.
Therefore:
\begin{align*}
  \DistrExp{\mu'}{T} - \DistrExp{\mu}{T}
  &= \sum_{\sigma \in \Sigma} \parens*{\mu'(\sigma) - \mu(\sigma)} T(\sigma)
  \le w^{+} \max_{\sigma \in \Sigma} T(\sigma)
  = \TVDist{\mu}{\mu'} \max_{\sigma \in \Sigma} T(\sigma)
  \qedhere
\end{align*}
\end{proof}

\subsection{Quantum Computing Background}
We first introduce the basic concepts of quantum computing, elaborating on \cref{sub:quantum_computing}, followed by the formalism and results of quantum information we use in our proofs.
We refer to the excellent textbooks by~\citet{nielsen2010quantum,wilde2013quantum,YM08} for more comprehensive introductions to quantum computing.

\paragraph*{Notation and Basics}
In this paper, a Hilbert space~$\mathcal H$ is a finite-dimensional complex vector space with an inner product.
Throughout the paper we use \emph{Dirac notation}:
we write~$\ket\psi \in \mathcal H$ for vectors, $\bra\phi$ for covectors, and~$\braket{\phi|\psi}$ for the inner product.
The \emph{norm} of a vector $\ket{\psi} \in \cH$ is given by $\norm{\psi} = \norm{\ket\psi} = \sqrt{\braket{\psi|\psi}}$.
Here, $\psi, \phi$ are arbitrary labels.
To any finite set $A$, we associate a \emph{Hilbert space} $\cH_A$, which has an orthonormal \emph{standard basis} (also called \emph{computational basis}) labelled by elements $a \in A$, denoted $\ket{a} \in \cH_A$.
A \emph{quantum variable}~$q$ with classical values in~$A$ is modelled the Hilbert space $\cH_A$.
Given two spaces $\cH_A$ and $\cH_B$, the combined space is defined by the tensor product $\cH_A \ot \cH_B$, which can be identified with~$\cH_{A \times B}$.
The identity operator on a Hilbert space~$\cH_A$ is denoted by $I_A = \sum_{a \in A} \proj{a}$, where we observe that $\proj a$ is the orthogonal projection onto the one-dimensional subspace~$\C \ket{a}$.
We write~$I$ when the Hilbert space is clear from the context.
The adjoint of a linear operator~$M$ is denoted by~$M^\dagger$.
An operator~$U$ is called unitary if~$UU^\dagger = U^\dagger U = I$.
An operator~$H$ is called Hermitian if~$M = M^\dagger$; it is called positive semidefinite if it is Hermitian and its eigenvalues are nonnegative.
The set of positive semidefinite operators is denoted $\cP(\cH)$.
The \emph{operator norm} of an operator $M$ is denoted $\norm{M}$,
and \emph{trace norm} is denoted $\norm{M}_1$.
Given an operator~$M$  that acts on some space~$\cH_A$, we can extend it to any larger space $\cH_A \ot \cH_B$ as $M_A = M \ot I$.
Given an operator~$M$ that acts on some space~$\cH_A \ot \cH_B$,
we denote its \emph{partial trace} over $B$ as $\tr_B(M)$, which is an operator that acts on $\cH_A$, defined as $\tr_B(M) = \sum_{b \in B} (I \ot \bra{b}) M (I \ot \ket{b})$.
A linear function~$\cE$ mapping operators on one Hilbert space to operators on another is called a \emph{superoperator}.
It is called \emph{trace-preserving} if~$\tr M = \tr \cE(M)$ for every operator~$M$.
A superoperator $\cE$ on $\cH$ is called \emph{positive} if~$\cE(M)$ is PSD for every PSD~$M$,
and \emph{completely positive} if the superoperator~$\cE \ot \cI_{\cH'}$ is positive for every Hilbert space~$\cH'$, where~$\cI_{\cH'}$ denotes the identity superoperator.
A \emph{quantum channel} $\cE \colon \cL(\cH_X) \to \cL(\cH_Y)$ is a completely-positive and trace-preserving superoperator.
We denote by~$\cE^\dagger$ the adjoint of a superoperator with respect to the Hilbert-Schmidt inner product, which satisfies the defining property that $\tr (A~\cE(B)) = \tr(\cE^\dagger(A)~B)$ for all operators~$A$,~$B$.
The adjoint is completely positive iff~$\cE$ is completely positive.
The natural distance measure on superoperators~$\cE$ is the \emph{diamond norm}, defined by $\norm{\cE}_\diamond = \max_{\cH',\rho} \norm{(\cE \ot \cI_{\cH'})(\rho)}_1$.

\paragraph*{Quantum States}
Recall that a \emph{pure state} of a quantum variable with Hilbert space~$\cH$ is specified by a unit vector~$\ket{\psi} \in \cH$.
More generally, one can consider \emph{mixed states}, which are given by positive-semidefinite operators~$\rho \in \cP(\cH)$ with trace equal to one (often called density operators or density matrices).
In this terminology, a pure state is a special case of a mixed state: any unit vector~$\ket\psi$ determines a mixed state~$\rho = \proj\psi$ of rank one, and any rank-one mixed state arises in this way.
Moreover, every probability distribution~$\mu \in \Distr{A}$ determines a mixed quantum state~$\rho = \sum_a \mu(a) \proj{a}$; such quantum states are called \emph{classical}.
Mixed states are the standard way to treat of quantum and probability theory in a unified way.
Furthermore, they naturally arise when considering subsets of quantum variables:
if $\rho$ is a state of two quantum variables~$A$ and~$B$, then $\rho_A = \tr_B(\rho)$ describes the state of quantum variable~$A$;
importantly, the latter can be mixed even if the former is pure.

\paragraph*{Quantum Operations}
We can apply unitary operators to quantum states.
If we apply a unitary~$U$ on~$\cH$ to a pure state $\ket{\psi}$, we obtain the pure state~$U\ket{\psi}$.
For example, the Hadamard matrix $H = \frac1{\sqrt2} \begin{psmallmatrix} 1 & 1 \\ 1 & -1 \end{psmallmatrix}$
is a unitary acting on the Hilbert space of a qubit~$\cH=\C^2 = \cH_{\{0,1\}}$, and on applying it to the input state $\ket0$, we get $\ket{+} := H \ket{0} = \frac1{\sqrt2}(\ket0 + \ket1)$.
More generally, an application of a unitary operator~$U$ is modeled by the quantum channel~$\cU$ that sends any input mixed state~$\rho$ to the output mixed state~$\cU(\rho) = U \rho U^\dagger$.

We can also measure quantum states.
A \emph{measurement} (here, \emph{standard basis measurement}) is a quantum operation, that on an input pure state $\ket{\psi}$,
outputs a basis label $\sigma$ with probability $\abs{\braket{\sigma|\psi}}^2$,
and the state of the quantum system in the end becomes (``collapses to'') $\ket{\sigma}$.
More generally, a standard basis measurement can be modeled by the quantum channel~$\cM$ that sends any input mixed state~$\rho$ to the output mixed state~$\cM(\rho) = \sum_\sigma \braket{\sigma|\rho|\sigma} \proj\sigma$.

Finally, we can always add quantum variables in some well-defined initialize state ($\rho \mapsto \rho \ot \proj0$) as well as discard quantum variables (modeled by the partial trace).

\paragraph*{Quantum Channels}
The above quantum operations all corresponds to quantum channels.
Conversely, any quantum channel can be implemented by applying an isometry to (or a unitary on) a larger system, i.e.\ using an auxiliary space, which is subsequently discarded.
This important result is known as Stinespring's theorem.
The appropriate notion of an extension is formalized in the following definition.

\begin{definition}[Isometric or Unitary Extension]\label{def:iso uni ext}
An \emph{isometric extension} of a channel $\cE \colon \cL(\cH_X) \to \cL(\cH_Y)$ is an isometry $V \colon \cH_X \to \cH_Y \ot \cH_Z$ such that $\cE(\rho) = \tr_{Z} [V \rho V^\dagger]$.
A \emph{unitary extension} is a unitary $U \colon \cH_X \ot \cH_Z \to \cH_Y \ot \cH_{Z'}$ such that $U(I_X \ot \ket0_{Z})$ is an isometric extension of~$\cE$.
\end{definition}

Isometric extensions are not unique, since the state of the auxiliary variable~$Z$ can be modified by an arbitrary unitary.
However, this is the only source of ambiguity.
In fact, the following result by \citet{Kretschmann2008normbounds,kretschmann2007cont} shows that nearby channels have nearby isometric extensions, and vice versa.
Namely, the diamond norm distance of two channels and the operator norm of their isometric extensions can be bounded in terms of each other.

\begin{theorem}[{\cite[Theorem~1]{Kretschmann2008normbounds}}]
\label{lemma:error-channel-to-stinespring}
For any two channels $\cE_1, \cE_2 : \cL(\cH_X) \to \cL(\cH_Y)$ with isometric extensions $V_1, V_2 \colon \cH_X \to \cH_Y \ot \cH_Z$, we have
\[
    \min_{U} \norm{(I \ot U)V_1 - V_2}^2
    \le
    \norm{\cE_1 - \cE_2}_{\diamond}
    \le
    2 \min_{U} \norm{(I \ot U)V_1 - V_2}.
\]
\end{theorem}

\subsection{Lifting Probabilistic Computing to Quantum Computing}\label{sub:quantum_channels_and_probabilistic_functions}
The goal of this section is to prove \cref{thm:quantum-approx-subroutine-subst}, which states that if a unitary quantum algorithm approximately implements a certain functionality when given ideal quantum queries to classical subroutines, then it still does so approximately when given imperfect quantum queries to these subroutines.
The proof requires some background from quantum information theory.

\paragraph*{Lifting Probabilistic States and Functions}
First, as explained above, we can lift any probability distribution to a so-called ``classical'' quantum state:

\begin{definition}[Quantum state associated to a probability distribution]
  To every distribution $\mu \in \Distr{X}$, we associate a mixed quantum state $\rho_\mu = \sum_{x \in X} \mu(x) \proj{x}$.
  Such states are called \emph{classical}.
\end{definition}

Similarly, we can lift any probabilistic function to a quantum channel:

\begin{definition}[Quantum channel associated to a probabilistic function]
\label{def:qchan-of-prob}
To every probabilistic function $F \colon X \to \Distr{Y}$,
we associate a quantum channel $\quantChan{F} \colon \cL(\cH_X) \to \cL(\cH_Y)$ defined as
\begin{equation}\label{eq:E_F action}
  \quantChan{F}(\rho) = \sum_{x\in X} \braket{x|\rho|x} \parens*{ \sum_{y \in Y} F(x)(y) \proj{y} }.
\end{equation}
\end{definition}

\noindent
Applying this channel on an input basis state $\proj{x}$ produces a classical output state corresponding to the distribution $\mu = F(x)$, i.e.\ $\quantChan{F}(\proj{x}) = \sum_{y} \mu(y) \proj{y} = \rho_\mu$.
More generally, given any input quantum state, $\quantChan{F}$ first measures it to obtain some~$x \in X$ and then proceeds as above.

For probabilistic functions, the distance~$\Delta$ coincides with the induced trace norm distance, as well as the diamond norm distance of the corresponding quantum channels:

\begin{lemma}
\label{lem:equivalence-of-channel-norms}
For any two probabilistic functions $F, F' : X \to \Distr{Y}$,
\[
\probDistance{F}{F'}
= \max_\rho \tfrac12 \norm{\quantChan{F}(\rho) - \quantChan{F'}(\rho)}_1
= \tfrac12 \norm{\quantChan{F} - \quantChan{F'}}_\diamond.
\]
\end{lemma}

We also consider a weaker notion of realizing a probabilistic function by a quantum channel, where we only demand the output state be correct on \emph{classical} input states.
This arises naturally in our setting of quantizing classical programs, but does not specify the channel action uniquely and is less well-behaved under composition.

\begin{definition}[Quantum channel implementing a probabilistic function]
\label{def:qchan-implementing}
A quantum channel $\cF \colon \cL(\cH_X) \to \cL(\cH_Y)$ \emph{implements} a probabilistic function~$F \colon X \to \Distr{Y}$ if~\eqref{eq:E_F action} holds for every classical input state~$\rho$.
Equivalently, for every~$x \in X$ we have
\[
  \cF(\proj{x}) = \sum_{y \in Y} F(x)(y) \proj{y}.
\]
\end{definition}

\indent
Note that $\cF$ implements a probabilistic function~$F$ if, and only if, $\cF \circ \cM_X = \cE_F$, where $\cM_X$ denotes the quantum channel corresponds to a standard basis measurement of quantum variable~$X$.

\paragraph*{Unitary Extensions and Implementations}
Recall that any quantum channel has a unitary extension (\cref{def:iso uni ext}).
We use the term ``unitary extension'' of a probabilistic function to mean a unitary extension of the quantum channel associated to it in \cref{def:qchan-of-prob}:

\begin{definition}[Unitary extension]
A unitary $U = U_{XE \to YE'}$ is called a \emph{unitary extension of} a probabilistic function $F : X \to \Distr{Y}$ if $U$ is a unitary extension of the quantum channel~$\quantChan{F}$.

It is called an \emph{$\eps$-close unitary extension} of~$F$ if it is a unitary extension of some $F' : X \to \Distr{Y}$ such that $\probDistance{F}{F'} \le \eps$.
\end{definition}

We can then bound the TV distance of probabilistic functions in terms of the operator-norm distance of arbitrary unitary extensions:

\begin{lemma}
\label{lemma:error-unitary-to-tv}
Let $U, U'$ be unitary extensions of probabilistic functions $F, F' : X \to \Distr{Y}$.
Then,
\[
  \probDistance{F}{F'}
  \le
  \norm{U - U'}.
\]
\end{lemma}
\begin{proof}
We have that $\probDistance{F}{F'} = \frac12 \norm{\cE_F - \cE_{F'}}_\diamond$ by \cref{lem:equivalence-of-channel-norms}.
We use the upper bound in \cref{lemma:error-channel-to-stinespring} to obtain $\norm{\cE_F - \cE_{F'}}_\diamond \le 2\norm{U - U'}$, which proves the desired result.
\end{proof}

Conversely, nearby probabilistic functions admit nearby unitary extensions:

\begin{lemma}
\label{lemma:error-tv-to-unitary}
For every two probabilistic functions $F, F' : X \to \Distr{Y}$, and for every unitary extension $U$ of $F$, there exists a unitary extension $U'$ of $F'$ such that
\[
  \norm{U - U'}
  \le
  \sqrt{2 \cdot \probDistance{F}{F'}}
\]
\end{lemma}
\begin{proof}
We have that $\probDistance{F}{F'} = \frac12 \norm{\cE_F - \cE_{F'}}_\diamond$ by \cref{lem:equivalence-of-channel-norms}.
We then use the lower bound in \cref{lemma:error-channel-to-stinespring} to find an isometric extension $V'$ for $F'$, such that
\[
  \norm{U(I \ot \ket0) - V'}
  \le \sqrt{\norm{\cE_F - \cE_{F'}}_\diamond}
  = \sqrt{2\probDistance{F}{F'}}.
\]
Then we can extend $V'$ to $U'$ s.th $\norm{U - U'} = \norm{V - V'} \le \sqrt{2\probDistance{F}{F'}}$, proving the result.
\end{proof}

We use the above results to bound the total error of quantum algorithms to implement some desired functionality when given an approximate unitary extension of a subroutine.

\ThmApproxPrimUnitaryImpl*
\begin{proof}
\label{proof:ThmApproxPrimUnitaryImpl}
By \cref{lemma:error-tv-to-unitary}, there exists a unitary extension $U$ of $f$ such that
$\norm{U - \tilde{U}} \le \sqrt{2\tilde\eps}$.
Hence $\norm{W[U, U^\dagger] - W[\tilde{U}, \tilde{U}^\dagger]} \le L \sqrt{2\eps}$ by the triangle inequality.
By assumption, we know that $W[U, U^\dagger]$ is a unitary extension of some probabilistic function~$g$,
and $W[\tilde{U}, \tilde{U}^\dagger]$ is a unitary extension of some probabilistic function~$\tilde{g}$.
Therefore using \cref{lemma:error-unitary-to-tv}, $\probDistance{g}{\tilde{g}} \le L \sqrt{2\tilde\eps}$.
On the other hand, because~$W[U, U^\dagger]$ is an $\eps$-close unitary extension of~$P[f]$, we know that $\probDistance{g}{P[f]} \leq \eps$.
Together with the triangle inequality, we obtain that $\probDistance{\tilde{g}}{P[f]}\le \eps+L\sqrt{2\tilde\eps}$.
Hence $W[\tilde{U}, \tilde{U}^\dagger]$ is an $(\eps+L\sqrt{2\tilde\eps})$-close unitary extension of~$P[f]$, concluding the proof of the theorem.
\end{proof}

Above, we also defined when a quantum channel implements a probabilistic function in a weaker sense (\cref{def:qchan-implementing}).
We will say that a unitary ``implements'' a probabilistic function if it is the unitary extension of such a channel.

\begin{definition}[Unitary implementation]
A unitary $U = U_{XE \to YE'}$ \emph{implements} a probabilistic function $F : X \to \Distr{Y}$ if $U$ is a unitary extension of a channel implementing~$F$ (\cref{def:qchan-implementing}).
It \emph{implements $F$ up to error $\eps$} if it implements some $F' : X \to \Distr{Y}$ such that $\probDistance{F}{F'} \le \eps$.
\end{definition}

While weaker than the notion of an unitary extension of probabilistic function, these notions can be related.
On the one hand, any unitary extension is also an implementation.
Hence:

\begin{lemma}
\label{lemma:unitary:extension-to-impl}
If a unitary $U = U_{XE \to YE'}$ is an $\eps$-close unitary extension of $F:X \to \Distr{Y}$, then $U$ implements $F$ up to error $\eps$.
\end{lemma}
\begin{proof}
By definition, $U$ is a unitary extension of some $F'$ such that $\probDistance{F}{F'} \le \eps$.
Therefore $U$ also implements $F'$, and therefore implements $F$ up to error $\eps$.
\end{proof}

On the other hand, we can convert any unitary implementation into a unitary extension by composing it with the unitary extension of a measurement channel (that is, with a $\COPY$ gate).
This also works robustly:

\begin{lemma}
\label{lemma:unitary:impl-to-extension}
If a unitary $U = U_{XE \to YE'}$ implements a probabilistic function $F:X \to \Distr{Y}$ up to error $\eps$, then $U' = U'_{X(EX') \to Y(E'X')} = (U \ot I_{X'})(\COPY{}_{XX'} \ot I_E)$ is an $\eps$-close unitary extension of~$F$.
\end{lemma}
\begin{proof}
By definition, $U$ implements some $F'$ such that $\probDistance{F}{F'} \le \eps$.
We claim that $U'$ is a unitary extension of $F'$.
To see this, note that if $U$ is the unitary extension of any quantum channel~$\cF'$ implementing $F'$, then
\begin{align*}
    &  \tr_{E'X'} \parens*{ U' \parens*{ \rho \ot \proj 0_E \ot \proj 0_{X'}} (U')^\dagger }
\\ &= \tr_{E'} \parens*{ U \parens*{ \tr_{X'} \parens*{ \COPY{}_{XX'} \parens*{ \rho \ot \proj 0_{X'} } \COPY{}_{XX'}^\dagger } \ot \proj 0_E } U^\dagger }
\\ &= \cF'(\mathcal M_X(\rho)) = \cE_{F'}(\rho),
\end{align*}
meaning that $U'$ is a unitary extension of the channel~$\cE_{F'}$, that is, of~$F'$, and therefore an $\eps$-close unitary extension of $F$.
\end{proof}

\section{Source Language \ProtoLang{}}
\label{app:protolang}
This appendix contains detailed typing rules and semantics for \ProtoLang{} omitted in \cref{sec:framework:targetlang}.

\subsection{Typing}
\label{app:protolang:typing}

\paragraph{Typing Contexts}
A \emph{typing context} $\Gamma = \{ x_i : \tau_i \}$ is a mapping from variable names to types.
We write~$x \in \Gamma$ if the typing context contains the variable $x$,
and its corresponding type is denoted $\Gamma[x]$.
We denote the tuple of variables of $\Gamma$ as $\Vars(\Gamma) = \{ x_i \}$.
Concatenating two typing contexts $\Gamma_1$ and $\Gamma_2$ is denoted~$\Gamma_1;\Gamma_2$.

\paragraph{Typing Judgements}
Our typing rules are stated under a global function context $\Phi$ and a global typing context $\Gamma$.
A well-typed deterministic expression $e$ of type $\tau$ is denoted $\wellTypedExpr{\Gamma}{e}{\tau}$,
and a well-typed distribution expression $\mu$ with values of type $\tau$ is denoted $\wellTypedExpr{}{\mu}{\tau}$.
Similarly, a well-typed statement is denoted $\wellTypedStmt{\Phi}{\Gamma}{s}$.
A well-typed function function $f$ with inputs $\vec{\tau}$ and outputs $\vec{\tau}'$ is denoted $\wellTypedFun{\Phi}{f}{\vec{\tau}}{\vec{\tau}'}$
We present the typing rules for \ProtoLang{} in \cref{fig:protolang:typing-rules}.

\begin{figure}[t]
\NotationSize{}

\NotationBox[Typing Expressions]{$\wellTypedExpr{\Gamma}{e}{\tau}$}

\[
  \inferrule[TE-Const]
    { v \in \Domain{\tau} }
    {\wellTypedExpr{\Gamma}{v}{\tau}}
  \qquad
  \inferrule[TE-Var]
    {\Gamma[x] = \tau}
    {\wellTypedExpr{\Gamma}{x}{\tau}}
  \qquad
  \inferrule[TE-Op]
  { \wellTypedExpr{}{\BasicOp{n}}{\vec{\tau} \to \tau'}
  \\ \Gamma[\vecx] = \vec{\tau}
  }
  {\wellTypedExpr{\Gamma}{\BasicOp{n}(\vecx)}{\tau'}}
\]

\[
  \inferrule[TE-ArrIndex]
  {\Gamma[x] = \Arr{n}{\tau}
  \\ i \in \{0, \ldots, n - 1\}
  }
  {\wellTypedExpr{\Gamma}{\ArrIndex{x}{i}}{\tau}}
  \qquad
  \inferrule[TE-ArrUpdate]
  {\Gamma[x] = \Arr{n}{\tau}
  \\ i \in \{0, \ldots, n - 1\}
  \\ \Gamma[y] = \tau
  }
  {\wellTypedExpr{\Gamma}{\ArrUpdate{x}{i}{y}}{\Arr{n}{\tau}}}
\]

\NotationBox[Typing Distributions]{$\wellTypedExpr{}{\mu}{\DistrType{\tau}}$}

\[
  \inferrule[TD-Uniform]
  {  }
  {\wellTypedExpr{}{\Uniform_\tau}{\DistrType{\tau}}}
  \qquad
  \inferrule[TD-Bernoulli]
  { 0 \le p \le 1 }
  {\wellTypedExpr{}{\Bernoulli[p]}{\DistrType{\Bool}}}
\]

\NotationBox[Typing Statements]{$\wellTypedStmt{\Phi}{\Gamma}{s}$}

\[
  \inferrule[T-Expr]
    {\Gamma[x] = \tau
    \\ \wellTypedExpr{\Gamma}{e}{\tau}
    }
    {\wellTypedStmt{\Phi}{\Gamma}{x <- e}}
  \qquad
  \inferrule[T-Distr]
    {\Gamma[x] = \tau
    \\ \wellTypedExpr{}{\mu}{\DistrType{\tau}}
    }
    {\wellTypedStmt{\Phi}{\Gamma}{\protosample{x}{\mu}}}
  \qquad
    \inferrule[T-Seq]
    {\wellTypedStmt{\Phi}{\Gamma}{s_1}
    \\ \wellTypedStmt{\Phi}{\Gamma}{s_2}
    }
    {\wellTypedStmt{\Phi}{\Gamma}{s_1;s_2}}
  \qquad
  \inferrule[T-Ifte]
  {  \Gamma[b] = \Bool{}
  \\ \wellTypedStmt{\Phi}{\Gamma}{s_t}
  \\ \wellTypedStmt{\Phi}{\Gamma}{s_f}
  }
  {\wellTypedStmt{\Phi}{\Gamma}{\protoif{b}{s_t}{s_f}}}
\]

\[
    \inferrule[T-FunCall]
    {\wellTypedFun{\Phi,\Gamma}{\Phi[f]}{\vec{\tau}}{{\vec{\tau}'}}
    \\ \Gamma[\vec{x}] = \vec{\tau}
    \\ \Gamma[\vec{y}] = \vec{\tau}'
    }
    {\wellTypedStmt{\Phi}{\Gamma}{\vecy \is f(\vecx)}}
  \qquad
    \inferrule[T-PrimCall]
    { \Gamma[\vec{y}] = \vec{\tau}'
    \\ \wellTypedExpr{}{\cP[\vec\lambda]}{{\vec\tau'}}
    }
    {\wellTypedStmt{\Phi}{\Gamma}{\vecy \is \cP[\vec\lambda]}}
\]

\NotationBox[Typing Partially-applied Function Expressions]{$\wellTypedExpr{}{\lambda}{\vec\tau \to \vec\tau'}$}

\[
  \inferrule[T-PartialFunExpr]
  { \wellTypedFun{}{\Phi[f]}{(\vec\tau;\vec\tau')}{{\vec\tau''}}
  \\ \Gamma[\vecx] = \vec\tau
  }
  { \wellTypedFun{}{f(\vecx, \BlankArg^{*})}{\vec\tau'}{{\vec\tau''}} }
\]

\NotationBox[Typing Functions]{$\wellTypedExpr{}{\cF}{\vec\tau \to \vec\tau'}$}

\[
  \inferrule[T-Fun]
  { \FreeVars{(s)} \subseteq \veca
  \\ \vecr \subseteq \Vars(s)
  \\ \wellTypedStmt{}{}{s}
  \\\\ \Gamma[\veca] = \vec\tau
  \\ \Gamma[\vecr] = \vec\tau'
  }
  { \wellTypedFun{}{\protodef{f}{\veca}{}{s}{\vecr}}{\vec\tau}{\DistrType{\vec\tau'}} }
\]

\NotationBox[Typing Primitive Expressions]{$\wellTypedExpr{}{\cP[\vec\lambda]}{\vec\tau}$}

\[
  \inferrule[T-Primitive]
  { \wellTypedExpr{}{\cP}{(\vec\tau_1 \to \vec\rho_1) \times \cdots \times (\vec\tau_k \to \vec\rho_k) \to \vec\tau}
  \\ \forall i.~ \wellTypedFun{}{\lambda_i}{\vec\tau_i}{\vec\rho_i}
  }
  { \wellTypedExpr{}{\cP[\lambda_1, \ldots, \lambda_k]}{\vec\tau}
  }
\]

\caption{Typing rules for \ProtoLang{} w.r.t.\ function context $\Phi$ and typing context $\Gamma$.}
\label{fig:protolang:typing-rules}
\end{figure}

\subsection{Semantics}
\label{app:protolang:semantics}
We give a probabilistic denotational semantics for \ProtoLang{}.
To do so, we first discuss the state space and the interpretation of \texttt{declare}d functions, and using these, we describe the semantics of program statements.

\paragraph{Values and States}
The set of values that a variable of type $t$ takes is denoted by~$\Domain{t}$.
Similarly, a typing context $\Gamma$ has a value space denoted $\Domain{\Gamma}$ which is the set of labelled tuples of values of each variable in the context, that is $\Domain{\Gamma} = \prod_{x \in \Gamma} \Domain{\Gamma[x]}$.

\paragraph{Denotational Semantics}
The semantics of \ProtoLang{} programs is defined w.r.t an \emph{evaluation context} $\tuple{\Phi, \Gamma, \hat{F}}$:
a tuple consisting of a function context~$\Phi$,
a typing context $\Gamma$,
and an interpretation context~$\hat{F}$ mapping each names of external functions (i.e. $f$ such that $\Phi[f] = \protoext{f}$) to their interpretations $\hat{F}[f] = \hat{f}$.
We use the notation $\evalProb{\cdot}$ for the probabilistic denotational semantics.
\Cref{def:proto:denotational-semantics} describes the full denotational semantics of all language constructs.
For expressions, $\evalExprDet{e}(\sigma)$ denotes the deterministic evaluation of the expression $e$ in state $\sigma$, that is, the value obtained by substituting the values of each variable in $x$ with the value $\sigma(x)$.
For a sequence $s_1;s_2$, we first evaluate $s_1$, and then $s_2$.
For function calls, we extract the function arguments and bind them to the parameter names of the function, evaluate its body, and finally extract the results from the function output and bind them to the variables on the left.

\begin{figure}
\NotationSize{}

\NotationBox[Value Domain of Basic Types]{$\Domain{\tau} \subseteq \Vals{}$}
\begin{align*}
  \Domain{\Fin{N}} &= \{0, \ldots, N-1\}
\\
  \Domain{\BitVec{n}} &= \{0, 1\}^n
\\
  \Domain{\Arr{n}{\tau}} &= \Domain{\tau}^n
\end{align*}

\begin{subfigure}[t]{0.45\textwidth}
\NotationBox[Evaluating Expressions]{$\evalExprDet{e} : \Sigma \to \Domain{\tau}$}
\begin{align*}
  \evalExprDet{v}(\sigma) &= v
\\
  \evalExprDet{x}(\sigma) &= \sigma(x)
\\
 \evalExprDet{\BasicOp{n}(\vecx)}(\sigma) &= \evalExprDet{\BasicOp{n}}(\sigma(\vecx))
\\
  \evalExprDet{\ArrIndex{x}{i}}(\sigma) &= \sigma(x).i
\\
  \evalExprDet{\ArrUpdate{x}{i}{y}}(\sigma) &= \sigma(x)[\sigma(y) / i]
\end{align*}
\end{subfigure}
\hspace{0.08\textwidth}
\begin{subfigure}[t]{0.45\textwidth}
\NotationBox[Evaluating Distributions]{$\evalProb{\mu} : \Distr{\Domain{\tau}}$}
\begin{align*}
  \evalProb{\Uniform{}_\tau} &= \sum_{v \in \Domain{\tau}} \frac1{\abs{\Domain{\tau}}} \detstate{v}
\\
  \evalProb{\Bernoulli{}[p]} &= p \detstate{1} + (1-p)\detstate{0}
\end{align*}
\end{subfigure}

\NotationBox[Evaluating Statements]{$\evalProb{s} : \Sigma \to \Distr{\Sigma}$}
\begin{align*}
  \evalProb{x \is e}(\sigma)
  &= \detstate{\sigma[\evalExprDet{e}(\sigma)/x]}
\\
  \evalProb{\protosample{x}{\mu}}(\sigma)
  &= \DistrExp{v \sim \evalProb{\mu}}{\sigma[v/x]}
\\
  \evalProb{s_1;s_2}(\sigma)
  &= \DistrExp{\evalProb{s_1}(\sigma)}{\evalProb{s_2}{}}
\\
  \evalProb{\protoif{b}{s_t}{s_f}}(\sigma)
  &= {\begin{cases}
    \evalProb{s_t}(\sigma) & \text{if}~\sigma(b) = 1 \\
    \evalProb{s_f}(\sigma) & \text{if}~\sigma(b) = 0
  \end{cases}}
\\
  \evalProb{\vecy \is f(\vecx)}(\sigma)
  &= \DistrExp{\vec{v} \sim \mu}{\sigma[\vec{v} / \vecy ]}
  ~~\text{where}~ \mu = \evalProb{\Phi[f]}(\sigma(\vecx))
\\
  \evalProb{\vecy <- \Primitive{}[\lambda_1, \ldots, \lambda_k]}(\sigma)
  &= \DistrExp{\vec{v} \sim \mu}{\sigma[\vec{v} / \vecy ]}
  ~~\text{where}~ \mu = \evalProb{\Primitive}(\evalProb{\lambda_1}(\sigma), \ldots, \evalProb{\lambda_k}(\sigma))
\end{align*}

\NotationBox[Evaluating Partial Function Expressions]{$\evalProb{\lambda}{} : \Sigma \to \Vals \to \Distr{\Vals}$}
\begin{align*}
  \evalProb{f(\vecx, \BlankArg{}^{*})}(\sigma)(\vecv)
  &= \evalProb{\Phi[f]}(\sigma(\vecx), \vecv)
\end{align*}

\NotationBox[Evaluating Functions]{$\evalProb{\cF}{} : \Vals \to \Distr{\Vals}$}
\begin{align*}
  \evalProb{\protodef{f}{\veca}{}{s'}{\vecr}}(\vecv) &= \evalProb{s'}(\{ \veca : \vecv \})(\vecr)
  \\
  \evalProb{\protoext{f}}(\vecv) &= \detstate{\InterpCtx[f](\vecv)}
\end{align*}

\caption{Denotational semantics of \ProtoLang{}.}
\label{def:proto:denotational-semantics}
\end{figure}

\section{Target Quantum Language \CQPL{}}
\label{app:cqpl}

This appendix contains detailed typing rules and semantics for \CQPL{} omitted in \cref{sec:framework:protolang}.

\subsection{Typing}
\label{app:cqpl:typing}
\CQPL{} is a statically typed language, and we assume each variable has a base type,
and each operator (classical and unitary) and function has a type signature.
The typing judgements are defined w.r.t.\ a global procedure context $\Pi$ and a global typing context $\Gamma$.
The typing rules for \CQPL{} is given in \cref{fig:cqpl:typing:unitary,fig:cqpl:typing:cq,fig:cqpl:typing:unitary:ops}.

\begin{figure}[t]
\NotationSize{}
\NotationBox[Typing Unitary Operators]{$\wellTypedExpr{}{U}{\vec{\tau}}$}

\[
  \inferrule[TU-Gate]
  { U \in \cL(\cH_{\tau}) }
  { \wellTypedExpr{}{U}{\tau} }
  \qquad
  \inferrule[TU-Swap]
  { }
  { \wellTypedExpr{}{\SWAPGate{}}{(\vec{\tau}; \vec{\tau})} }
  \qquad
  \inferrule[TU-Copy]
  { }
  { \wellTypedExpr{}{\COPY}{(\vec{\tau}; \vec{\tau})} }
  \qquad
  \inferrule[TU-Ctrl]
  { \wellTypedExpr{}{U}{\vec{\tau}} }
  { \wellTypedExpr{}{\ctrlU{U}}{(\Bool;\vec{\tau})} }
  \qquad
  \inferrule[TU-Adj]
  { \wellTypedExpr{}{U}{\vec{\tau}} }
  { \wellTypedExpr{}{\texttt{Adj-}{U}}{\vec{\tau}} }
\]
\[
  \inferrule[TU-PhaseOnZero]
  {  }
  { \wellTypedExpr{}{\PhaseOnZero{}{\varphi}}{\vec{\tau}} }
  \qquad
  \inferrule[TU-Expr]
  { \Gamma[\FreeVars{}(e)] = \vec{\tau}
  \\ \wellTypedExpr{\Gamma}{e}{\tau'}
  }
  { \wellTypedExpr{}{U_e}{(\vec{\tau};\tau')} }
  \qquad
  \inferrule[TU-Distr]
  { \wellTypedExpr{}{\mu}{\tau}
  }
  { \wellTypedExpr{}{U_\mu}{\tau} }
\]

\caption{Typing rules for unitary operators in \CQPL{}}
\label{fig:cqpl:typing:unitary:ops}
\end{figure}

\begin{figure}[t]
\NotationSize{}
\NotationBox[Typing Unitary Statements]{$\wellTypedQPL{\Pi}{\Gamma}{w}$}

\[
    \inferrule[TW-Skip]
    {  }
    {\wellTypedQPL{\Pi}{\Gamma}{\qpskip{}}}
    \quad\quad
    \inferrule[TW-Seq]
    {\wellTypedQPL{\Pi}{\Gamma}{w_1}
    \\ \wellTypedQPL{\Pi}{\Gamma}{w_2}}
    {\wellTypedQPL{\Pi}{\Gamma}{w_1;w_2}}
    \quad\quad
    \inferrule[TW-Unitary]
    { \wellTypedExpr{}{U}{\Gamma[\vecq]}
    }
    {\wellTypedQPL{\Pi}{\Gamma}{\qpunitary{\vecq}{U}}}
    \qquad
    \inferrule[TW-Call]
    { \wellTypedExpr{}{\Pi[g]}{\vec{\tau}}
    \\ \Gamma[\vecq] = \vec{\tau}
    }
    { \wellTypedQPL{\Pi}{\Gamma}{\qpcallu{g}{\vecq}}
    ~~~~\text{and}~~~
    \wellTypedQPL{\Pi}{\Gamma}{\qpcalldagger{g}{\vecq}}
    }
\]

\caption{Typing rules for unitary statements in \CQPL{}}
\label{fig:cqpl:typing:unitary}
\end{figure}

\begin{figure}[t]
\NotationSize{}
\NotationBox[Typing Classical Statements]{$\wellTypedQPL{\Pi}{\Gamma}{c}$}

\[
  \inferrule[TC-Skip]
  {  }
  {\wellTypedQPL{\Pi}{\Gamma}{\qpskip{}}}
\quad\quad
  \inferrule[TC-Assign]
  {\wellTypedExpr{\Gamma}{e}{\Gamma[x]}}
  {\wellTypedQPL{\Pi}{\Gamma}{\qpassign{x}{e}}}
\quad\quad
  \inferrule[TC-Random]
  {\wellTypedExpr{}{\mu}{\Gamma[x]}}
  {\wellTypedQPL{\Pi}{\Gamma}{\qprandom{x}{\mu}}}
\quad\quad
    \inferrule[TC-Seq]
  {\wellTypedQPL{\Pi}{\Gamma}{c_1}
  \\ \wellTypedQPL{\Pi}{\Gamma}{c_2}}
  {\wellTypedQPL{\Pi}{\Gamma}{c_1;c_2}}
\]

\[
  \inferrule[TC-Ifte]
  { \Gamma[b] = \Bool{}
  \\ \wellTypedQPL{\Pi}{\Gamma}{c}
  }
  {\wellTypedQPL{\Pi}{\Gamma}{\qpif{b}{c}}}
  \qquad
  \inferrule[TC-Call]
  { \wellTypedExpr{}{\Pi[g]}{\vec{\tau}}
  \\ \Gamma[\vecx] = \vec{\tau}}
  {\wellTypedQPL{\Pi}{\Gamma}{\qpcall{h}{\vecx}}}
  \qquad
  \inferrule[TC-CallMeas]
  { \wellTypedExpr{}{\Pi[g]}{\vec{\tau}}
  \\ \Gamma[\vecx] = \vec{\tau}
  }
  {\wellTypedQPL{\Pi}{\Gamma}{\qpcallandmeas{g}{\vecx}}}
\]

\caption{Typing rules for classical statements in \CQPL{}}
\label{fig:cqpl:typing:cq}
\end{figure}

\subsection{Semantics}
\label{app:cqpl:semantics}
In this section, we provide a denotational semantics for \CQPL{} programs.
We first describe the unitary semantics of unitary statements, followed by the probabilistic semantics of the classical statements (which in turn uses the unitary semantics).
We can also use an operational semantics, and show that the two semantics are equivalent.

\paragraph{External Procedure Interpretations}
Each external \CQPL{} classical procedure $h$ is interpreted by an abstract function $\hat{h} : \Vals \to \Vals$.
Similarly, each external unitary procedure $g$ is interpreted by a unitary operation $U_g \in \cL(\cH)$,
where $\cH$ is the hilbert space with basis indexed by $\Vals$.

\paragraph{Evaluation Context}
The semantics is defined w.r.t.\ an \emph{evaluation context} $\tuple{\Pi, \Gamma, \CInterpCtx{}, \UInterpCtx{}}$,
where
$\Pi$ is a procedure context,
$\Gamma$ is a typing context,
$\CInterpCtx{}$ is a \emph{classical interpretation} context,
mapping a name $h$ of a declared classical procedure to its interpretation $\CInterpCtx{}[h] = \hat{h}$,
and $\UInterpCtx$ is a \emph{unitary interpretation} context,
mapping a name $g$ of a declared unitary procedure to its interpretation $\UInterpCtx{}[g] = U_g$.

\paragraph{Unitary Semantics}
First, we present the semantics of unitary \CQPL{} statements in terms of unitary operators on appropriate Hilbert spaces.
The \emph{denotational semantics} of well-typed unitary statement $w$ is a unitary operator on $\cH$, denoted:
\[
    \evalUQPL{w} \in \cL(\cH)
\]
This is inductively defined in \cref{fig:cqpl:semantics:unitary}.
The semantics of $\qpskip{}$ is given by the identity operator.
The semantics of $\qpunitary{\vec{q}}{U}$ is given by the unitary operator~$U$ acting on quantum variables~$\vecq$ (and as the identity on all other quantum variables).
A sequence statement amounts to the composition of the individual unitaries.
Calling a declared procedure applies the unitary interpretation of the procedure on the input variables.
Calling a defined procedure applies the semantics of the procedure body on the input variables.

\begin{figure}[t]
\NotationSize{}

\NotationBox[Evaluating Unitary Operators]{$\evalUnitaryOp{U}{} \in \cL(\cH)$}

\begin{align*}
  \evalUnitaryOp{U}{} = U
  ~\text{for basic gates}~U
\end{align*}
\begin{align*}
  \evalUnitaryOp{\SWAPGate{}}{} = \sum_{a, b} \ketbra{b, a}{a, b}
  \qquad
  \evalUnitaryOp{\COPY}{} = \sum_{a, b} \ketbra{a, b \oplus a}{a, b}
\end{align*}
\begin{align*}
  \evalUnitaryOp{U_e}{} = U_{\Bracks{e}}
  \qquad
  \evalUnitaryOp{U_\mu}{} = U_{\Bracks{\mu}}
  \qquad
  \evalUnitaryOp{\PhaseOnZero{}{\phi}}{} = I - (1 - e^{i\phi}) \proj0
\end{align*}
\begin{align*}
  \evalUnitaryOp{\texttt{Adj-}{U}}{} = (\evalUnitaryOp{{U}}{})^\dagger
  \qquad
  \evalUnitaryOp{\ctrlU{U}}{} = \proj{0} \ot I + \proj{1} \ot \evalUnitaryOp{U}{}
\end{align*}

\NotationBox[Evaluating Unitary Statements]{$\evalUQPL{w} \in \cL(\cH)$}
\begin{align*}
  \evalUQPL{\qpskip}
    &= I
  \\
  \evalUQPL{\qpunitary{\vec{q}}{U}}
    &= \evalUnitaryOp{U}{\vecq}
  \\
  \evalUQPL{\qpcallu{g}{\vecq}}
    &= \evalUQPL{\Pi[g]}_{\vecq}
  \\
  \evalUQPL{\qpcalldagger{g}{\vecq}}
    &= \parens*{\evalUQPL{\Pi[g]}_{\vecq}}^\dagger
  \\
  \evalUQPL{w_1;w_2}
    &= \evalUQPL{w_2} \circ
       \evalUQPL{w_1}
\end{align*}

\NotationBox[Evaluating Unitary Procedures]{$\evalUQPL{g} \in \cL(\cH)$}
\begin{align*}
  \evalUQPL{\uqplprocdef{g}{\vecx}{w'}}
    &= \evalUQPL{w'}_{\vecx}
  \\
  \evalUQPL{\extuproc{g}}
    &= \UInterpCtx[g]
\end{align*}
\caption{Semantics of unitary fragment \CQPL{} (\cref{def:cqpl:semantics:unitary}).
We can pick any fixed choice of unitary extensions for $U_{\Bracks{e}}$ and $U_{\Bracks{\mu}}$ (cf.\ \cref{eq:U_f,eq:U_mu}).
For example, \cref{eq:strong} is one standard choice for the former.}
\label{fig:cqpl:semantics:unitary}
\label{def:cqpl:semantics:unitary}
\end{figure}

\paragraph{Probabilistic Semantics}
We now define the denotational semantics of classical statements in \CQPL{}, which is given by functions mapping states to distributions of states.
The \emph{probabilistic denotational semantics} of \CQPL{} is defined in \cref{fig:cqpl:semantics:general}.
The statement $\qpskip$ does nothing.
The statement $\qpassign{x}{e}$ updates the state of $x$ with $\Bracks{e}(\sigma)$, while $\qprandom{x}{\mu}$ samples from distribution $\mu$ into $x$.
We use the monadic structure of probability distributions to sequence operations.
In case $\vecx$ has fewer variables than the input arguments of $g$, then we set the remaining inputs to $\ket{0}$.
$\qpifte{b}{s_t}{s_f}$ runs $s_t$ when $b$ is true, otherwise $s_f$.
To evaluate $\qpcallandmeas{g}{\vecx}$, we use the unitary semantics of $g$ and the rules for quantum measurement outcomes.

\begin{figure}[t]
\NotationSize{}
\NotationBox[Evaluating Probabilistic Statements]{$\evalCQPL{s} : \Sigma \to \Distr{\Sigma}$}

\begin{align*}
  \evalCQPL{\qpskip}(\sigma)
  &= \detstate{\sigma}
  \\
  \evalCQPL{\qpassign{x}{e}}(\sigma)
  &= \detstate{\sigma[\Bracks{e}(\sigma)/x]}
  \\
  \evalCQPL{\qprandom{x}{\mu}}(\sigma)
  &= \sum_{v \in \Domain{t}} \Bracks{\mu}(v) \detstate{\sigma[v/x]}
  \\
  \evalCQPL{s_1;s_2}(\sigma)
  &= \DistrExp{\evalCQPL{s_1}(\sigma)}{\evalCQPL{s_2}}
  \\
  \evalCQPL{\qpcall{h}{\vecx}}(\sigma)
    &= \evalCQPL{\Pi[h]}(\sigma(\vecx))
  \\
  \evalCQPL{\qpcallandmeas{g}{\vecx}}(\sigma)
  &= \evalCQPL{\Pi[g]}(\sigma(\vecx))
  \\
  \evalCQPL{\qpifte{b}{s_t}{s_f}}(\sigma)
  &= {\begin{cases}
        \evalCQPL{s_t}(\sigma) & \sigma(b) = 1
    \\  \evalCQPL{s_f}(\sigma) & \sigma(b) = 0
    \end{cases}}
\end{align*}

\NotationBox[Evaluating Probabilistic Procedures]{$\evalCQPL{h} : \Vals \to \Distr{\Vals}$}
\begin{align*}
  \evalCQPL{\cqplprocdef{h}{\vecx}{}{s}}(\vecv)
    &= \evalCQPL{s}(\{ \vecx:\vecv \})(\vecx)
  \\
  \evalCQPL{\extcproc{h}}(\vecv)
    &= \detstate{\CInterpCtx{}[h](\vecv)}
\end{align*}

\caption{Semantics of probabilistic fragment of \CQPL{}}
\label{fig:cqpl:semantics:general}
\end{figure}

\paragraph{Probabilistic Semantics for Unitary Programs}
We also associate a probabilistic semantics to the unitary fragment.
This is defined as the probabilistic function corresponding to the quantum channel (see \cref{def:qchan-of-prob}) obtained by applying the unitary semantics to a zero-initalized auxiliary space, and tracing out the auxiliary space at the end.
We denote this using the shorthands $\evalCQPL{g}$ and $\evalCQPL{w}$ for unitary procedures and statements respectively.
These are defined in \cref{fig:cqpl:semantics:unitary:probabilistic}.

\begin{figure}
\NotationSize{}
\NotationBox[Prob. Semantics of $\SetOfUnitaryCommands{}$]{$\evalCQPL{w} : \Sigma \to \Distr{\Sigma}$}
\begin{align*}
  \evalCQPL{w}(\sigma)(\sigma')
    &= \norm{(\bra{\sigma'} \ot I) \evalUQPL{w} \ket{\sigma, 0}}^2
\end{align*}

\NotationBox[Prob. Semantics of Unitary Procs.]{$\evalCQPL{g} : \Vals{} \to \Distr{\Vals{}}$}
\begin{align*}
  \evalCQPL{g}(\vecv)(\vecv')
    &= \norm{(\bra{\vecv'} \ot I) \evalUQPL{g} \ket{\vecv, 0}}^2
\end{align*}

\caption{Probabilistic semantics of unitary fragment of \CQPL{}}
\label{fig:cqpl:semantics:unitary:probabilistic}
\end{figure}

\subsection{Cost}
\label{cqpl:cost}
In this section, we define the costs of \CQPL{} programs based on the cost model discussed in \cref{sec:cost:model}.
We can also use an instrumented operational cost semantics which is equivalent to the above.

\paragraph{Unitary cost}
We first define the worst-case cost of unitary fragment of \CQPL{} in \cref{fig:cqpl:cost:unitary}.
Built-in unitaries do not incur any cost.
The cost of a sequence of two statements in the sum of their individual costs.
The cost of calling a declared \texttt{uproc} is its tick value, while the cost of calling a defined \texttt{uproc} (or its adjoint) is the cost of the body of the procedure.

\begin{figure}[t]
\NotationSize{}
\begin{subfigure}[t]{0.45\textwidth}
\NotationBox[Unitary Statement Cost]{$\uqplcost{w} : \CostExpr{}$}
\begin{align*}
  \uqplcost{\qpskip{}} &= 0
  \\ \uqplcost{\qpunitary{\vec{q}}{U}} &= 0
  \\ \uqplcost{w_1;w_2} &= \uqplcost{w_1} + \uqplcost{w_2}
  \\ \uqplcost{\qpcallu{g}{\vec{q}}}
  &= \uqplcost{\Pi[g]}
  \\ \uqplcost{\qpcalldagger{g}{\vec{q}}}
  &= \uqplcost{\Pi[g]}
\end{align*}
\end{subfigure}
\hspace{0.09\textwidth}
\begin{subfigure}[t]{0.45\textwidth}
\NotationBox[Unitary Procedure Cost]{$\uqplcost{g} : \CostExpr{}$}
\begin{align*}
  \uqplcost{\uqplprocdef{g}{\vecq}{w'}} &= \uqplcost{w'}
  \\
  \uqplcost{\extuproc{g}} &= \TickOf{g}
\end{align*}
\end{subfigure}
\caption{Cost of unitary statements of \CQPL{}.}
\label{fig:cqpl:cost:unitary}
\end{figure}

\paragraph{Fine-grained cost expressions}
We define a fine-grained notion of cost expressions: mappings of the form $\{ f(\vecv) \mapsto n \}$, which denotes $n$ queries to procedure $f$ with inputs $\vecv$.
The set of fine-grained cost expressions is denoted $\FineGrainedCostExpr{}$.
Similar to cost expressions, we add and compare these expressions point-wise.
These can be converted to ordinary cost expressions $\CostExpr{}$ by dropping the recorded inputs and aggregating the total number of calls to each procedure; we denote this conversion function as $\SimplifyCost : \FineGrainedCostExpr{} \to \CostExpr{}$.
We will use these fine-grained cost expressions to later define the cost specification of primitives~(\cref{spec:quantum:exp-cost}).

\paragraph{Expected quantum cost}
We now define a cost over probabilistic statements and procedures in \CQPL{}.
We call this a \emph{quantum cost}, as the classical statements can invoke unitary procedures.
Unlike for purely unitary programs, this cost can can depend on the state of the program and the function interpretations and the control flow.
Therefore we will define an expected cost for such statements, which maps program states to fine-grained cost expressions.
This is denoted $\FineExpCostName$ and defined inductively in \cref{fig:cqpl:cost:general:fine-grained}.
As before, built-in expressions have zero cost.
The cost of a sequence $c_1;c_2$ on state $\sigma$ is the sum of the cost of $c_1$ on $\sigma$,
and the expectation of the cost of $c_2$ on the output distribution of $c_1$ on $\sigma$.
The cost of a \CallUProcAndMeas{} is the unitary cost (\UQPLCost{}) of the unitary procedure it calls.
The cost of a branch $\qpifte{b}{s_t}{s_f}$ is the cost of $s_t$ on input $\sigma$ when $b = 1$, otherwise the cost of $s_f$ on input $\sigma$.

\begin{figure}[t]
\NotationSize{}
\NotationBox[Fine-grained Expected Cost of Statements]{$\cqplexpcostfine{c} : \Sigma \to \FineGrainedCostExpr{}$}
\begin{align*}
  \cqplexpcostfine{\qpskip{}}(\sigma) &= 0
  \\ \cqplexpcostfine{\qpassign{x}{e}}(\sigma) &= 0
  \\ \cqplexpcostfine{\qprandom{x}{\mu}}(\sigma) &= 0
  \\
  \cqplexpcostfine{c_1;c_2}(\sigma)
    &= \cqplexpcostfine{c_1}(\sigma)
    + \DistrExp{\evalCQPL{c_1}(\sigma)}{\cqplexpcostfine{c_2}}
  \\
  \cqplexpcostfine{\qpcall{h}{\vecx}}(\sigma)
    &= \cqplexpcostfine{\Pi[h]}(\sigma(\vecx))
  \\
  \cqplexpcostfine{\qpcallandmeas{g}{\vecx}}(\sigma)
    &= \uqplcost{\Pi[g]}
  \\
  \cqplexpcostfine{\qpifte{b}{s_t}{s_f}}(\sigma)
  &= \begin{cases}
     \cqplexpcostfine{s_t}(\sigma) & \sigma(b) = 1
  \\ \cqplexpcostfine{s_f}(\sigma) & \sigma(b) = 0
  \end{cases}
\end{align*}
\NotationBox[Fine-grained Expected Cost of Procedures]{$\cqplexpcostfine{h} : \Vals \to \FineGrainedCostExpr{}$}
\begin{align*}
  \cqplexpcostfine{\cqplprocdef{h}{\veca}{}{c'}}(\vecv)
    &= \cqplexpcostfine{c'}(\{ \veca : \vecv \})
  \\
  \cqplexpcostfine{\extcproc{h}}(\vecv) &= \TickOf[{\vecv}]{h}
\end{align*}
\caption{Fine-grained expected cost \FineExpCostName{} for \CQPL{}.}
\label{fig:cqpl:cost:general:fine-grained}
\end{figure}

We also define a expected cost $\CQPLCost{}$ for both functions and statements,
which are obtained by simplifying the fine-grained cost $\FineExpCostName{}$,
as defined in \cref{fig:cqpl:cost:general}.

\begin{figure}[t]
\NotationSize{}
\NotationBox[Expected Cost of Statements]{$\cqplexpcost{c} : \Sigma \to \CostExpr{}$}
\begin{align*}
  \cqplexpcost{s}(\sigma) = \SimplifyCost(\cqplexpcostfine{s}(\sigma))
\end{align*}
\NotationBox[Expected Cost of Procedures]{$\cqplexpcost{h} : \Vals \to \CostExpr{}$}
\begin{align*}
  \cqplexpcost{h}(\vecv) = \SimplifyCost(\cqplexpcostfine{h}(\vecv))
\end{align*}
\caption{Expected cost \ExpCostName{} for \CQPL{}.}
\label{fig:cqpl:cost:general}
\end{figure}

\section{Proofs for Formal Guarantees}\label{app:proofs}

This appendix provides the proofs for our compiler and cost correctness theorems.

\subsection{Compiled Interpretation Contexts}
The source language \ProtoLang{} has externally interpreted functions provided by the mapping $\InterpCtx{}$.
The compilers produce a corresponding \CQPL{} interpretations for these, which are then directly used by the target semantics.

First, to evaluate the external functions compiled by \QuantumCompiler{} (which are \kwbasic{ext proc}s), we construct a classical interpretation context from the source function interpretation context $\InterpCtx{} = \{ f_i : F_i \}_i$.
We denote this $\CInterpCtx{}_{\InterpCtx{}}$, defined as $\CInterpCtx{}_{\InterpCtx{}} := \{ f_i : \hat{h}_i \}_i$, where $\hat{h}_i(\vecv;\vecr) = \vecv; F_i(\vecv)$.

Similarly, to evaluate the external functions compiled by \UnitaryCompiler{} (which are \kwbasic{ext uproc}s), we construct a unitary interpretation context from the source function interpretation context $\InterpCtx{} = \{ f_i : F_i \}_i$.
We denote this $\UInterpCtx{}_{\InterpCtx{}}$, defined as $\UInterpCtx{}_{\InterpCtx{}} := \{ f^U_i : U_i \}_i$, where $U_i$ is a unitary extension of~$F_i$ (see \cref{eq:U_f,def:iso uni ext}).

All our results are stated w.r.t.\ the above implicitly constructed interpretation contexts $\UInterpCtx{}_{\InterpCtx{}}, \CInterpCtx{}_{\InterpCtx{}}$.

\subsection{Specification for Primitives}

We now state the requirements that the compilation of each primitive should satisfy.
We state these as specifications of the semantics and query cost over the unitary and quantum compilations, assuming they are given access to ideal implementations of their function arguments.

We first state the specification on the semantics of the unitary compilation:
\begin{definition}[Correctness of Unitary Primitive Implementation]
\label{spec:unitary:semantics}
Consider a primitive call statement $\vecy <- \cP_\eps[\vec\lambda]$,
where $\lambda_i = f_i(\vecx^{(i)}, \BlankArg{}^{*})$.
Assume we have access to arbitrary perfect implementations $f^U_i$ for each $f_i$, i.e.\ $\evalUQPL{\Pi[f^U_i]}$ is a unitary extension of $\evalProb{\Phi[f_i]}$.
Then the \CQPL{} unitary procedure $\UAlgPrim[{\cP}]{\eps}[f^U_1, \ldots]$
should satisfy:
\[
  \evalUQPL{
    \UAlgPrim[\cP]{\eps}[f^U_1, \ldots]
  }
  ~\text{implements}~
  \evalProb{\vecy <- \cP_{\eps}[\vec\lambda]}
  ~\text{up to error}~ \eps.
\]
\end{definition}

We now state the specification on the unitary query costs:
\begin{definition}[Cost of Unitary Primitive Implementation]
\label{spec:unitary:cost}
Consider a primitive call statement $\vecy <- \cP_\eps[\vec\lambda]$,
where $\lambda_i = f_i(\vecx^{(i)}, \BlankArg{}^{*})$.
Further, let $\Pi[f^U_i] = \extuproc{f^U_i}$.
Then the \CQPL{} unitary procedure $\UAlgPrim[{\cP}]{\eps}[f^U_1, \ldots]$ should satisfy
\[
    \uqplcost{\UAlgPrim[{\cP}]{\eps}[g_1, \ldots]}
    \le \{ f^U_i \mapsto \UAlgPrimQueries{\cP}{\eps}{i} \}_i
\]
\end{definition}

We now state the specification on the semantics of the quantum compilation:
\begin{definition}[Correctness of Quantum Primitive Implementation]
\label{spec:quantum:semantics}
Consider a primitive $\cP_\eps[\vec\lambda]$,
where $\lambda_i = f_i(\vecx^{(i)}, \BlankArg{}^{*})$.
Assume we have access to arbitrary perfect implementations $f^U_i, f_i$ for each $f_i$: 
\[
  \evalCQPL{\qpcallandmeas{f^U_i}{\vecx^{(i)}, \veca^{(i)}, \vecr^{(i)}}}
  = \evalCQPL{\qpcall{f_i}{\vecx^{(i)}, \veca^{(i)}, \vecr^{(i)}}}
  = \evalProb{\vecr^{(i)} <- f(\vecx^{(i)}, \veca^{(i)})}
\]
Then the \CQPL{} classical procedure $\QAlgPrim[{\cP}]{\eps}[f^U_1, f_1, \ldots]$ must satisfy:
\[
  \probDistance{
    \evalCQPL{
      \qpcall{
        \QAlgPrim[\cP]{\eps}[f^U_1, f_1, \ldots]
      }{
        \vecx^{(1)}, \ldots, \vecy
      }
  }
  }{
    \evalProb{\vecy <- \cP_{\eps}[\vec\lambda]}
  }
  \le \eps.
\]
\end{definition}

We now state the specification for the expected cost of the quantum compilation.
As this cost depends on the input, as well as the interpretation of the functions, the specification is also quantified by the same.
\begin{definition}[Expected Cost of Quantum Primitive Implementation]
\label{spec:quantum:exp-cost}
Consider a primitive $\cP_\eps[\vec\lambda]$,
where $\lambda_i = f_i(\vecx^{(i)}, \BlankArg{}^{*})$.
For each $i$, assume $\Pi$ contains external procedures $\extcproc{h_i}$ and $\extuproc{g_i}$ with interpretations that match the semantics of the corresponding source functions~$f_i$:
\[
  \evalCQPL{\qpcallandmeas{g_i}{\vecx^{(i)}, \veca^{(i)}, \vecr^{(i)}}}
  = \evalCQPL{\qpcall{h_i}{\vecx^{(i)}, \veca^{(i)}, \vecr^{(i)}}}
  = \evalProb{\vecr^{(i)} <- f_i(\vecx^{(i)}, \veca^{(i)})}
\]
Then the \CQPL{} classical procedure $\QAlgPrim[{\cP}]{\eps}[g_1, h_1, \ldots]$ should satisfy
\begin{gather*}
  \cqplexpcostfine{
    \qpcall{
      \QAlgPrim[\cP]{\eps}[g_1, h_1, \ldots]
    }{
      \vecx^{(1)}, \ldots, \vecy
    }
  }(\sigma)
  \\ \le
  \{ g_i \mapsto \QAlgPrimQueriesExpU{\cP}{\eps}{i}{\cS} \}_i
  +
  \{
     h_i(\vecv) \mapsto \QAlgPrimQueriesExpQ{\cP}{\eps}{i}{\cS, \vecv}
  \}_{i, \vecv}
\end{gather*}
for every input $\sigma$ satisfying the promise of the primitive,
where $\cS = \{ \evalProb{\lambda_i}(\sigma) \}_i$.
\end{definition}

\subsection{Proofs for Unitary Compilation and Cost Analysis}\label{app:ucost}
This appendix contains the proofs and additional theorems for the unitary compilation $\UnitaryCompiler{}$ and cost analysis $\CostMetricU$.

\subsubsection{Unitary compilation preserves typing}

\begin{restatable}[Unitary Compilation is Well-Typed]{theorem}{ThmUnitaryCompileWellTyped}
\label{thm:uqpl:compile:well-typed}
For every well-typed \ProtoLang{} statement $s$,
its unitary compilation $\compileUQPL{s}$ is well-typed.
\end{restatable}
\begin{proof}
By induction on $s$.
\end{proof}

\subsubsection{Unitary compilation preserves semantics}
To prove that the \UCostCompiler{} preserves semantics as stated in \cref{thm:uqpl:compile:semantics:simpler}, we first prove composable semantics preservation result on statements~(\cref{thm:uqpl:compile:semantics}) and functions~(\cref{thm:uqpl:compile:semantics:fun}).
The proofs of these two theorems are mutually inductive; this is valid because well-formed programs must have finite recursion depth by assumption.

\begin{restatable}[Unitary Error Analysis (Statements)]{theorem}{ThmUnitaryCompilePreservesSemantics}
\label{thm:uqpl:compile:semantics}
For every well-formed \ProtoLang{} statement $s$,
$\evalUQPL{\compileUQPL{s}}$ implements $\evalProb{s}$ up to error $\progerrprobU{s}$.
\end{restatable}
\begin{proof}
\label{proof:ThmUnitaryCompilePreservesSemantics}
We prove this by structural induction on $s$:

\begin{proofcase}{$s = x <- e$}
Then $\compileUQPL{s} = \qpunitary{\FreeVars(e), x'}{U_e}; \qpunitary{x,x'}{\SWAPGate}$.
As $x'$ is a fresh variable, it is zero-initialized.
Therefore $\evalUQPL{\compileUQPL{s}}$ implements $\evalProb{s}$.
\end{proofcase}

\begin{proofcase}{$s = \protosample{x}{\mu}$}
Then $\compileUQPL{s} = \qpunitary{x'}{U_\mu}; \qpunitary{x,x'}{\SWAPGate}; \qpunitary{x,x''}{\COPY{}}$, where $x', x''$ are fresh, zero-initialized variables.
The copy ensures that the channel on $\Sigma \to \Sigma$ is kept classical when discarding the auxiliary variables.
Therefore $\evalUQPL{\compileUQPL{s}}$ implements $\evalProb{s}$.
\end{proofcase}

\begin{proofcase}{$s = \vecy <- f(\vecx)$}
Then $\compileUQPL{s} = \qpcallu{f^U}{\vecx, \vecy', \vecz'};\qpunitary{\vecy, \vecy'}{\SWAPGate}$, where $\vecy', \vecz'$ are fresh, zero-initialized variables.
Let $\eps = \progerrprobU{\Phi[f]} = \progerrprobU{s}$.
Then by induction using \cref{thm:uqpl:compile:semantics:fun}, we know that the unitary $\evalUQPL{\compileUQPL{\Phi[f]}}$ is a $\eps$-close unitary extension of $\evalProb{\Phi[f]}$.
Therefore the compiled program runs $f^U$ on the inputs $\vecx$ and zero-initialized $\vecy', \vecz'$ and swaps in the results.
Therefore it implements $\evalProb{s}$ up to error $\eps$.
\end{proofcase}

\begin{proofcase}{$s = s_1;s_2$}
Then $\compileUQPL{s} = w = w_1; w_2$ where $w_1 = \compileUQPL{s_1}$ and $w_2 = \compileUQPL{s_2}$.
Let $\eps_1 = \progerrprobU{s_1}$ and $\eps_2 = \progerrprobU{s_2}$.
Note that the compilation is \emph{compositional}, therefore always produces separate auxiliary variables; lets call these auxiliary variables $\vecz_1, \vecz_2$.
By the inductive hypothesis we know that $\evalUQPL{w_i}$ implements $\evalProb{s_i}$ up to error $\eps_i$ for every $i \in \{1,2\}$.
By definition, there exists some $F'_i : \Sigma \to \Sigma$ such that $\evalUQPL{w_i}$ implements $F_i$ and $\probDistance{F'_i}{\evalProb{s_i}} \le \eps_i$.
Using the semantics, we have
\[
  \evalUQPL{w} = (\evalUQPL{w_2} \ot I_{\vecz_1}) (\evalUQPL{w_1} \ot I_{\vecz_2}).
\]
therefore $\evalUQPL{w}$ implements $F'_2 \circ F'_1$.
Therefore using \cref{lem:sequence-approx-prob}, $\evalUQPL{w}$ implements $\evalProb{s}$ up to error $\eps_1 + \eps_2 = \progerrprobU{s}$.
\end{proofcase}

\begin{proofcase}{$s = \protoif{b}{s_t}{s_f}$}
The proof is similar to the sequence case.
The total error is only due to the individual compilations $\compileUQPL{s_t}$ and $\compileUQPL{s_f}$.
After each branch, we swap the results out and restore the initial values of the variables in $\Sigma$, ensuring that the following branch runs on the initial classical input.
Finally, we pick the result based on the control bit $b$, and swap it.
\end{proofcase}

\begin{proofcase}{$s = \vecy <- \cP_\eps[\vec\lambda]$}
Let $\lambda_i = f_i(\vecx^{(i)}, \BlankArg{}^{*})$.
Let $w := \compileUQPL{s}$, then
\[
    w =
    \qpcall{\UAlgPrim{\eps}[f^U_1, \ldots]}{\vecx^{(1)}, \ldots, \vecy', \vecz'};~
    \qpunitary{\vecy, \vecy'}{\SWAPGate{}},
\]
where $\vecy', \vecz'$ are fresh zero-initialized variables.
Then by induction using \cref{thm:uqpl:compile:semantics:fun}, the unitary $\evalCQPL{\Pi[f^U_i]}$ is a $\eps_i$-close unitary extension from inputs to outputs of $\evalProb{\Phi[f_i]}$, where $\eps_i = \progerrprobU{\Phi[f_i]}$.

We also know that by the specification of \cref{spec:unitary:cost}, for each $i$, the algorithm at most $\UAlgPrimQueries{\cP}{\eps}{i}$ calls to $f^U_i$ (and its inverse).
Then we can use \cref{thm:quantum-approx-subroutine-subst} repeatedly for each $i$, to replace calls to $f^U_i$ with its ideal implementation, to obtain a total error of
\[
    \eps + \sum_i L_i \sqrt{2 \progerrprobU{f_i}} \le \progerrprob{s}
    \qedhere
\]
\end{proofcase}
\end{proof}

\begin{theorem}[Unitary Error Analysis (Functions)]
\label{thm:uqpl:compile:semantics:fun}
For every well-formed \ProtoLang{} function $f$, and function context $\Phi$,
the unitary semantics of its unitary compilation $\evalUQPL{\compileUQPL{\Phi[f]}}$ is a $(\progerrprobU{\Phi[f]})$-close unitary extension of its probabilistic semantics $\evalProb{\Phi[f]}$.
\end{theorem}
\begin{proof}
There are two cases to consider:

\begin{proofcase}{$\Phi[f] = \protoext{f}$}
Let $F = \InterpCtx{}[f]$.
Then the semantics of the compiled procedure $f^U$ is $\UInterpCtx{}[f^U]$, which by definition is a unitary extension of~$F$, and hence has error~$0$.
\end{proofcase}

\begin{proofcase}{$\Phi[f] = \protodef{f}{\veca}{}{s}{\vecr}$}
Then the compiled unitary procedure is
\[
  \uqplprocdef{f^U}{\veca, \vecr, \vecz}{ \qpunitary{\veca, \veca'}{\COPY{}};~ \compileUQPL{s} }
\]
By \cref{thm:uqpl:compile:semantics}, we know that $\evalUQPL{\compileUQPL{s}}$ implements $\evalProb{s}$ up to error $\progerrprobU{s}$.
Therefore \cref{lemma:unitary:impl-to-extension} proves this case, as $\progerrprobU{s} = \progerrprobU{\Phi[f]}$.
\end{proofcase}
\end{proof}

We now restate and prove that the \UCostCompiler{} preserves semantics~(\cref{thm:uqpl:compile:semantics:simpler}):
\ThmUnitaryCompilePreservesSemanticsSimpler*
\begin{proof}
By \cref{thm:uqpl:compile:semantics}.
\end{proof}

\subsubsection{Correctness of unitary cost analysis}
Compiling a \ProtoLang{} program produces a \CQPL{} program whose cost (\UQPLCost{}) is upper-bounded by the cost function (\CostMetricU{}) of the source program.

\ThmUnitaryCompilePreservesCost*
\begin{proof}
\label{proof:ThmUnitaryCompilePreservesCost}
By induction on $s$.
\end{proof}

\subsection{Proofs for Quantum Compilation and Expected Cost Analysis}\label{app:cost}
This appendix contains the proofs and additional theorems for the unitary compilation $\QuantumCompiler{}$ and cost analyses $\CostMetricQ$ and $\CostMetricHavoc{}$.

\subsubsection{Quantum compilation preserves typing}
\begin{restatable}[\CostCompiler{} preserves typing]{theorem}{ThmQuantumCompileWellTyped}
\label{thm:compile:well-typed}
For every well-typed \ProtoLang{} statement $\wellTypedStmt{}{}{s}$, its quantum compilation is also well-typed $\wellTypedQPL{\Pi}{\Gamma}{\compileCQPL{s}}$.
\end{restatable}
\begin{proof}
By induction on $s$.
\end{proof}

\subsubsection{Quantum compilation preserves semantics}

\ThmQuantumCompilePreservesSemantics*
\begin{proof}
\label{proof:ThmQuantumCompilePreservesSemantics}
We prove this by induction on $s$.

\begin{proofcase}{$s = x <- e$, $s = \protosample{x}{\mu}$}
The compiled program computes exactly the output of the source program, so the distance is $0$.
\end{proofcase}

\begin{proofcase}{$s = \vecy <- f(\vecx)$ and $\Phi[f] = \protoext{f}$}
Then the semantics of the compiled statement is given by $\CInterpCtx{}[f]$,
which by definition is $\CInterpCtx{}[f](\veca;\vecr) = \veca;f(\vecr)$.
Therefore the output states match and the error is zero.
\end{proofcase}

\begin{proofcase}{$s = \vecy <- f(\vecx)$ and $\Phi[f] = \protodef{f}{\veca}{}{s'}{\vecr}$}
Then by the induction hypothesis, the semantics of the body of the compiled proc $f^U$ and the semantics of body of $f$ have a distance of atmost $\progerrprob{f}$, which is equal to $\progerrprob{s}$.
\end{proofcase}

\begin{proofcase}{$s = \vecy <- f(\vecx)$}
For a function call, we simply invoke the function body on the state of the arguments, and substitute them back.
The error in $f$ is the same as its body, and therefore the inequality holds.
An intuitive way to see this is by inlining the function body.
\end{proofcase}

\begin{proofcase}{$s = s_1;s_2$}
Therefore $\compileCQPL{s_1;s_2} = \compileCQPL{s_1}; \compileCQPL{s_2}$.
Then the induction hypotheses are
\[
    \probDistance
        {\evalCQPL{\compileCQPL{s_1}}}
        {\evalProb{s_1}}
    \le \progerrprob{s_1}
    \quad\text{and}\quad
    \probDistance{\evalCQPL{\compileCQPL{s_2}}}{\evalProb{s_2}} \le \progerrprob{s_2}.
\]
Therefore \cref{lem:sequence-approx-prob} upper-bounds the total distance by $\progerrprob{s_1} + \progerrprob{s_2} = \progerrprob{s_1;s_2}$.
\end{proofcase}

\begin{proofcase}{$s = \vecy \is \cP_\eps[\vec\lambda]$}
We use the triangle inequality with an intermediate program that uses a perfect implementation of each function argument. 
The first error term is at most $\eps$, the failure probability of the primitive according to the specification of \cref{spec:quantum:semantics}.
The second term is bounded by the total error of all compiled function calls,
in particular, we use \cref{lemma:error-tv-to-unitary} to bound the error of unitary calls,
and the sequence rule (\cref{lem:sequence-approx-prob}) to sum the individual TV distances.
\end{proofcase}
\end{proof}

\subsubsection{Correctness of quantum expected cost analysis}

We state a useful result that the expected cost is always bounded by the havoc cost analysis.
This guarantees that all our compiled programs are certainly terminating.
\begin{restatable}[]{theorem}{ThmQuantumCompilePreservesMaxCost}
\label{thm:cqpl:compile:max-cost}
For every well-formed \ProtoLang{} statement $s$, and well-formed input $\sigma$,
\[
  \cqplexpcost{\compileCQPL{s}}(\sigma) \le \costqmax{s},
\]
\end{restatable}
\begin{proof}
By induction on $s$.
\end{proof}

We now restate and prove the bound on the expected cost of a quantum program produced by our quantum compiler.
\ThmQuantumCompilePreservesCost*
\begin{proof}
\label{proof:ThmQuantumCompilePreservesCost}
We prove this by induction on $s$.

\begin{proofcase}{$s = x <- e$, $s = \protosample{x}{\mu}$}
Both \CostMetricQ{} and \CQPLCost{} are equal.
\end{proofcase}

\begin{proofcase}{$s = \vecy <- f(\vecx)$ and $\Phi[f] = \protoext{f}$}
Both $\CostMetricQ{}[s]$ and $\CQPLCost{}[s]$ are equal to $\ClassicalTick{f}$,
and $\progerrprob{s}$ is zero.
\end{proofcase}

\begin{proofcase}{$s = \vecy <- f(\vecx)$ and $\Phi[f] = \protodef{f}{\veca}{}{s'}{\vecr}$}
Both \CostMetricQ{} and \CostCompiler{} simply execute the function body with the same parameters,
so this case holds by the induction hypothesis on the function body.
\end{proofcase}

\begin{proofcase}{$s = s_1; s_2$}
Therefore $\compileCQPL{s_1;s_2} = c_1; c_2$ where $c_1 = \compileCQPL{s_1}$, and $c_2 = \compileCQPL{s_2}$.
\newcommand{\Act}[1]{E_{#1}}
\newcommand{\Est}[1]{\hat{E}_{#1}}
\newcommand{\EstMax}[1]{\hat{E}^{\max}_{#1}}
We abbreviate the following common expressions for brevity:
$\Act1(\sigma) := \cqplexpcost{c_1}(\sigma)$,
$\Est1(\sigma) := \costq{s_1}{\sigma}$,
$\Act2(\sigma') := \cqplexpcost{c_2}(\sigma')$,
and 
$\Est2(\sigma') := \costq{s_2}{\sigma'}$.
We abbreviate the worst case costs as $\EstMax1 := \costqmax{s_1}$ and $\EstMax2 := \costqmax{s_2}$.
Also define $\eps_1 = \progerrprob{s_1}$ and $\eps_2 = \progerrprob{s_2}$ and $\eps = \progerrprob{s_1;s_2} = \eps_1 + \eps_2$.
Then from the induction hypothesis we have
\[
  \Act1(\sigma) \le \Est1(\sigma) + \eps_1 \EstMax1
\]
for every $\sigma$,
and
\[
  \Act2(\sigma') \le \Est2(\sigma') + \eps_2 \EstMax2
\]
for every $\sigma'$. 
Also, \cref{thm:cqpl:compile:max-cost} upper-bounds the actual expected costs by the worst case costs:
\[
\Act1(\sigma) \le \EstMax1
\qquad \text{and} \qquad
\Act2(\sigma') \le \EstMax2
\]

We can similarly abbreviate the cost of the sequence as
$\Act{}(\sigma) := \cqplexpcost{c_1;c_2}(\sigma)$, 
$\Est{}(\sigma) := \costq{s_1;s_2}{\sigma}$, and
$\EstMax{} := \costqmax{s_1;s_2}$.
Using the inductive definitions, we can express the cost of the compiled program as
\[
  \Act{}(\sigma) = \Act1(\sigma) + \DistrExp{\evalCQPL{c_1}(\sigma)}{\Act2},
\]
and the cost of the source program as
\[
  \Est{}(\sigma) = \Est1(\sigma) + \DistrExp{\evalProb{s_1}(\sigma)}{\Est2}.
\]
And the worst case cost is simply $\EstMax{} = \EstMax1 + \EstMax2$.
The expected cost of $c_1$ is bounded by
\[
  \Act1(\sigma) 
  \le \Est1(\sigma) + \eps_1 \EstMax1
  \le \Est1(\sigma) + \eps \EstMax1
\]
The expected cost of $c_2$ is bounded using \cref{lem:prob:deviation-of-expectation} as
\begin{align*}
  \DistrExp{\evalCQPL{c_1}(\sigma)}{\Act2}
  &\le 
  \DistrExp{\evalProb{s_1}(\sigma)}{\Act2}
  + \TVDist{\evalCQPL{c_1}(\sigma)}{\evalProb{s_1}(\sigma)} \EstMax2
  \\ &\le 
  \DistrExp{\evalProb{s_1}(\sigma)}{\Act2}
  + \eps_1 \EstMax2
  \\ &\le
  \DistrExp{\sigma' \sim \evalProb{s_1}(\sigma)}{\Est2(\sigma') + \eps_2 \EstMax2}
  + \eps_1 \EstMax2
  \\ &=
  \DistrExp{\evalProb{s_1}(\sigma)}{\Est2}
  + \eps_2 \EstMax2
  + \eps_1 \EstMax2.
  \\ &=
  \DistrExp{\evalProb{s_1}(\sigma)}{\Est2}
  + \eps \EstMax2
\end{align*}
where we used the induction hypotheses and $\eps_1 + \eps_2 = \eps$.
Therefore, by adding the above the bounds for each $c_1$ and $c_2$, we obtain the required inequality.
\end{proofcase}

\begin{proofcase}{$s = \vecy \is \cP_\eps[\vec\lambda]$}
We know that the primitive satisfies the cost specification of \cref{spec:quantum:exp-cost}, and the expected query cost formulas are upper-bounded by the havoc query cost formulas.
Therefore we can use the sequence proof above to repeatedly to bound the cost of each step, and therefore the total cost.
\end{proofcase}
\end{proof}

\section{Primitive $\PrimSimon{}$}
\label{app:prim-simon}
This appendix contains the detailed correctness proofs for the primitive $\PrimSimon{}$.
\Cref{fig:prim:simon} shows the detailed semantics and costs for this primitive.

\begin{figure}[t]
\small
\[ \def\arraystretch{1.5}
\begin{array}{rl}
  \textsc{Syntax.}&
  s \is \PrimSimon{}_{\pcoll}[\lambda]
  \\
  \textsc{Typing.}&
  \vdash \PrimSimon{} : (\BitVec{n} \to \BitVec{n}) \to \BitVec{n}
  \\
  \textsc{Promise.}&
  \textsf{given}~ f : \Domain{\BitVec{n}} \to \Domain{\BitVec{n}}: 
  \exists! s_f \in \Domain{\BitVec{n}} \setminus \{ 0 \} ~~\text{s.th.}~~
  \\ &\qquad
  f(x) = f(x \oplus s_f) ~\forall x \in \Domain{\BitVec{n}}
  \\ &\qquad
  \forall t \in \Domain{\BitVec{n}} \setminus \{0,s_f\}, ~ 
  \abs{\{ x \in \Domain{\BitVec{n}} \mid f(x \oplus t) = f(x) \}} \le \pcoll 2^n
  \\
  \textsc{Semantics.}&
  \evalProb{\PrimSimon_{\pcoll}}[f] = 
      s_f ~~ \text{(as defined above)}
  \\
  \textsc{Cost} (\QuantumCompiler{}).&
  \QAlgPrimQueriesExpU{\PrimSimon{}}{\pcoll, \eps}{}{\cdot}
  = \QAlgPrimQueriesU{\PrimSimon{}}{\pcoll, \eps}{}
  = 2 \cdot \SimonQueries{n}{\pcoll}{\eps}
  \\ &
  \QAlgPrimQueriesExpQ{\PrimSimon{}}{\cdot}{}{}
  = \QAlgPrimQueriesQ{\PrimSimon{}}{\cdot}{}
  = 0
  \\
  \textsc{Cost} (\UnitaryCompiler{}).&
  \UAlgPrimQueries{\PrimSimon{}}{\pcoll, \eps}{}
  = 2\cdot \SimonQueries{n}{\pcoll}{\eps}
\end{array}
\]
\caption{Primitive \PrimSimon{} where $\SimonQueries{n}{\pcoll}{\eps} = \parens*{n + \log_2(1/\eps)} / \log_2(2 / (1 + \pcoll))$.}
\label{fig:prim:simon}
\end{figure}

\subsection{Algorithm.}
We use the algorithm described by \citet{Kaplan2016}.
We run the Simon's subroutine on $f$ for $cn$ times to obtain a collection of vectors, and output a vector that is orthogonal to all of them.

\begin{definition}[Algorithm {$\QAlgPrim[\PrimSimon{}]{\pcoll, \eps}$}]
\label{alg:simon:quantum}
Assume we are given access to a unitary procedure $g$ with arguments $\vecx; x, y; \veca$, where $x, y$ have type $\BitVec{n}$.
Then the procedure $\QAlgPrim[\PrimSimon{}]{\pcoll, \eps}[g](\vecx, y)$ is defined in \cref{fig:algo:qsimon}.
\end{definition}

\begin{figure}
\begin{lstlisting}[language=cqpl]
// function argument
uproc g($\vecx$, x, y, $\veca$) ...

uproc SimonOneRound($\vecx$, x, y, y', $\veca$) do {
  x *= $H^{\ot n}$;
  call g($\vecx$, x, y, $\veca$);
  y, y' *= COPY;
  call$^\dagger$ g($\vecx$, x, y, $\veca$);
  x *= $H^{\ot n}$;
}

proc $\QAlgPrim[\PrimSimon{}]{\pcoll, \eps}$[g]($\vecx$, $y$) do {
  // Q = $Q(n, \pcoll, \eps)$
  for i : {1 ... Q} {
    meas SimonOneRound($\vecx$, u_i);
  }

  // post-processing:
  //     compute a bitstring $s$ that is orthogonal to every $u_i$ computed above,
  //     by solving the system of equations $s \cdot u_i = 0$.
  //     store this result in $y$.
  //     if no solution, then set $y = 0$.
}
\end{lstlisting}
\caption{\CQPL{} program for algorithm {$\QAlgPrim[\PrimSimon{}]{}$}.}
\label{fig:algo:qsimon}
\end{figure}

\begin{definition}[Algorithm {$\UAlgPrim[\PrimSimon{}]{\pcoll, \eps}$}]
\label{alg:simon:unitary}
Assume we are given access to a unitary procedure $g$ with all but the last two arguments bound to some $\vecx$, and auxiliary variables $\veca$,
and the last two arguments of type $\BitVec{n}$ each.
Then the unitary procedure $\UAlgPrim[\PrimSimon{}]{\pcoll, \eps}[g](\vecx, y, \veca')$ is defined in \cref{fig:algo:usimon}.
\end{definition}

\begin{figure}
\begin{lstlisting}[language=cqpl]
// function argument
uproc g($\vecx$, x, y, $\veca$) ...

uproc SimonOneRound($\vecx$, x, y, y', $\veca$) do {
  x *= $H^{\ot n}$;
  call g($\vecx$, x, y, $\veca$);
  y, y' *= COPY;
  call$^\dagger$ g($\vecx$, x, y, $\veca$);
  x *= $H^{\ot n}$;
}

proc $\UAlgPrim[\PrimSimon{}]{\pcoll, \eps}$[g]($\vecx$, y, $\veca$, $\vec{u}$, $\vecz$, $\vecz'$, aux_lin) do {
  // Q = $Q(n, \pcoll, \eps)$
  for i : {1 ... Q} {
    call SimonOneRound($\vecx$, $u_i$, $z_i$, $z'_i$, $\veca$);
  }

  // post-processing (unitarily, using auxiliary memory `aux_lin`):
  //     compute a bitstring $s$ that is orthogonal to every $u_i$ computed above,
  //     by solving the system of equations $s \cdot u_i = 0$.
  //     XOR this result into $y$.
  //     if no solution, then do not update $y$ (i.e. it stays 0)
}
\end{lstlisting}
\caption{\CQPL{} program for algorithm {$\UAlgPrim[\PrimSimon{}]{}$}.}
\label{fig:algo:usimon}
\end{figure}

\subsection{Correctness.}
We use the result from \cite[Theorem 1]{Kaplan2016} which proves that the algorithms above compute $s$ with probability atleast $1 - \eps$.

\begin{theorem}[Correctness of {$\QAlgPrim[\PrimSimon{}]{\pcoll, \eps}$}]
\label{thm:simon:quantum:semantics}
The algorithm {$\QAlgPrim[\PrimSimon{}]{\pcoll, \eps}$} (\cref{alg:simon:quantum}) satisfies the specification of \cref{spec:quantum:semantics}.
\end{theorem}
\begin{proof}
By the result of \cite[Theorem 1]{Kaplan2016}, using $cn$ queries, we obtain a failure probability of at most $\parens*{2 \parens*{\frac{1+\pcoll}{2}}^c}^n$, and we must pick $c$ such that the failure probability is at most $\eps$.
\begin{align*}
  \parens*{2 \parens*{\frac{1+\pcoll}{2}}^c}^n \le \eps
  &\iff 
  \parens*{\frac{1+\pcoll}{2}}^{cn} \le \frac{\eps}{2^n}
  \\ &\iff 
  \parens*{\frac{2}{1+\pcoll}}^{cn} \ge \frac{2^n}{\eps}
  \\ &\iff 
  cn \log_2\parens*{\frac{2}{1+\pcoll}} \ge n + \log_2(1/\eps)
  \\ &\iff 
  cn \ge \frac{n + \log_2(1/\eps)}{\log_2\parens*{\frac{2}{1+\pcoll}}}
\end{align*}
\end{proof}

\begin{theorem}[Correctness of {$\UAlgPrim[\PrimSimon{}]{\pcoll, \eps}$}]
\label{thm:simon:unitary:semantics}
The algorithm {$\UAlgPrim[\PrimSimon{}]{\pcoll, \eps}$} (\cref{alg:simon:unitary}) satisfies the specification of \cref{spec:unitary:semantics}.
\end{theorem}
\begin{proof}
Similar to the proof of \cref{thm:simon:quantum:semantics}.
\end{proof}

\subsection{Complexity.}
We reuse the complexity result from \cite[Theorem 1]{Kaplan2016}:
\begin{theorem}[Query Cost of Primitive \PrimSimon{}]
The algorithms $\UAlgPrim[\PrimSimon]{{\pcoll, \eps}}[g]$ and $\QAlgPrim[\PrimSimon]{{\pcoll, \eps}}[g]$ use at most the following queries to the unitary procedures $g$ and $g^\dagger$ each:
\[
  \SimonQueries{n}{\pcoll}{\eps} =
	\frac{n + \log_2(1/\eps)}{\log_2\parens*{\frac{2}{1+\pcoll}}}
\]
and therefore satisfy the specifications of \cref{spec:unitary:cost,spec:quantum:exp-cost} respectively.
\end{theorem}
\begin{proof}
By the structure of the programs given in \cref{fig:algo:qsimon,fig:algo:usimon}.
\end{proof}

\section{Primitive $\PrimAmplify{}$}\label{app:prim-amplify}
In this appendix we provide the detailed description for primitive \PrimAmplify{}, as well as the correctness proofs.

The primitive \PrimAmplify{} accepts a ``sampling function'' $f$, and a ``minimum solution probability'' $p_\text{min}$.
The function $f$ returns a sample $y$, as well as a boolean flag if the sample is \emph{good} (i.e. satisfies some required condition).
The probability $p_\text{min}$ is a lower-bound on the minimum probability that $f$ outputs a good sample if there is one.
Then the primitive \PrimAmplify{} produces a good sample with probability $1$ if there is one,
and when there are none, it outputs the same as $f$.

\begin{figure}[t]
\small
\[ \def\arraystretch{1.5}
\begin{array}{rl}
  \textsc{Syntax.}&
  y, b \is \protoamplify{\lambda}
  \\
  \textsc{Typing.}&
  \vdash \PrimAmplify{} : (() \to \DistrType{(\tau, \Bool)}) \to \DistrType{(\tau, \Bool)}
  \\
  \textsc{Promise.}&
    \pgood = 0 \lor \pgood \ge \pmin
    ~\text{where}~ 
    \mu = \evalProb{\lambda}(\sigma)(), ~\text{and}~
    \pgood = \Prob{\mu}{({*}, 1)}
  \\
  \textsc{Semantics.}&
  \evalProb{\PrimAmplify_{\pmin}}[\hat{f}] = \begin{cases}
      \frac{\mu_{\mid b=1}}{\pgood} & \pgood \ge \pmin \\
      \mu  & \pgood = 0 
  \end{cases}
  \\ &
  \quad
  \text{where}~ \hat{f} = \evalProb{\lambda}(\sigma), \mu = \hat{f}() ~\text{and}~ \pgood := \Prob{\mu}{({*}, 1)}
  \\
  \textsc{Cost} (\QuantumCompiler{}).&
  \QAlgPrimQueriesExpU{\PrimAmplify{}}{\pmin, \eps}{}{\hat{f}}
  = 2 \cdot E_\QSearch{}(\frac1{\pmin}, \frac{\pgood}{\pmin}, \eps)
  \\ &
  \QAlgPrimQueriesU{\PrimAmplify{}}{\pmin, \eps}{}
  = 2 \cdot W_\QSearch{}(\frac1{\pmin}, \eps)
  \\& \qquad\text{where}~
    E_\QSearch{}(N, T, \eps) = \begin{cases}
        F(N, T)\parens*{1 + \frac1{1 - \frac{F(N, T)}{\alpha\sqrt{N}}}} & T > 0
    \\  W_\QSearch{}(N, \eps) & T = 0
    \end{cases},
  \\&\qquad \text{\phantom{where}}~
    F(N, T) = \begin{cases}
        \frac{\alpha \sqrt{N}}{3 \sqrt{T}}  & T < N/4
    \\  2.0344 & T \ge N/4
    \end{cases},
  \\&\qquad \text{\phantom{where}}~
    W_\QSearch{}(N, \eps) = \alpha \ceil{\log_3(1/\eps)} \sqrt{N},
  \\&\qquad \text{\phantom{where}}~
    \alpha = 9.2.
  \\ &
  \QAlgPrimQueriesExpQ{\PrimAmplify{}}{\pmin}{\eps}{\cdot}
  = \QAlgPrimQueriesQ{\PrimAmplify{}}{\pmin}{\eps}
  = 0
  \\
  \textsc{Cost} (\UnitaryCompiler{}).&
  \UAlgPrimQueries{\PrimAmplify{}}{\pmin, \eps}{}
  = 2l + 1 
  \\&\qquad \text{where}~
    l = \ceil*{\frac12 \frac{\arcosh(1/\sqrt{\eps})}{\arcosh(1/\sqrt{1-\pmin})} - \frac12}
\end{array}
\]
\caption{Primitive \PrimAmplify{}}
\label{fig:prim:amplify}
\end{figure}

\subsection{Unitary Algorithm and Proofs}

We use the \emph{fixed-point amplitude amplification} algorithm by \citet{yoder14FPAA} to implement the unitary compilation of \PrimAmplify{}.

\begin{definition}[Algorithm {$\UAlgPrim[{\PrimAmplify}]{\pmin}$}]
\label{alg:amplify:unitary}
The \CQPL{} unitary procedure $\UAlgPrim[{\PrimAmplify}]{\pmin,\eps}[g]$ is defined in \cref{fig:algo:amplify:unitary}.
\end{definition}

\begin{figure}
\begin{lstlisting}[language=cqpl]
// function argument
uproc g($\vecx$, y, b, $\veca$) ...

uproc $\UAlgPrim[{\PrimAmplify}]{\pmin,\eps}$[g]($\vecx$, y, b, $\veca$) do {
  call g($\vecx$, y, b, $\veca$);

  for i in {1 ... l} {
    b *= Z($\beta_i$);

    call$^\dagger$ g($\vecx$, y, b, $\veca$);
    $\vecx$, y, b, $\veca$ *= PhaseOnZero($-\alpha_j$);
    call g($\vecx$, y, b, $\veca$);
  }
}
\end{lstlisting}
\caption{\CQPL{} program for algorithm {$\UAlgPrim[\PrimAmplify{}]{\pmin, \eps}$},
where 
$l = \ceil{\tfrac12 \arcosh(1/\sqrt{\eps}) / \arcosh(1/\sqrt{1-\pmin}) - \tfrac12}$,
$L = 2l + 1$,
$\gamma^{-1} = T_{1/L}(1 / \sqrt{\eps})$, and
$\alpha_j = -\beta_{l + 1 - j} = 2 \cot^{-1}(\tan(2\pi j / L) \sqrt{1 - \gamma^2})$.}
\label{fig:algo:amplify:unitary}
\end{figure}

\begin{theorem}[Correctness of {$\UAlgPrim[\PrimAmplify{}]{\pmin, \eps}$}]
\label{thm:amplify:unitary:semantics}
The algorithm {$\UAlgPrim[\PrimAmplify{}]{\pmin, \eps}$} (\cref{alg:amplify:unitary}) satisfies the semantic specification of \cref{spec:unitary:semantics}.
\end{theorem}
\begin{proof}
We use the result of \citet{yoder14FPAA}.
Pick $\delta=\sqrt{\eps}$.
This gives us $\gamma^{-1} = T_{1/L}(1/\sqrt{\eps})$, but we know $\gamma \ge \sqrt{1 - \pmin}$.
Therefore $T_{1/L}(1/\sqrt{\eps}) \le 1/\sqrt{1-\pmin}$.
We can now use the definition $T_{1/L}(x) = \arcosh(\frac1L \cosh(x))$ when $x \ge 1$ to obtain the required lower bound on $L$:
\[
  L = 2l + 1 \ge \frac{\arcosh(1/\sqrt{\eps})}{\arcosh(1/\sqrt{1-\pmin})}.
\]
\end{proof}
For an intuition, we see that for small $\eps$, and $\pmin = 1/N$ for some large $N$, we have $L \approx \sqrt{N} \ln(2/\eps)$.

\begin{theorem}[Cost of {$\UAlgPrim[\PrimAmplify{}]{\pmin, \eps}$}]
\label{thm:amplify:unitary:cost}
The algorithm {$\UAlgPrim[\PrimAmplify{}]{\pmin, \eps}$} (\cref{alg:amplify:unitary}) satisfies the cost specification of \cref{spec:unitary:cost}.
\end{theorem}
\begin{proof}
By the structure of the program in \cref{fig:algo:amplify:unitary}.
\end{proof}

\subsection{Quantum Algorithm and Proofs}
For the quantum algorithm, we use a modified version of the quantum search by \citet{boyer98qsearch}.

\begin{definition}[Algorithm {$\QAlgPrim[{\PrimAmplify}]{\pmin}$}]
\label{alg:amplify:quantum}
The \CQPL{} unitary procedure $\QAlgPrim[{\PrimAmplify}]{\pmin,\eps}[g]$ is defined in \cref{fig:algo:amplify:quantum}.
\end{definition}

\begin{figure}
\begin{lstlisting}[language=cqpl]
// function argument
uproc g($\vecx$, y, b, $\veca$) ...

// run k grover iterations
uproc grover$_k$($\vecx$, y, b, $\veca$) do {
  call g($\vecx$, y, b, $\veca$);
  repeat $k$ {
    b *= Z;

    call$^\dagger$ g($\vecx$, y, b, $\veca$);
    $\vecx$, y, b, $\veca$ *= PhaseOnZero($\pi$); // reflect about $\ket{0}$ (with global phase -1)
    call g($\vecx$, y, b, $\veca$);
  }
}

uproc $\QAlgPrim[{\PrimAmplify}]{\pmin,\eps}$[g]($\vecx$, y, b) do {
  not_done := 0 : Bool;
  repeat $N_\text{runs}$ {
    Q_sum := 0 : Fin<$Q_\text{max}$>;
    for j_lim : Fin<$Q_\text{max}$> in $\vec{J}$ {
      j :=$\texttt{\$}$ [1 .. j_lim] : Fin<$Q_\text{max}$>;
      Q_sum := Q_sum + j;
      not_done := not_done and (Q_sum <= j_lim);
      if (not_done) {
        meas grover$_j$($\vecx$, y, b); // run the grover iterations
        not_done := not_done and (not b);
      }
    }
  }
}
\end{lstlisting}
\caption{\CQPL{} program for algorithm {$\QAlgPrim[\PrimAmplify{}]{\pmin, \eps}$},
where 
$N_\text{runs} := \ceil{\log_3(1/\eps)}$,
$Q_\text{max} := \ceil{\alpha \sqrt{1/\pmin}}$ where $\alpha=9.2$.
We also have a finite list of iteration lengths $\vec{J} = \{ \floor{\min(\lambda^k m, \sqrt{1/\pmin})} \mid k \in \N^{+}_0 \}$, truncated to a total of $Q_\text{max}$,
where $\lambda = m = 6/5$.}
\label{fig:algo:amplify:quantum}
\end{figure}

\begin{theorem}[Correctness of {$\QAlgPrim[\PrimAmplify{}]{\pmin, \eps}$}]
\label{thm:amplify:quantum:semantics}
The algorithm {$\QAlgPrim[\PrimAmplify{}]{\pmin, \eps}$} (\cref{alg:amplify:quantum}) satisfies the specification of \cref{spec:quantum:semantics}.
\end{theorem}
\begin{proof}
We adapt the proof of Lemma 4 of \citet{Cade2023quantifyinggrover},
by replacing Hadamards by the unitary compilation of the sampler, and adapt the query formulas to parameters based on the sampler.
We pick the parameters $N = \frac1{\pmin}$ and $T = \frac{\pgood}{\pmin}$, to obtain a good sample with probability at least $1 - \eps$.
Intuitively, $\pmin$ is the minimum probability of finding a solution, which in the case of search is $1/N$.
And $\pgood$ is the fraction of good samples, which for search is $K/N$ (where $K$ is the number of solutions).
\end{proof}

\begin{theorem}[Cost of {$\QAlgPrim[\PrimAmplify{}]{\pmin, \eps}$}]
\label{thm:amplify:quantum:cost}
The algorithm {$\QAlgPrim[\PrimAmplify{}]{\pmin, \eps}$} (\cref{alg:amplify:quantum}) satisfies the cost specification of \cref{spec:quantum:exp-cost}.
\end{theorem}
\begin{proof}
By the proof of Lemma 4 of \citet{Cade2023quantifyinggrover}.
The factor of $2$ is due to uncomputation, as the program calls $g$ and $g^\dagger$ for each grover iteration.
\end{proof}

\section{Primitives $\PrimAny{}, \PrimAll{}, \PrimSearch{}$}\label{app:search-prims}

We implement the search-like primitives (\PrimAny{}, \PrimAll{}, \PrimSearch{}) by desugaring them to \PrimAmplify{}~(\cref{app:prim-amplify}).
In each case, we define a sampling function that draws a uniformly random element from $\tau$, and evaluates the predicate on it, and returns both the sample and the predicate output.
We can then use amplify on this sampling function to obtain a good sample (i.e. satisfying the predicate) with probability $1 - \eps$.
For ease of exposition, we describe their semantics and costs directly in \cref{fig:prim:search-like}, which are obtained from the semantics and costs of \PrimAmplify{}.

\begin{figure}[t]
\renewcommand{\arraystretch}{1.5}
\small
\begin{tabular}{rccc}
  \textsc{Primitive} ($\cP$) &
  \PrimAny & 
  \PrimAll &
  \PrimSearch \\
  \toprule

  \textsc{Syntax.} &
  $b <- \PrimAny[\lambda]$ &
  $b <- \PrimAll[\lambda]$ &
  $b, y <- \PrimSearch[\lambda]$ \\

  \textsc{Typing.} &
  $(\tau \to \Bool) \to \Bool$ &
  $(\tau \to \Bool) \to \Bool$ &
  $(\tau \to \Bool) \to \DistrType{(\Bool, \tau)}$ \\

  \textsc{Promise.}&
  \multicolumn{3}{c}{
    Function argument $\lambda$ is deterministic.
  }
  \\
  \midrule

  \textsc{Semantics.}&
  {$\begin{cases}
    \detstate{0} & \hat{f}(x) = \detstate{0} ~ \forall x \in \Domain{\tau}
    \\
    \detstate{1} & \text{else}
    \end{cases}
  $}
  &
  {$\begin{cases}
    \detstate{1} & \hat{f}(x) = \detstate{1} ~ \forall x \in \Domain{\tau}
    \\
    \detstate{0} & \text{else}
    \end{cases}
  $}
  &
  $\frac{1}{\abs{S}} \sum_{x \in S} \detstate{(b, x)}$
  \\
  &&& 
  \makecell[l]{
    where $b = \begin{cases}
    0 & \hat{f}(x) = \detstate{0} ~ \forall x \in \Domain{\tau}
    \\
    1 & \text{else}
    \end{cases}$
    \\ and   $S = S_{\hat{f}} = \{ x \in \Domain{\tau} \mid \hat{f}(x) = \detstate{b} \}$
  }
  \\
  \midrule

  \textsc{Cost (\QuantumCompiler{}).}&
  \multicolumn{3}{c}{\makecell[c]{
    $\QAlgPrimQueriesExpU{\cP}{\eps}{}{\hat{f}}
    = 2 \cdot E_\QSearch{}(\abs{\Domain{\tau}}, \abs{S_{\hat{f}}}, \eps)$~(\cref{fig:prim:amplify})
    \\[1.5ex]
    $\QAlgPrimQueriesU{\cP{}}{\eps}{}
    = 2 \cdot W_\QSearch{}(\abs{\Domain{\tau}}, \eps)$~(\cref{fig:prim:amplify})
    \\[1.5ex]
    $\QAlgPrimQueriesExpQ{\cP}{\eps}{}{\cdot}
    = \QAlgPrimQueriesQ{\cP}{\eps}{}
    = 0$
  }}
  \\
  \midrule

  \textsc{Cost (\UnitaryCompiler{}).}&
  \multicolumn{3}{c}{
    $\UAlgPrimQueries{\cP{}}{\eps}{} = \UAlgPrimQueries{\PrimAmplify{}}{p,\eps}{}$~(\cref{fig:prim:amplify})
    where $p = 1 / \abs{\Domain{\tau}}$.
  }
  \\
  \bottomrule
\end{tabular}
\vspace{0.5cm}
\caption{Search-like Primitives. The query costs use the same formulas used by \PrimAmplify{}~(\cref{fig:prim:amplify}).}
\label{fig:prim:search-like}
\end{figure}

\paragraph{Desugaring \PrimSearch{}}
The statement $b, y <- \PrimSearch[f(\vecx, \BlankArg{})]$ desugars to the following \ProtoLang{} program:
\begin{align*}
  & \protodef{f'}{\vecx}{}{
    \protosample{y}{\Uniform{}_\tau};~
    b <- f(\vecx, y)
  }{y, b}
\\& y, b <- \PrimAmplify_{1/\abs{\Domain{\tau}}}[f'(\vecx)]
\end{align*}

\paragraph{Desugaring \PrimAny{}}
The statement $b <- \PrimAny[f(\vecx, \BlankArg{})]$ desugars to the following \ProtoLang{} program:
\begin{align*}
  & \protodef{f'}{\vecx}{}{
    \protosample{y}{\Uniform{}_\tau};~
    b <- f(\vecx, y)
  }{y, b}
\\& \protodef{f_a}{\vecx}{}{
    y, b <- \PrimAmplify_{1/\abs{\Domain{\tau}}}[f'(\vecx)]
  }{b}
\\& b <- f_a(\vecx)
\end{align*}

\paragraph{Desugaring \PrimAll{}}
To implement $\PrimAll{}$, we look for an element that \emph{does not satisfy} the predicate.
If we find such an element, the output will be $0$, and otherwise $1$.
Therefore statement $b <- \PrimAll[f(\vecx, \BlankArg{})]$ desugars to the following \ProtoLang{} program:
\begin{align*}
  & \protodef{f'}{\vecx}{}{
    \protosample{y}{\Uniform{}_\tau};~
    b <- f(\vecx, y);~
    b' <- \NotOp{}~ b
  }{y, b'}
\\& \protodef{f_a}{\vecx}{}{
    y, b' <- \PrimAmplify_{1/\abs{\Domain{\tau}}}[f'(\vecx)];~
    b <- \NotOp{}~ b'
  }{b}
\\& b <- f_a(\vecx)
\end{align*}

\section{Comparing Quantum and Classical Search}
\label{app:classical-search}
This section describes additional variants for search with their detailed costs.
These variants have the same syntax, typing and semantics as \PrimAny{}.

\subsection{Deterministic classical search}
The primitive $\PrimCAny{}$ implements search by a linear scan.
\Cref{alg:any:det} in \cref{fig:classical search} describes the quantum compilation $\QAlgPrim{\PrimCAny{}}$,
and 
\Cref{alg:any:ubrute} describes the unitary compilation $\UAlgPrim{\PrimCAny}$.
The query cost equations are described in \cref{fig:prim:any:det}.

\begin{figure}[t]
\small
\[ \def\arraystretch{1.5}
\begin{array}{rl}
  \textsc{Cost} (\QuantumCompiler{}).&
  \QAlgPrimQueriesExpU{\PrimCAny{}}{\eps}{}{\cdot}
  = \QAlgPrimQueriesU{\PrimCAny{}}{\eps}{}
  = 0
  \\ &
  \QAlgPrimQueriesExpQ{\PrimCAny{}}{\eps}{}{\hat{f}, v} = 1
  \\ &
  \QAlgPrimQueriesQ{\PrimCAny{}}{\eps}{} = N
  \\
  \textsc{Cost} (\UnitaryCompiler{}).&
  \UAlgPrimQueries{\PrimCAny{}}{\eps}{}
  = N
\end{array}
\]
\caption{Cost of primitive \PrimCAny{} where $N = \abs{\Domain{\tau}}$.}
\label{fig:prim:any:det}
\end{figure}

\subsection{Randomized classical search}
The primitive \PrimRAny{} implements a randomized search by sampling with replacement, with a cut-off.
\Cref{alg:any:rand} describes the quantum compilation $\QAlgPrim{\PrimRAny{}}$.
and \cref{alg:any:ubrute} describes the unitary compilation $\QAlgPrim{\PrimCAny{}}$.
The query costs equations are described in \cref{fig:prim:any:rand}.

\begin{figure}[t]
\small
\[ \def\arraystretch{1.5}
\begin{array}{rl}
  \textsc{Cost} (\QuantumCompiler{}).&
  \QAlgPrimQueriesExpU{\PrimRAny{}}{\eps}{}{\cdot}
  = \QAlgPrimQueriesU{\PrimRAny{}}{\eps}{}
  = 0
  \\ &
  \QAlgPrimQueriesExpQ{\PrimRAny{}}{\eps}{}{\hat{f}, v}
  = \begin{cases}
    \ceil{\ln(1/\eps)} & \text{if}~ K = 0
  \\
    \frac1{K} & \text{if}~ K > 0 ~\text{and}~ \hat{f}(v) = \detstate{1}
  \\
    \frac{N}{K (N-K)} & \text{if}~ K > 0 ~\text{and}~ \hat{f}(v) = \detstate{0}
  \end{cases}
  \\ &\quad\text{where}~ K = \abs{\{v \in \Domain{\tau} \mid \hat{f}(v) = \detstate{1} \}}
  \\ &
  \QAlgPrimQueriesQ{\PrimRAny{}}{\eps}{}
  = N \ceil{\ln(1/\eps)}
  \\
  \textsc{Cost} (\UnitaryCompiler{}).&
  \UAlgPrimQueries{\PrimRAny{}}{\eps}{}
  = N
\end{array}
\]
\caption{Cost of primitive \PrimRAny{} where $N = \abs{\Domain{\tau}}$.}
\label{fig:prim:any:rand}
\end{figure}

\paragraph{Expected Complexity.}
For a space of size $N$ and failure probability $\eps$, the cut-off is $\mathsf{Q}_{\max} = N \ceil{\ln(1/\eps)}$.
We derive the expected number of samples $\mathsf{Q}$ in the case there are $K$ solutions,
using the indicators for failing after $t$ samples (meaning sample $t + 1$ is needed):
\begin{align*}
  \mathbb{E}(\mathsf{Q})
  = \sum_{t = 0}^{\mathsf{Q}_{\max} - 1} \parens*{1 - \frac K N}^{t}
  = \frac{1 - (1 - K/N)^{\mathsf{Q}_{\max}}}{K/N}
  = \frac{N}{K} \parens*{1 - \parens*{1 - \frac{K}{N}}^{\mathsf{Q}_{\max}}}
\end{align*}

Using $\parens*{1 - p}^{1/p} \le 1/e$ (for $0 < p < 1$),
we can bound the expected queries as
\[
  \frac{N}{K}(1 - \eps^K) \le
  \mathbb{E}(\mathsf{Q})
  \le \frac{N}{K}
\]
Therefore we make an expected $N/K$ queries to non-solutions, and one query to a solution.
As each solution is equally likely (i.e.\ indistinguishable), the expected number of queries to each solution is $1/K$.
Similarly, each non-solution is also equally likely to be sampled, and therefore is queried an expected $N/(K (N-K))$ times.
If there are no solutions, then we sample each element an expected $\ceil{\ln(1/\eps)}$ times.

\begin{figure}[t]
  \begin{subfigure}[t]{0.48\textwidth}
\begin{lstlisting}[language=cqpl, label={alg:any:det}, caption={Deterministic classical search}]
proc DetAny[N, g]($\veca : \vec{\tau}$, $b$: Bool) {
  b := 0;
  for x in Fin<N> {
    if (b = 0) {
      call g($\veca$, x, b);
    }
  }
}
\end{lstlisting}
\begin{lstlisting}[language=cqpl, label={alg:any:rand}, caption={Randomized classical search}]
proc RandAny[N, g, $\eps$]($\veca : \vec{\tau}$, $b$: Bool) {
  repeat $N \ceil{\ln(1/\eps)}$ {
    if (b = 0) {
      x :=$\texttt{\$}$ Fin<N>;
      call g($\veca$, x, b);
    }
  }
}
\end{lstlisting}
  \end{subfigure}
  \begin{subfigure}[t]{0.48\textwidth}
\begin{lstlisting}[language=cqpl, label={alg:any:ubrute}, caption={Unitary classical search}]
uproc UClassicalAny[N, g]
  (qs, $b$: Bool, x: Fin<N>,
   $\{ b_i : \Bool \mid i \in [N] \}$, auxg) {
  with {
    for #i in Fin<N> {
      with { x *= U[() => #i]; }
      do {
        call g(qs, x, $b_{\#i}$, auxg);
      }
    }
  } do {
    $b_0, \ldots, b_{N-1}, b$ *= U[($\vec{\alpha}$) => OR_N($\vec{\alpha}$)];
  }
}
\end{lstlisting}
  \end{subfigure}
\caption{\CQPL{} programs for the various classical search algorithms.}
\label{fig:classical search}
\end{figure}

\section{Implementation}
\label{app:Implementation}

This appendix provides a more detailed exposition of the features of our Haskell implementation.

\subsection{Extensibility}
\OurFramework{} supports adding new primitives with ease.
We enable this by implementing a growing polymorphic AST inspired by prior Haskell work on extensible ASTs~\cite{datatypesalacarte2008,treesthatgrow2017}:
\begin{lstlisting}[style=haskellstyle, numbers=none]
    data Expr $\cP$ = ... | PrimCall $\cP$                  data Program $\cP$ = ...
\end{lstlisting}
where $\cP$ is the current extension in use.
An extension could be a single primitive or a collection of primitives, or some annotated (collection of) primitive.

\paragraph{Typeclasses.}
We use \textit{typeclasses} to allow adding functionality to each new primitive, and generics to automatically derive functionality for a collection of primitives.
For example, to provide the semantics, we use a typeclass
\begin{lstlisting}[style=haskellstyle, numbers=none]
    class Eval $\cP$ where eval :: (MonadEval $\cP'$ m, Eval $\cP'$) => $\cP$ -> $\Sigma$ -> m $\Sigma$
\end{lstlisting}
where \texttt{MonadEval} states that an underlying program of extension $\cP'$ is being evaluated in monad \texttt{m}.
We use generics to automatically derive instances for sum types.

\paragraph{Cost Analysis.}
We introduce type classes \texttt{UCost}, \texttt{HavocCost}, \texttt{ExpCost} for source-level cost analysis,
We describe the expected cost class below.
\begin{lstlisting}[style=haskellstyle, numbers=none]
    class (Eval $\cP$, UnitaryCost $\cP$, HavocCost $\cP$) => ExpCost $\cP$ where expCost :: $\cP$ -> $\Sigma$ -> Cost
\end{lstlisting}

To perform a cost analysis, we first annotate each primitive with a failure probability $\eps$. We use a new data to wrap our primitive, and extend give a simple polymorphic instance to pass the evaluation through the annotation:
\begin{lstlisting}[style=haskellstyle, numbers=none]
    data AnnFail $\cP$ = AnnFail Double $\cP$
    instance (Eval $\cP$) => Eval (AnnFail $\cP$) where eval (AnnFail _ prim) $\sigma$ = eval prim
\end{lstlisting}
and provide a cost analysis for our primitive using the an instance of \texttt{ExpCost}:
\begin{lstlisting}[style=haskellstyle, numbers=none]
    instance (Eval $\cP$) => ExpCost (AnnFail $\cP$) where expCost (AnnFail $\eps$ p) $\sigma$ = ...
\end{lstlisting}

\paragraph{Compilation.}
We also implement the language \CQPL{} in our package,
and provide typeclasses to for defining the compositional compilers \UnitaryCompiler{} and \QuantumCompiler{}.
\begin{lstlisting}[style=haskellstyle, numbers=none]
    class CompileU $\cP$ where compileU :: (MonadCompile m) => $\cP$ -> m UProc
    class (CompileU $\cP$) => CompileQ $\cP$ where compileQ :: (MonadCompile m) => $\cP$ -> m CProc
\end{lstlisting}
and define instances for compiling annotated primitives to produce the corresponding algorithm:
\begin{lstlisting}[style=haskellstyle, numbers=none]
    instance CompileU (AnnFail $\cP$) where ...
\end{lstlisting}

\subsection{Analysing and Optimizing Errors}

We considered programs above where each primitive was annotated with its (maximum allowed) failure probability.
Then our framework uses these annotations to compile as well as perform cost analyses of such programs.

\paragraph{Error Analysis.}
We provide typeclasses \texttt{FailProb} and \texttt{NormError} for source-level error analyses $\ErrProb$ and $\ErrProbU{}$ respectively.
\begin{lstlisting}[style=haskellstyle,numbers=none]
    instance FailProb (AnnFail $\cP$) where failProb (AnnFail $\eps$ p) = $\eps$ + ...
    instance FailProbUnitary (AnnFail $\cP$) where failProbU (AnnFail $\eps$ p) = $\eps$ + ...
\end{lstlisting}

\paragraph{Optimizing using Symbolic Errors.}
Often, we have a total error budget, and would like to choose the individual errors to satisfy it.
This can be tedious to compute by hand, but \OurFramework{} can automate such a task by using a multi-stage analysis:
\begin{enumerate}
  \item Annotate the program with symbolic error-budgets, using symbols $\eps_i$ for each primitive.
  \item Use the symbolic error analysis to compute an expression for the total error $\eps$.
  \item Solve for individual $\eps_i$ such that $\eps$ is bounded by the error-budget.
  \item Substitute back the concrete values for $\eps_i$ into the program.
\end{enumerate}
To allow this, we polymorphize our annotation to allow an arbitrary failure probability type, and generalize our instance to support any floating point.
\begin{lstlisting}[style=haskellstyle, numbers=none]
    data AnnFail prob $\cP$ = AnnFail prob $\cP$
    instance (Floating prob) => FailProb (AnnFail prob $\cP$) where failProb (AnnFail $\eps$ p) = $\eps$ + ...
    instance (Floating prob) => HavocCost (AnnFail prob $\cP$) where failProb (AnnFail $\eps$ p) = ...
\end{lstlisting}

We then use this analysis with a symbolic probability type \texttt{Sym Double} to first compute an expression for the overall error for a program,
and substitute the final epsilons back using helper functions:
\begin{lstlisting}[style=haskellstyle, numbers=none]
    annSym :: Program $\cP$ -> Program (AnnFail (Sym Double) $\cP$)
    subst :: (Sym Double -> Double) -> Program (AnnFail (Sym Double) $\cP$) -> Program (AnnFail Double $\cP$)
\end{lstlisting}

Our current implementation supports a few basic strategies to split the error budgets equally to each primitive call and its arguments.
We leave it to future work to compute better epsilons by solving the constraints subject to minimizing cost expressions, for example using a solver.

\end{document}